\documentclass[fleqn,usenatbib]{mnras}

\usepackage{newtxtext,newtxmath}

\usepackage[T1]{fontenc}

\DeclareRobustCommand{\VAN}[3]{#2}
\let\VANthebibliography\thebibliography
\def\thebibliography{\DeclareRobustCommand{\VAN}[3]{##3}\VANthebibliography}

\usepackage{graphicx}	\usepackage{amsmath}	\usepackage{rotating}
\usepackage{}

\title[Measurements of the $z=4-10$ X-ray Luminosity Function]{Measurements of the $\mathbf{z=4-10}$ X-ray Luminosity Function: the high space density of moderate-luminosity, obscured AGN}

\author[C. L. Barlow-Hall et al.]{
C. L. Barlow-Hall,$^{1}$\thanks{E-mail: c.barlow-hall@roe.ac.uk}
and
J. Aird,$^{1,2}$
\\
$^{1}$Institute for Astronomy, The University of Edinburgh, Royal Observatory, Blackford Hill, Edinburgh EH9 3HJ, UK\\
$^{2}$School of Physics \& Astronomy, University of Leicester, University Road, Leicester LE1 7RJ, UK\\
}

\date{Accepted XXX. Received YYY; in original form ZZZ}

\pubyear{2025}

\begin{document}
\label{firstpage}
\pagerange{\pageref{firstpage}--\pageref{lastpage}}
\maketitle

\begin{abstract}
Supermassive Black Holes (SMBHs) are often theorised to undergo significant growth within the early Universe, however, the X-ray Luminosity Function (XLF), often used as the principal tracer of the SMBH accretion density, lacks observational constraints above $z\sim6$, until now. We present new measurements of the $z=4-10$ XLF at intermediate luminosities, taking advantage of recent deep near-infrared imaging from UltraVISTA that enables us to identify galaxies and AGN at high redshifts within which we identify X-ray sources using the Chandra COSMOS data.
We first performed a cross-match to a deep Chandra source list, for which the X-ray sensitivity can be accurately quantified, before exploiting the available X-ray data further through direct extraction of X-ray counts at the positions of COSMOS2020 galaxies. With the resulting $z=4-10$ X-ray AGN sample, comprised of 21 blind detections and a further 11 directly extracted detections, we have measured the early space density of AGN, at moderate-luminosities where the majority of early SMBH growth occurred. These measurements reveal higher space-densities than expected, based on the extrapolation of XLF models from lower redshifts. Whilst our measured space densities at $z=4-5$ are consistent with model predictions, at $z=5-7$ we find space densities of the order of 10$\times$ the extrapolated model predictions and could be as high as $220\times$ the model extrapolations at $z=7-10$. In addition, we find evidence that a large fraction of the early AGN population are heavily obscured, with an obscured fraction of $0.982^{+0.007}_{-0.008}$; correcting for this obscuration further increases the measured space densities. Comparing to recent JWST results, these measurements begin to bridge the gap between the bright-end of the quasar luminosity function and the latest JWST observations of very early, low-luminosity AGN, indicating that a larger fraction of the first galaxies likely play host to a rapidly growing SMBH than previously thought.

\end{abstract}


\section{Introduction}\label{sec:intro}
Supermassive Black Holes (SMBHs) are some of the most violent and esoteric objects in the Universe, yet despite the consensus that they played a key role in shaping the Universe we see today \citep[e.g.][]{Somerville2015, Heckman2014, Kormendy1995} their formation and subsequent growth remains a very open question. When SMBHs undergo rapid mass accretion they produce strong emissions across a wide range of wavelengths, making them observable at the centre of their host galaxies at a given wavelength \citep[e.g. see][]{Padovani2017} as what are known as Active Galactic Nuclei (AGN). The necessity of feedback, due to these AGN, to limit the growth of high mass galaxies is clearly demonstrated by Cosmological Hydrodynamical simulations \citep[e.g.][]{Somerville2015}, whilst studies such as those of \citet{Boyle1998, Ueda2014, Aird2015} have shown the evolution of the rate of Black Hole mass density growth closely maps the trend observed in the rate of stellar density build up. It is thus accepted that SMBHs co-evolve with their host galaxy \citep[see][]{Heckman2014}, undergoing periods of AGN activity, with most galaxies expected to play host to a SMBH \citep[][]{Rees1984, Kormendy2013}.
Nevertheless, how the seeds of these black holes form in the very early Universe and which processes drive their early growth to produce the SMBHs that we observe remains unclear 
\citep[e.g. see][for a review]{Inayoshi2020}.

Observations of AGN at high redshift indicate that they contain SMBHs with masses of the order of $10^{8-9}\mathrm{M_\odot}$ \citep[e.g.][]{Zubovas2021} indicating that significant mass growth, of several orders of magnitude, must occur within the early Universe and thus during the epoch of inital stellar mass assembly of galaxies.
Reaching the observed masses from even the most massive black hole seeds in the available cosmic time requires near-continuous growth at the highest possible rates \citep[i.e. $\sim$the Eddington limit][]{Trakhtenbrot2016,Orofino2018,Zhu2020}.
As noted by \citet[][and references therein]{Sharma2024}, however, no SMBH can be expected to accrete at a constant or smoothly increasing rate, and thus even higher levels of mass growth must occur during active periods.

The growth of AGN across cosmic time, as a population, is traced by the Quasar\footnote{In this paper, we use the term ``Quasar'' to refer to all radiatively efficient AGN i.e. that capture the bulk of SMBH mass assembly \citep[][]{Alexander2012}, regardless of luminosity or obscuration level, and thus the QLF, ideally, should capture the overall SMBH accretion density.} Luminosity Function (QLF); a measure of the space density of AGN at a given redshift and luminosity \citep[][]{Page1996, Boyle2000, Kalfountzou2014}, and typically has the general form of a double power law. The QLF has been measured using Optical/Ultra Violet (UV) \citep[e.g.][]{Ren2021, Ross2013, McGreer2013, Schmidt1968, Boyle2000, Fan2001}, Infrared (IR) \citep[e.g.][]{Lacy2015, Assef2011} and X-ray selected AGN \citep[e.g.][]{Aird2010, Aird2015, Ueda2014, Gilli2009, Pouliasis2024, Georgakakis2015, Miyaji2001, Ananna2019}, all of which show the general trend of a rising space density from $z\sim0$ up to cosmic noon before it declines again above $z\sim3$ \citep[as found by][]{Schmidt1995, Aird2010, Vito2014, Aird2015}. X-ray surveys are particularly useful for determining the QLF, known as the X-ray Luminosity Function (XLF), thanks to the uncontaminated AGN selection using X-rays and the well defined sensitivity of X-ray surveys \citep[][]{Padovani2017}. X-ray emissions from an AGN are also a close tracer of the instantaneous growth of the SMBH, as they are produced by a hot electron corona very close to the SMBH \citep[see][and references therein]{Brandt2015}. Notably multiple studies have found that the XLF can probe the regime of X-ray bright, optically normal galaxies, thus identifying lower luminosity and obscured AGN that would otherwise go undiscovered at rest-frame UV/Optical \citep[see e.g.][for reviews]{Brandt2015, Brandt2005, Lusso2023}. 
For the most heavily obscured AGN, when the line-of-sight column density reaches the Compton-thick regime (equivalent hydrogen column density, $N_\mathrm{H}\gtrsim 10^{24}$~cm$^{-2}$) even hard X-ray emission will be substantially suppressed and the emergent X-ray spectrum is dominated by a Compton ``reflection'' component \citep[e.g.][]{Ikeda2009,Brightman2011}; while such Compton-thick sources may still be detected in sufficiently deep X-ray imaging it becomes more challenging to infer their intrinsic luminosities and thus measure the overall XLF including these populations \citep[see e.g.][]{Buchner2015,Ananna2019}.

Many studies have used parametric models to describe the evolution of the XLF, mainly applying evolutionary trends to the standard double power-law which characterises the population, and constrained these models using X-ray surveys \citep[such as e.g.][]{Page1996, Gilli2007, Ueda2014, Aird2015}. However, due to both the X-ray flux limit and the drop in the space density toward higher redshifts, samples of X-ray selected AGN at high redshift remain very limited (e.g. \citealt{Luo2017} identified just two $z>5$ sources in the {\it Chandra} Deep Field South 7~Ms sample, while \citealt{Marchesi2016} identified four $z>6$ sources within the $\sim$2~deg$^2$ {\it Chandra} COSMOS-Legacy survey). 
For faint sources we need deep observations, while as the space density is known to decrease toward higher redshifts a large survey area is required. As such, samples of X-ray selected AGN at $z\gtrsim3$ remain relatively small and parametric models of the evolution of the XLF toward higher redshifts remain poorly constrained at $z\gtrsim5$ \citep[see][for examples of XLF fit to $3\leq z\leq6$]{Vito2014, Ueda2014, Vito2018, Georgakakis2015}. Most recently \citet{Pouliasis2024} extended their model of the XLF out to $z\sim6$ based on X-ray selected AGN from a combination of Chandra and XMM-Newton imaging in order to reach the depth and survey area required at these redshifts, yet still lacked sources at higher redshifts.

Despite the limited X-ray detections of AGN above $z=6$, recent observations with JWST have revealed many AGN candidates out to $z\sim10$, predominantly identified through detection of broad H$\alpha$ and H$\beta$ emission lines \citep[e.g.][]{Kocevski2024, Maiolino2023, Greene2023, Matthee2024} as well as the detection of high-ionisation lines and application of narrow-line diagnostics \citep[e.g.][]{Scholtz2023, Gonzalez2025}. Approximately 20\% \citep[][]{Hainline2024} of the AGN candidates found by JWST are the so-called Little Red Dots, which appear very red in colour with a characteristic V-shaped spectral energy distribution (SED) and compact morphology \citep[][]{Matthee2024, Greene2023, Kocevski2024}. Whilst X-rays provide a fairly unambiguous signature for the identification of AGN, however the vast majority of these JWST AGN lack X-ray detections \citep[see][for a notable exception]{Bogdan2023}. Some studies, such as those of \citet{Maiolino2025}, have suggested this apparent X-ray weakness is due to a higher density of obscuring gas within these high redshift sources \citep[see also][]{Inayoshi2024} or the geometry of the system \citep[][]{Madau2024}, with \citet{Brooks2024} finding evidence for this potentially X-ray obscuring material.

With the advent of JWST the samples of very high redshift galaxies and of AGN within high redshift galaxies is increasing, indicating a far higher space density of AGN at $z>6$ than extrapolated models previously predicted, including the XLF models. It should be noted, however, that JWST is imaging only very small survey areas and probing the faint-end of the QLF, a regime where current XLF models are very poorly constrained. Recently, \citet{Barlow-Hall2023} placed constraints upon the bright-end of the XLF at $z\sim6$ based on a large compilation of \textit{Swift} X-ray telescope observations \citep[the Extragalactic Serendipitous Swift Survey, ExSeSS: see][]{Delaney2023}, finding a steep bright-end slope consistent with the bright-end of the model XLF extrapolations \citep[see also][for constraints from \textit{eROSITA}]{Wolf2021}. However, the space density of AGN within the regime between the faint sources found by JWST and the bright sources found by large sky surveys 
remains unconstrained. Here we utilise the COSMOS field, providing a relatively large survey area with deep X-ray imagining, in order to constrain the moderate luminosity XLF and bridge the gap in AGN space density measurements from large area shallow surveys that find the brightest sources to the small area deep fields which are able to probe the large numbers of very faint sources.

The \textit{Chandra} X-ray Observatory \citep{Weisskopf2002} is one of the most sensitive X-ray observatories to date, probing fluxes of $\leq10^{-16}\,\mathrm{erg\,s^{-1}\,cm^{-2}}$ in the 0.5--2\,keV energy band in the deepest observations, a factor of $\geq100$ deeper than achieved in ExSeSS. In order to obtain this depth of imaging, \textit{Chandra} has observed a much smaller area of sky compared to \textit{Swift} and other less sensitive survey telescopes. However, the depth granted by \textit{Chandra} is crucial to probing moderate luminosity AGN at high redshifts. 
Although this X-ray data existed prior to the work presented here, it is with recent deep NIR Ultra VISTA imaging across large areas that allows for the identification of high redshift galaxies and AGN (along with reliable photometric redshift measurements) enabling new catalogues such as COSMOS2020 \citep{Weaver2022} to cover the $z=4-10$ regime \citep[see also][which have found an excess of high-redshift galaxies only visible in NIR observations]{Bowler2020, Finkelstein2022, Donnan2023a, Donnan2024}. Thus, we are now able to identify reliable high-redshift counterparts to the faintest X-ray sources revealed in the Chandra-COSMOS Legacy imaging.

The form of the COSMOS2020 catalogue and the X-ray data used in this work are detailed in \S\ref{chpt3:data}. The cross-matching of COSMOS2020 and X-ray sources performed in order to obtain the blind source sample is outlined in \S\ref{chpt3:data:nway}, and \S\ref{chpt3:data:extraction} details the creation of the blind and extracted sample through the direct extraction of X-ray counts for all COSMOS2020 sources. We then detail the measurement of the XLF using these X-ray AGN samples in \S\ref{chpt3:xlf}, and test the extremes of these measured constraints arising from the differing photometric redshift estimates from COSMOS2020 (\S\ref{chpt3:xlf:redshift}). The impact of obscuration on the high redshift XLF is then investigated in \S\ref{chpt3:obscuration}. Finally, the implications of these new constraints on the XLF are discussed in \S\ref{chpt3:discussion} and we summarise our findings in \S\ref{chpt3:conclusion}.

Given the extent of details on the sample and method of XLF measurement, the casual reader may wish to skip to \S\ref{chpt3:obscuration:correct}, and the discussion in \S\ref{chpt3:discussion}. More specifically the comparison between our final X-ray constraints on the bolometric luminosity function of AGN and recent JWST measurements are shown in Figure~\ref{chpt3:fig:jwst_qlf} and the measured evolution of the Black Hole Accretion Rate Density is shown in Figure~\ref{fig:bhad}.
Throughout this work we adopt a Flat $\Lambda$CDM cosmology with $H_0=70$km\,s$^{-1}$Mpc$^{-1}$, $\Omega_M=0.3$ and $\Omega_{\Lambda}=0.7$, and all errors given are the $1\sigma$ uncertainties on the values.

\section{The COSMOS Field Data Sets}\label{chpt3:data}

The COSMOS field, centred at $\mathrm{10^{h}00^{m}27^{s}.92+02^{\circ}12'03''.50}$ (J2000), was originally defined by \citet{Scoville2007} for large-area deep (0.05\,arcsec) Hubble Space Telescope (HST) imaging in order to address a number of science goals; from large dark matter distributions and galaxy assembly to early galaxies and the evolution of AGN. This $\sim2$\,degree$^2$ equatorial field was chosen as its location ensured it was visible to all astronomical instruments, at the time, thus allowing for detailed multi-wavelength studies of the sources within the field. The COSMOS field also benefits from having a relatively low and uniform Galactic extinction, compared to other equatorial fields, whilst being devoid of bright X-ray, UV and radio sources and thus reducing the contamination of the signals of faint sources within the field.
Following its definition, the COSMOS field has been surveyed using a wide range of facilities spanning the full wavelength range (including \textit{HST}, \textit{Spitzer}, GALEX, \textit{XMM-Newton}, \textit{Chandra}, Subaru, ALMA, VLA, VISTA, ESO-VLT, UKIRT, NOAO, CFHT, and more), with deep observations performed by almost all major astronomical facilities \citep[see e.g.][]{Sanders2007, Capak2007, Taniguchi2015, LeFevre2020}. The wealth of data available within the COSMOS field makes it incredibly important to extragalactic astronomy. Crucially for the identification of galaxies out to very high redshifts, the COSMOS field has very deep imaging at Near and Mid-Infrared wavelengths from the VISTA telescope \citep{McCracken2012, Moneti2023, Ashby2018}. 

\subsection{UV-to-IR photometry and photometric redshifts from COSMOS2020}\label{chpt3:data:cosmos2020}
Since COSMOS was first defined, several photometric catalogues of this field have been publicly released as deeper data has become available \citep{Capak2007, Ilbert2009, Ilbert2013, Muzzin2013, Laigle2016}. The COSMOS2020 catalogue of \citet{Weaver2022} is the most up to date release, including deep UV \citep{Sawicki2019} and ultra deep optical imaging \citep{Aihara2019} as well as Near IR data from the UltraVISTA survey DR4 \citep{Moneti2023, McCracken2012}. For COSMOS2020, all legacy datasets used in previous iterations of the COSMOS catalogue \citep[e.g. COSMOS2015: ][]{Laigle2016} have also been reprocessed to align with newly available astrometry from Gaia DR1 \citep{GaiaCollaboration2016}. The legacy imaging has also been resampled to a pixel scaling of 0.15\,arcsec and stacked images aligned to the COSMOS Tangent point ($\mathrm{10^h00^m27.92^s+02^{\circ}12'03''.50}$). 
\citet{Weaver2022} identified sources within the wealth of deep COSMOS data using two different detection methods; implementing SExtractor \citep[][]{Bertin1996} as in \citet{Laigle2016}, and THE FARMER software of \citet{Weaver2023} which utilises THE TRACTOR photometry modelling code \citep[see][for details]{Lang2016}. These two versions of the COSMOS catalogue are referred to as the CLASSIC Catalogue and THE FARMER Catalogue, respectively. 

For this work we use THE FARMER catalogue, as it has been found to have a higher completeness than the Classic Catalogue \citep[e.g.][]{Shuntov2022}. It also provides more accurate photometry for the faint sources (which will comprise the majority of the high redshift sample we will use in this paper), is more robust against source blending and is able to identify a higher density of $z>6$ sources \citep[see e.g.][]{Brinch2023, Sillassen2022}.

To create THE FARMER catalogue, source detection is performed on a $\chi^2$ detection image that combines the $i$-, $z$-, $Y$- $J$- $H$- and $K_s$-band imaging and is generated using the SWarp package \citep{Szalay1999}.
Parametric models of the source shape are then used to perform photometry across all wavebands. 
THE FARMER software measures the photometric fluxes by scaling one of five parametric models (Point Source, Simple Galaxy, ExpGalaxy, DevGalaxy and Composite Galaxy\footnote{Point source models use the PSF directly, SimpleGalaxy uses a fixed exponential light profile and ExpGalaxy uses an exponential light profile parametrised by the source's image properties. The DevGalaxy model is similar to the ExpGalaxy, but uses the de Vaucouleurs light profile, whilst the CompositeGalaxy model combines both the ExpGalaxy and DevGalaxy models.}) assigned to each source \citep[see][for details]{Weaver2023}. Thus, this method of measuring the photometry does not require PSF homogenisation or the use of fixed apertures and the best quality photometry is not degraded to the lowest quality resolution achieved.
As these models cannot be applied to saturated bright stars or sources contaminated by stellar halos, \citet{Weaver2022} adopt the HSC-SSP PDR2 \citep{Coupon2018} bright-star masks in order to mask out all stars with a G-band magnitude of 18 or less in the Gaia DR2 star catalogue. In order to prevent the introduction of inhomogeneities to the constraints on the models for photometric extraction, THE FARMER catalogue is also limited to the UltraVISTA region such that there is coverage in all wavelength bands used. 
After applying these cuts, THE FARMER photometric catalogue of COSMOS sources obtained by \citet{Weaver2022} covers $1.792\,\mathrm{deg^{2}}$ of sky and contains a total of $964\,506$ sources, all determined to have reliable detections.

\citet{Weaver2022} determine the photometric redshifts of their sources using both the EAzY \citep{Brammer2008} and LePhare\footnote{http://www.cfht.hawaii.edu/~arnouts/lephare.html} \citep{Arnouts2002, Ilbert2006} photometric redshift codes, with the photometric measurements corrected for Galactic extinction based on the dust maps of \citet{Schlafly2011}.
We take the LePhare photometric redshift measurements as the redshifts of our final galaxy catalogue, as LePhare accounts for a possible AGN component in the photometric redshift determination.
LePhare is also found to have a lower outlier percentage when comparing the photometric redshift estimates to available spectroscopic measurements \citep{Weaver2022}.
LePhare is used to determine the photometric redshift and physical parameters of each source by fitting with a range of templates including 19 elliptical/spiral galaxy templates from \citet{Polletta2007}, 12 from models of blue star-forming galaxies and 2 galaxy templates with exponentially declining star formation histories \citep{Bruzual2003, Onodera2012}. LePhare is also used to perform fits using AGN templates consisting of both unobscured sources where the AGN dominates the light at most wavelengths and obscured sources with significant host galaxy contributions \citep[see][for details]{Salvato2009, Salvato2011}, as well as stellar templates from \citet{Pickles1998}. 
For sources at redshifts of $z>10$, the observed wavelength of the Lyman break will be shifted outside the wavelength range of the deep near-IR photometry in the COSMOS2020 catalogue and thus such extremely high-redshift sources are likely to be undetected.
As such, the redshift grid used by LePhare and thus our sample is limited to a maximum redshift of $z=10$.

Finally, we restrict the catalogue to only include sources that lie within the footprint of the \textit{Chandra} X-ray imaging, as deep X-ray data is not available for sources outside this region. The final, filtered catalogue consists of 816\,944 optical/IR selected sources. 
\subsection{Chandra Imaging}\label{chpt3:data:chandra}
The \textit{Chandra} X-ray Observatory is the highest angular resolution X-ray telescope to date. With an on-axis resolution of $0.5\,\mathrm{arcsec}$, Chandra can achieve the sensitivity needed to find moderate-luminosity high-redshift X-ray sources. 

The Chandra-COSMOS (C-COSMOS) program of \citet{Elvis2009} was the first contiguous survey of the COSMOS field and provided an unprecedented area and depth of X-ray imaging at the time. Utilising a strategy of overlapping pointings, with offsets of 8\,arcmin around the survey centre at $\mathrm{10^h00^m24^s+02^\circ10'55''}$, provided a near-uniform sensitivity across the centre of the COSMOS field. The final C-COSMOS survey covered $\sim0.9\,\mathrm{deg^2}$ to an effective exposure of 160\,ks over the central $\sim0.5\,\mathrm{deg^2}$ and 80\,ks over the remaining $\sim0.4\,\mathrm{deg^2}$. Later, \citet{Civano2016} combined the C-COSMOS survey with 2.8\,Ms of additional Chandra imaging (performed over four periods: Nov 2012-Jan 2013, Mar 2013-July 2013, Oct 2013-Jan 2014 and Mar 2014) to produce the Chandra COSMOS-Legacy Survey,  
providing uniform coverage of the $1.7\,\mathrm{deg^2}$ COSMOS/HST field to an effective exposure of 160ks and a total area coverage of $\sim2.2\,\mathrm{deg^2}$,
80\% of which has a resolution of 2-4\,arcsec.

For this study we use the deep Chandra imaging provided by the Chandra COSMOS-Legacy survey \citep{Civano2016}. We adopt the custom reanalysis of these Chandra data following the data reduction and source detection process detailed in \citep{Laird2009, Nandra2015, Kocevski2018} to produce the clean images and blind source catalogue used in this study and in previous works \citep{Aird2017, Aird2018, Aird2019, Laloux2023, Laloux2024}. Performing this reanalysis provides access to all resulting data products, required for this work, and ensures that we can control and quantify the sensitivity of the Chandra imaging when producing the source catalogues.

The actual data reduction of the Chandra COSMOS-legacy data used in this work was performed by \citet{Georgakakis2017} using the pipeline described by \citet{Nandra2015} and \citet{Kocevski2018}. Here we will briefly summarise the reduction of the Chandra imaging, the subsequent source identification, and the methods used to determine the sensitivity and assess the X-ray completeness.

\subsubsection{Data Reduction and Initial Source Detection}\label{chpt3:data:chandra:reduction}
The raw data from Chandra is cleaned and processed using the CIAO (Chandra Interactive Analysis of Observations) suite v.4.8 \citep{Fruscione2009}. 
Correcting for the telescope dithering, event files are then cleaned by removing Hot Pixels, image streaking and Cosmic-Ray afterglows. Anomalous peaks in the observed X-ray background are identified from a light curve of the measured background as periods when the rate is $\geq1.4$ times the quiescent background (as given by the count rate limit required for an excess variance in the light curve of 0), and subsequently cut from the event file. With all anomalous pixels and events removed, the number of events (photon detections) of a given measured energy (0.5--7\,keV, 0.5--2\,keV, 2--7\,keV and 4--7\,keV for Full, Soft, Hard and Ultra Hard images respectively) at each pixel are summed across all observations to produce the final image file. For a given energy band the corresponding effective exposure map is generated, accounting for exposure time, detector quantum efficiency, and effective area (weighted by energy assuming a power-law spectrum of photon index $\Gamma=1.4$), as well as bad pixels, chip gaps and edges, and the dithering motion of the telescope. Finally the Point Spread Function (PSF) at each pixel position is determined through simulations of Chandra's High Resolution Mirror Assembly (HRMA), performed using the MARX Simulator of \citet{Wise1997}, then weighted by the effective exposure of each individual observation to generate a PSF map providing the radius corresponding to an enclosed energy fraction (EEF) of 70\% and 90\%.

Once the raw data is processed a source detection method is applied to the resulting image, in order to produce a blind\footnote{By ``blind catalogue'' we mean based only on the X-ray data and without using any a priori knowledge of source positions from other wavelengths.} catalogue of X-ray sources. 
Initial source detections are performed on the full resolution images, using the WAVDETECT source detection code from the CIAO suite \citep{Fruscione2009}. This initial detection is done with a very low significance threshold, given by a false probability threshold of $10^{-4}$, and limited to the image region where the exposure is $\geq10$\% of the maximum exposure. This low significance threshold prevents real sources being missed, however there will be a number of spurious sources also introduced as a result. 
For all candidate sources the total photon counts, $T$, are measured in an aperture of radius given by the 70\% EEF of the PSF. The local background level is measured in a 100\,pixel wide annulus of inner radius $1.5\times$ the 95\% EEF PSF, at each source position, then re-scaled to the source detection area to estimate the expected background in the source aperture, $b$. 
For each source, we can then calculate the net, effective exposure-corrected count rate,
\begin{equation}
    c = \frac{T-b}{p_{70}t_\mathrm{exp}}
    \label{chpt3:eq:ctr}
\end{equation}
where the factor of $p_{70}=0.7$ is applied in order to correct for the choice of an aperture extraction radius corresponding to the 70\% EEF at the source position and $t_\mathrm{exp}$ is the effective exposure.
We also calculate the false probability, $P(\geq T|b)$, that gives the probability that the source is a spurious detection due to a fluctuation of the background, using the equation,
\begin{equation}
    P(\geq T|b)=1-P(< T|b)=1-\sum^{T-1}_{i=0}\frac{b^i}{i!}e^{-b}.
    \label{chpt3:eq:falseProb}
\end{equation}
A threshold false probability value of $\lesssim4\times10^{-6}$ is then applied to the candidate sources, in order to remove those believed to be spurious detections and leave only the reliable source sample.

In order to search for potentially missed sources, all reliable detections are masked out and a second pass of this detection method is performed. The background and false probability of each candidate source are subsequently recalculated, and the same false probability threshold is applied once more. As there is negligible difference in the detections list following this second detection stage, the process is not run again. For sources within 5\,pixels of one another only the source with the highest measured counts is kept, in order to prevent duplicated sources within the final catalogue. Once the source list is found, the background map is created by removing all source counts from the image and replacing based on the local background \citep[see][]{Georgakakis2008}.

This source detection process is carried out on the images from all four energy bands (Full, Soft, Hard and Ultra Hard). The resulting reliable source lists are then merged and for any sources that are not independently detected in all bands the missing counts, background and exposure values are extracted from the corresponding imaging and maps at the source position. 
Count rates, $c$, in a given band may be converted to fluxes assuming a certain spectral model. 
However, in this work we convert directly from count rates to rest-frame 2--10~keV luminosity ($L_\mathrm{X}$) for a given redshift.

In this work we use the catalogue resulting from this source extraction process, containing a total of 3\,627 X-ray sources, referred to as the blind source catalogue. The images, background maps, exposure maps and PSF maps generated through the above reduction process are also used in \S\ref{chpt3:data:extraction} for the extraction of X-ray information at the positions of all sources in the reduced COSMOS2020 catalogue. 

\subsubsection{The Chandra Sensitivity across the COSMOS field}\label{chpt3:data:chandra:area}
The X-ray Luminosity Function (XLF) determination requires the sensitivity of the X-ray survey to be known and quantified. We do this by generating a sensitivity map \citep[following the process detailed in][]{Georgakakis2008}, determining the number of counts required to produce a significant detection, i.e. a detection above the specified false probability threshold given the expected background count rate, at each pixel position in the image. From the sensitivity map, we then determine the area curve; the total area of sky where the available Chandra imaging is sensitive to sources above a given X-ray flux threshold. 

For any given pixel position in an image there is small probability that a random fluctuation will result in sufficient counts for a spurious detection (i.e. due to a fluctuation of the counts from the background) or result in a faint source appearing brighter and thus being detected when otherwise it would not be. 
The probability that the total \emph{observed} counts exceeds the given detection threshold ($L$ counts, determined for a given false probability threshold using equation \ref{chpt3:eq:falseProb}) is thus described by the cumulative probability,

\begin{equation}
    P(\geq L|c,b)=1-P(<L|\lambda)=\frac{1}{\Gamma(L)}\int_0^\lambda {e^{-t}t^{L-1}dt}
    \label{chpt3:eq:areacalc}
\end{equation}

where $\Gamma(L)$ is the gamma function and $\lambda=b+ct_\mathrm{exp}p_{70}$ where $ct_\mathrm{exp}p$ is the expected number of counts from a source with count rate $c$.  
The area curve is the sum of the cumulative probability distributions across all pixels in the image, determined for a given source count rate, multiplied by the sky area of a pixel. 
\begin{figure}
    \centering
    \includegraphics[width=\linewidth]{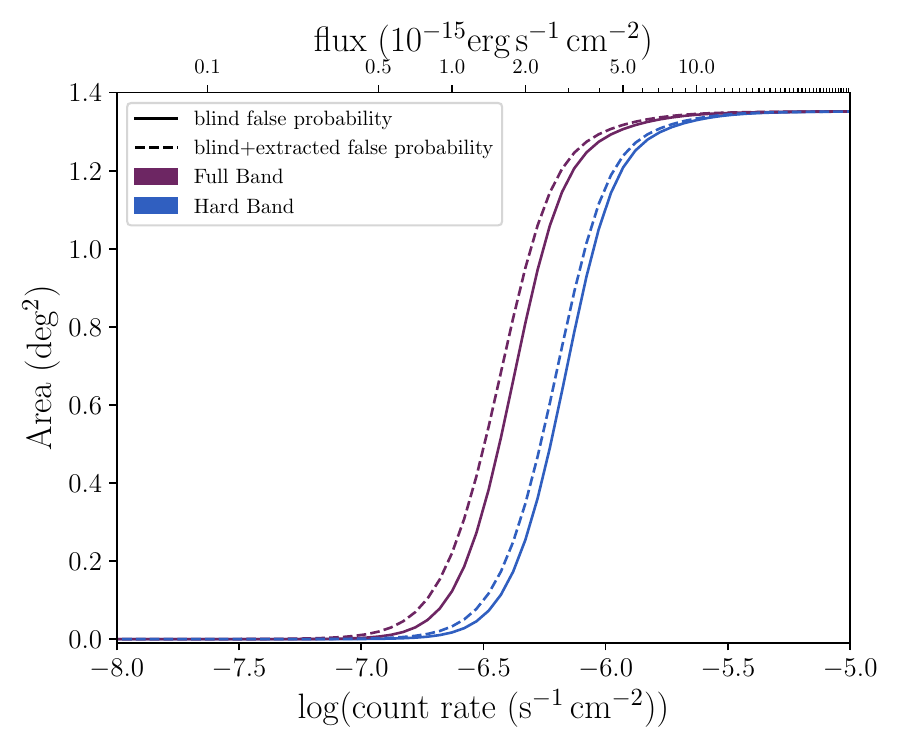}
    \caption[Chandra area curves]{Hard band (2--7~keV, blue) and full band (0.5--7~keV, purple) area curves, giving the sky area coverage as a function of the effective area-corrected count rate, determined with the blind sample false probability threshold of $4\times10^{-6}$ (solid line) and the blind+extracted sample threshold of $3.38\times10^{-5}$ (dashed line). All use of these area curves is based on the count rate (converted to intrinsic X-ray luminosities at a given $z$ under differing spectral assumptions); however, the upper axis shows the corresponding 2--10~keV and 0.5--10~keV observed-frame fluxes for the hard and full bands (respectively) using a fixed conversion that assumes a power-law spectrum with $\Gamma=1.4$.}
    \label{chpt3:fig:chandraarea}
\end{figure}

Any masking applied to the Chandra imaging will reduce the sky area for our study. Thus, prior to the area curve calculation, 
we apply the bright-star mask and limit the survey area to the UltraVISTA region (see \S\ref{chpt3:data:chandra:reduction}).

Sensitivity maps, and corresponding area curves, are calculated for the Full and Hard band imaging for both a false probability threshold of $4\times10^{-6}$, corresponding to the blind source catalogue detailed above, and for the less stringent threshold adopted when considering the X-ray data at known COSMOS2020 source positions (see \S\ref{chpt3:data:nway} below). Figure \ref{chpt3:fig:chandraarea} shows the resulting area curves, as given by a false probability threshold of $4\times10^{-6}$, for the Full and Hard energy bands used for the XLF determination (\S\ref{chpt3:xlf} and \S\ref{chpt3:obscuration}). From these area curves, the full survey area of our X-ray catalogue is 1.35\,deg$^2$, with a 10\% area coverage at a count rate of $2.41\times10^{-7}\,\mathrm{s^{-1}\,cm^{-2}}$ and $3.89\times10^{-7}\,\mathrm{s^{-1}\,cm^{-2}}$ for the Full and Hard band imaging respectively (corresponding to a flux of $1.02\times10^{-15}\,\mathrm{erg\,s^{-1}\,cm^{-2}}$ and $1.66\times10^{-15}\,\mathrm{erg\,s^{-1}\,cm^{-2}}$ respectively, for a power law spectrum of $\Gamma=1.4$).

\subsection{X-ray and redshift catalogue cross-matching}\label{chpt3:data:nway}
In order to perform redshift-dependent studies of the X-ray AGN population, we identify counterparts in the filtered COSMOS2020 sample (see \S\ref{chpt3:data:cosmos2020}) to sources within the blind X-ray catalogue (detailed in \S\ref{chpt3:data:chandra:reduction}), producing a sample of X-ray sources with optical/IR photometry. The counterparts to our blind catalogue are determined using the Bayesian Cross-Matching code NWay \citep[][]{Buchner2021}.

As detailed in \citet{Salvato2018}, Nway determines the probability ($p_i$) of each multi-wavelength source being the true counterpart to the X-ray detection and the probability that the true counterpart of each X-ray source is in the multi-wavelength sample ($p_\mathrm{any}$), given the separation between the X-ray source and potential counterparts, positional uncertainties and total area of sky. Weighting is then applied to these probabilities, given specified (or internally generated) magnitude and colour priors for the cross-match.

Before identifying counterparts to our blind X-ray sample, any X-ray sources within the bright-star masks applied to the COSMOS2020 catalogues---or within a distance from the mask corresponding to the X-ray positional uncertainty--- are removed from the sample, in order to ensure the final cross-matched sample contains only X-ray sources for which, with reasonable confidence, the optical/IR counterparts do not lie within the masked regions of the COSMOS2020 sample. Apertures of radius $r=0.82\,\mathrm{arcsec}$, given by the mean positional uncertainty of all blind X-ray sources, are placed at the X-ray source positions, and all sources for which this aperture intersects the bright star mask regions are removed from the X-ray sample producing a sample of 2792 X-ray sources. In addition, as the COSMOS2020 sample used in this study is limited to the UltraVISTA sky area, we similarly limit the blind X-ray source sample to this area. Thus the resulting area of both our masked COSMOS2020 and blind X-ray source samples is $1.35\,\mathrm{degrees^{2}}$, containing 816\,944 and 2\,408 sources respectively.

We consider potential COSMOS2020 counterparts, to the blind X-ray sources, within a $10\,\mathrm{arcsec}$ search radius of each X-ray position. This search radius exceeds 3 times the maximum X-ray positional uncertainty of the sample and thus ensures that $>99\%$ of the true counterparts are included (we note that the typical offset between an X-ray source position and the final, best counterpart is typically $\sim0.47\,\mathrm{arcsec}$, i.e. much lower than this initial search radius). 

We perform NWay cross-matches using a number of internally generated priors.
These priors are based on the magnitudes in the UltraVISTA $Ks$-band, IRAC channel 1 and IRAC channel 2, chosen as they are good tracers of AGN activity at high redshifts \citep[see e.g.][]{Reines2016} as well as the IRAC channel 1-channel 2 colour. For each run, we adopt the counterpart to each X-ray source with the highest $p_i$ value as the best match and apply a threshold on $p_\mathrm{any}$ to select X-ray sources with a reliable in the COSMOS2020 catalogue. 

In order to assess the fraction of incorrect associations from a given NWay run, and determine the appropriate $p_\mathrm{any}$ threshold, we repeat the cross-matching process with NWay using a catalogue of fake X-ray sources with randomised positions. The resulting catalogue of fake sources assigned COSMOS2020 counterparts provides an estimate of the incorrect associations that will be present in the sample of cross-matched blind X-ray sources. 

The fake sample is generated by applying a random positional shift, of $20-120,\mathrm{arcsec}$ to the positions of all sources in the unmasked blind X-ray sample. The same masking that is applied to the real sample of X-ray sources is also applied to this fake sample, in order to remove any fake sources that fall within the masked regions of the COSMOS2020 sample, where the photometry is unreliable. Following the same approach as for the real X-ray sample, we assign counterparts to the fake X-ray sample with the magnitude priors generated, by Nway, during the cross-matching of the real X-ray sample to COSMOS2020. The resulting sample is then, similarly, limited to only the potential counterparts with the largest $p_i$ value, reducing the sample to only the best cross-matches identified for each fake X-ray source.

Using the real and fake cross-matched samples, we assess the quality of the resulting sample of counterparts to our blind X-ray sources. We determine the completeness (the fraction of the real sample which have a correctly matched counterpart) of the sample of counterparts to the sample of real X-ray sources, as given by,

\begin{equation}
    \mathrm{Completeness} = \frac{N(p_\mathrm{any} > p_\mathrm{any,threshold})}{N_\mathrm{total}}
\end{equation}

where $N_\mathrm{total}$ is the total number of X-ray sources in the sample, and $N(p_\mathrm{any} > p_\mathrm{any,threshold})$ is the number of matched sources with $p_\mathrm{any}$ (the probability that any of the potential counterparts in the COSMOS2020 sample are the true counterpart) greater than a given threshold value ($p_\mathrm{any,threshold}$). The fraction of matched sources that are likely to be incorrectly matched, known as the false positive fraction, is similarly calculated from the fake match sample by taking the ratio of the number of matches to the fake sample with $p_\mathrm{any}$ exceeding the same $p_\mathrm{any,threshold}$ to the total number of X-ray sources. 

The completeness and false positive fraction are calculated for a range of $p_\mathrm{any,threshold}$ values, for each of the matched samples resulting from different combinations of magnitude and colour based priors 
We find the 
cross-match performed with the Ks-band and IRAC colour magnitude priors produces the lowest False Positive Fraction of all tested combination of magnitude priors, while also achieving a Completeness of over 0.9. Thus, we select the cross-match performed with the combination of a Ks-band magnitude and IRAC channel1-channel2 colour magnitude priors for the final cross-matched sample. 
 
For this study we choose a false positive fraction of 10\%, as an acceptable failure rate of the cross-matching. This corresponds to a sample completeness of $0.97$ and $p_\mathrm{any,threshold}=0.388$. When considering the high-redshift samples only, given by the best fit photometric redshift values (as defined in \S\ref{chpt3:xlf:sample} below), this $p_\mathrm{any,threshold}$ corresponds to a False Positive fraction of $1.0\times10^{-3}\%$, $3.3\times10^{-4}\%$ and $1.1\times10^{-4}\%$ for the samples with photometric redshifts of $z\geq4$, $z\geq5$ and $z\geq6$ respectively. These very small False Positive Fractions in the high-redshift samples are due to a much lower surface density of high-redshift sources compared to the full COSMOS2020 sample. As such, a random alignment between an X-ray source position and a high-redshift galaxy, which is subsequently falsely identified as the best counterpart to the X-ray source, is very unlikely to occur.

\subsection{X-ray counts extraction at galaxy positions}\label{chpt3:data:extraction}
High-redshift AGN appear observationally very faint, thus there are likely to be fainter X-ray AGN within the high-redshift galaxy sample that fall just below the X-ray detection thresholds adopted in the creation of the blind sample. As such these very faint AGN will not be present within the blind source catalogue. In order to further exploit the available X-ray data of high-redshift AGN, and push the limits of this study to the fainter X-ray regime, we extract X-ray counts directly from the cleaned Chandra imaging (detailed in \S\ref{chpt3:data:chandra:reduction}) at the positions of the COSMOS2020 sources within our sample. 

Prior to extracting X-ray counts at the COSMOS2020 positions, any COSMOS2020 sources in our sample that lie within 2 times the 90\% EEF radius for the Chandra PSF of a bright X-ray source\footnote{For these purposes, we assume all sources in the blind X-ray source sample count as ``bright X-ray sources''.} position are removed.
This ensures that no counts due to the emission from a bright, nearby source (identified through the blind X-ray detection) are erroneously included when attempting to identify additional, fainter X-ray sources.

For all sources within the COSMOS2020 sample, with those near bright X-ray sources removed, the 70\% EEF radius for the Chandra PSF at each source position is identified from the PSF maps generated in \S\ref{chpt3:data:chandra:reduction} At the position of each COSMOS2020 source in our sample we place apertures of radius given by the 70\% EEF radius, and measure the total counts and total background within these apertures from the Chandra image and background map respectively\footnote{The value of pixels within the source aperture are only counted if their centre lies within the source aperture. No additional weighting to the pixel values is applied.}. Similarly, the average effective exposure within the aperture, for each source, is determined from the exposure map. The net, effective exposure-corrected count rates are calculated as in equation \ref{chpt3:eq:ctr}.

\begin{figure}
    \centering
    \includegraphics[width=\linewidth]{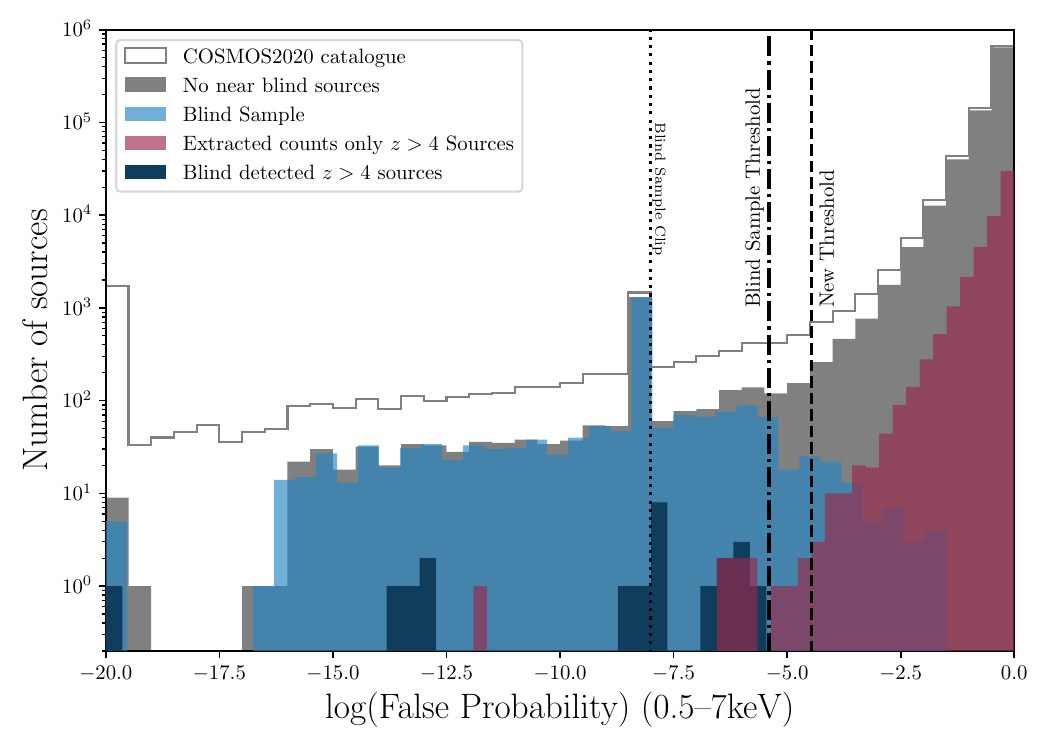}
    \caption[Full-band sample false probability histograms]{Histograms of the false probabilities of sources from the full-band Chandra-COSMOS data, showing values directly extracted from the Chandra imaging or the blind detections where available. The full sample of COSMOS2020 sources is shown by the grey outline, with those remaining once sources around the blind X-ray detected sources are removed shown in solid grey. The blind only sample, obtained using the full-band false probability threshold of $4\times10^{-6}$, is shown in blue (the histogram extends beyond the $4\times10^{-6}$ threshold due to the detection of sources in the soft, hard or ultrahard bands, but not in the full-band shown here).     The high redshift sources within this sample, for both the blind only sources (navy) and the newly extracted sources only (pink) are shown. The updated threshold applied to the extracted sample is indicated by the dashed line, with the threshold of $4\times10^{-6}$ shown by the dot-dashed line, for comparison. For this histogram, we set a minimum false probability of $10^{-20}$, resulting in a peak of sources where all those with a false probability below this are set to a value of $10^{-20}$. Similarly any sources with extremely low (but non-zero) false probabilities within the blind detected source catalogue are set to a false probability of $1\times10^{-8}$ (as indicated by the thin dotted line) in order to distinguish them from a false probability of 0, this results in an artificial peak in source numbers at this false probability.}
    \label{chpt3:fig:falseprob}
\end{figure}

For each extracted count rate of the COSMOS2020 sources, the false probability is calculated using equation \ref{chpt3:eq:falseProb}. As can be seen by the pink histograms in figure \ref{chpt3:fig:falseprob}, the frequency of sources increases exponentially towards a false probability of 1, i.e the majority of the extracted count rates for our COSMOS2020 sources are consistent with being produced by random fluctuations in the background. 

In order to identify those sources for which the extracted count rates are unlikely to be due purely to a random background fluctuation, and thus obtain the sample of X-ray detections, we apply a false probability threshold to the extracted sample, as was done for the blind X-ray catalogue. However, as only positions of known COSMOS2020 sources are considered, and thus where a galaxy or AGN was known to be a priori, a less stringent threshold than was used in the formation of the blind catalogue can be applied. This enables fainter X-ray sources to be identified within the Chandra imaging, and increases the size of the sample of high-redshift, X-ray detected AGN without introducing a significant number of additional false X-ray sources. 

\begin{figure}
    \centering
    \includegraphics[width=\linewidth]{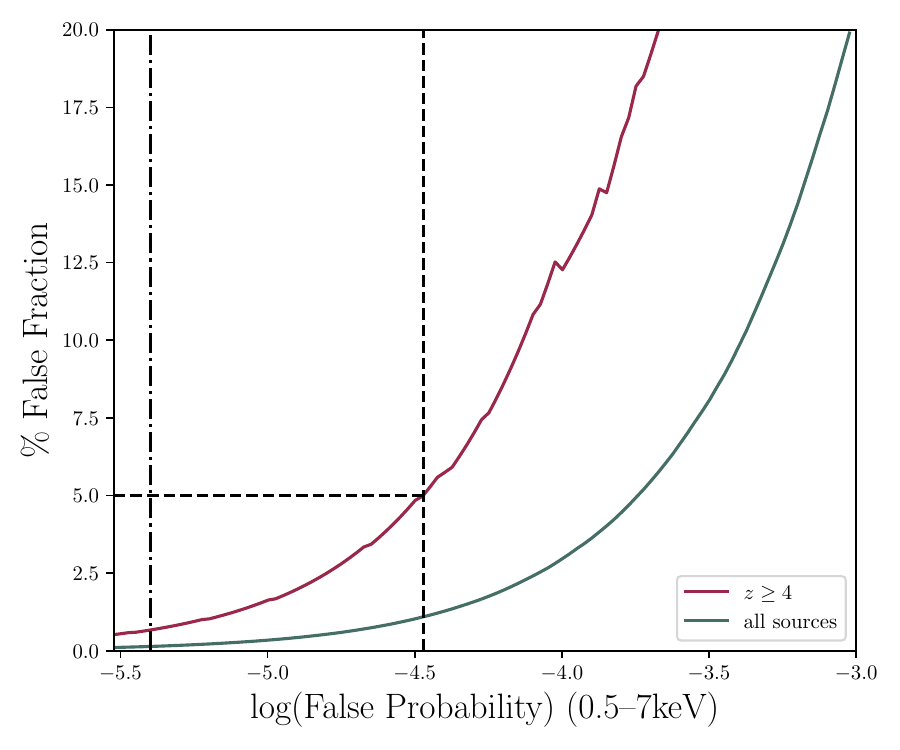}
    \caption[False fraction vs. false probability]{The change in false fraction with increasing false probability threshold, shown for all sources (sage) and $z\geq4$ sources only (pink) in the full-band blind+extracted sample. The blind catalogue false probability fraction of $4\times10^{-6}$ is shown by the dot-dashed line, with the new threshold false probability for the blind+extracted sample ($3.38\times10^{-5}$) shown by the dashed line, corresponding to a False Fraction of 5\% with respect to the $z\geq4$ sources only (shown by the horizontal dashed line for reference).}
    \label{chpt3:fig:falsefrac}
\end{figure}

The threshold false probability is determined iteratively. For a range of false probability thresholds, the fraction of the X-ray detected sample which have been falsely detected (the False Fraction) is calculated, using the following relation,

\begin{equation}
    {\rm False Fraction} = \frac{\lambda_\mathrm{false}}{N(f_\mathrm{prob}<f_\mathrm{prob,threshold})}
    \label{chpt3:eq:FalseF}
\end{equation}

where $N(f_\mathrm{prob}<f_\mathrm{prob,threshold})$ is the number of X-ray detected sources in the sample, with $f_\mathrm{prob}$ less than the threshold, $f_\mathrm{prob,threshold}$, and $\lambda_\mathrm{false}$ is the corresponding expected number of false detections, given by, 

\begin{equation}
    \lambda_\mathrm{false} = N_\mathrm{gal} f_\mathrm{prob,threshold}
    \label{chpt3:eq:FFlambda}
\end{equation}

where $N_\mathrm{gal}$ is the total number of sources in the sample that X-ray data was extracted for. This false fraction calculation is carried out for both the full near-IR--selected source sample and the sample of sources with a best fit photometric redshift of $z\geq4$ (see section \ref{chpt3:xlf:sample} for details on the redshift values). 
The resulting False Fraction (shown in figure \ref{chpt3:fig:falsefrac}) can be seen to rise at lower False Probability Threshold values for the $z\geq4$ sample (shown in blue) compared to the full sample (shown in red).
Thus, for a given False Probability Threshold a higher proportion of the $z\ge4$ sources are expected to be false detections compared to applying the same threshold to the entire galaxy sample. This occurs as, while the sample size of $z\geq4$ sources is smaller than for the full galaxy sample (i.e. $N_\mathrm{gal}$ and thus $\lambda_\mathrm{false}$ is lower), the count rates of true X-ray sources at higher redshifts are typically lower than amongst the full galaxy sample and thus the number satisfying a given $f_\mathrm{prob,threshold}$ is lower, decreasing the denominator in Equation~\ref{chpt3:eq:FalseF} and increasing the false fraction.

We thus adopt the $f_\mathrm{prob,threshold}$ value that corresponds to a False Fraction of 5\% in the $z\geq4$ sample, giving a threshold of $3.38\times10^{-5}$. This threshold is then applied to our combined sample of 882856 extracted and blind X-ray sources, to obtain our final sample of 2591 X-ray AGN containing 21 sources at $z\ge4$. 

\section{Initial measurements of the XLF}\label{chpt3:xlf}

In this section, we present our initial measurements of the XLF of AGN at $z=4-10$. 
In \S\ref{chpt3:xlf:sample} we define our primary sample of high-redshift X-ray detected AGN.
We present measurements of the XLF based purely on the blind X-ray sample in \S\ref{chpt3:xlf:blind} and from the combination of the blind and extracted X-ray samples in \S\ref{chpt3:xlf:extracted}.
In \S\ref{chpt3:xlf:redshift} we consider the impact of differing photometric redshift estimates on our XLF measurements.
Throughout this section, we neglect the impact of intrinsic obscuration on the estimated luminosities of our X-ray AGN and the resulting XLF measurements, deferring such studies to \S\ref{chpt3:obscuration} below.

\subsection{The High-Redshift sample and X-ray Luminosities}\label{chpt3:xlf:sample}
Following the processes detailed in \S\ref{chpt3:data}, we obtained a sample of 2591 \textit{Chandra} X-ray sources with counterparts in the FARMER version of the COSMOS2020 photometric redshift catalogue comprised of 2265 detections from the blind search of \textit{Chandra} Legacy imaging and an additional 326 sources obtained through direct extraction of X-ray counts from the \textit{Chandra} images at known source positions. 
Here we summarise this X-ray sample, the high-redshift cut and properties used in the following sections.
 
To define our sample of high-redshift sources, we adopt the LePhare photometric redshift estimates based on fits performed by \citet{Weaver2022}. However, due to poor imaging or faint sources, a few of the sources in the parent near-IR--selected sample lack photometeric measurements. For these sources LePhare was unable to obtain a fit based on the available photometry (these sources are marked as failed fits by LePhare). To ensure only sources with reliable photometric redshift measurements are included in our high-z sample, we remove the 261 sources for which LePhare was unable to reliably fit a photometric redshift measurement (due to a lack of photometric measurements, contamination by nearby sources or edge effects in the imaging)\footnote{This corresponds to a failed fraction of 10\%. Whilst it would be unphysical for all 261 failed fits to correspond to high redshift sources ($\sim8\times$ more sources than our primary sample), this failed fraction suggest there may be $\sim3$ sources missed from our final $z\geq4$ primary sample. If so our final measurements of the space density would be higher than currently, and thus further indicative of a larger space density than predicted by model extrapolations (see also section \ref{chpt3:discussion}).}. Following this the X-ray source sample is comprised of 2035 blind X-ray detected sources and 295 directly extracted sources, for a total sample size of 2330. 

As detailed above (in \S\ref{chpt3:data:cosmos2020}), the LePhare code was used to perform redshift fits using galaxy templates and again using templates with an AGN spectral component. For sources with multiple redshift measurements from LePhare the better fitting photometric redshift (that with the smaller $\chi^2$), either using galaxy or AGN spectral energy distribution templates, is taken to be the \emph{best} estimate of the source redshift. 
As such 1239 of our X-ray sources have best-fitting photometric redshifts from galaxy spectral models and 1091 sources have best-fitting photometric redshifts based on AGN templates. Applying a redshift cut of $z_\mathrm{best}\geq4.0$ to this sample we obtain the primary high-redshift X-ray sample of 32 sources, with which we perform initial measurements of the observed X-ray Luminosity Function (see \S\ref{chpt3:xlf:blind} and \S\ref{chpt3:xlf:extracted} below). 

For all sources within the high-redshift samples, the measured net count rate in a given band (i.e. the full 0.5--7\,keV and hard 2--7\,keV bands) is converted directly into a rest-frame 2-10\,keV luminosity (denoted as $L_\mathrm{X}$). Following \citet{Aird2015}, we initially assume an X-ray spectral model consisting of a power-law with a photon index $\Gamma=1.9$, a reflection component modelled using the pexrav model \citep[][which assumes an infinite plane reflector]{Magdziarz1995} of relative strength 1.0 and take the Galactic absorption ($N_{\mathrm{H, Gal}}$) to be that of the COSMOS field (average of $10^{20.25}\mathrm{cm^{-2}}$, \citealt{HI4PICollaboration2016}).\footnote{
We note that the spectral model used here is relatively simple and does not correctly model intrinsically obscured X-ray sources. To investigate the impact of obscuration on the XLF measurements in Section \ref{chpt3:obscuration}, we also use a more complex spectral model from \citet{Balokovic2018} which includes the impact of both photoelectric absorption and Compton scattering of high energy photons by a dusty ``torus'' component.}
This model is then folded through the response function of {\it Chandra} to determine the conversion between the net, effective-exposure corrected count rate measured with {\it Chandra} ($c$ from Equation \ref{chpt3:eq:ctr}) and the rest-frame 2-10~keV luminosity ($L_\mathrm{X}$) of each source.
\begin{equation}
    L_\mathrm{X}=c\times\kappa(z, N_\mathrm{H, Gal}=10^{20.25},\Gamma=1.9)\mathrm{,}
    \label{chpt3:eq:Lx}
\end{equation}
where $\kappa(z, N_{\mathrm{H, Gal}}, \Gamma=1.9)$ is the redshift-dependent conversion factor. 
To calculate these conversion factors we adopt the Cycle 16 ACIS-I on-axis response function, corresponding to the approximate middle of the Legacy survey observing period. 
Variations of this response over the observation period of the C-COSMOS and COSMOS Legacy surveys (Cycles 9--17), mainly due to degradation of the sensitivity at low energies due to radiation damage and molecular contamination of the detector \citep[see e.g.][\footnote{See also The Chandra Proposers' Observatory Guide v27.0 \url{https://cxc.harvard.edu/proposer/POG/html/chap6.html}}]{Grant2024, ODell2007a, Cameron2002}, as well as changes in relative response at different off-axis angles are captured by the effective exposure maps. 
We note that the effective exposure maps---and thus the count rates, $c$---are calculated assuming a simpler $\Gamma=1.4$ spectral model (see Section~\ref{chpt3:data:chandra:reduction}), but our conversion factors incorporate the differing spectral assumptions when converting from a count rate directly into luminosities.

\begin{figure}
    \centering
    \includegraphics[width=\linewidth]{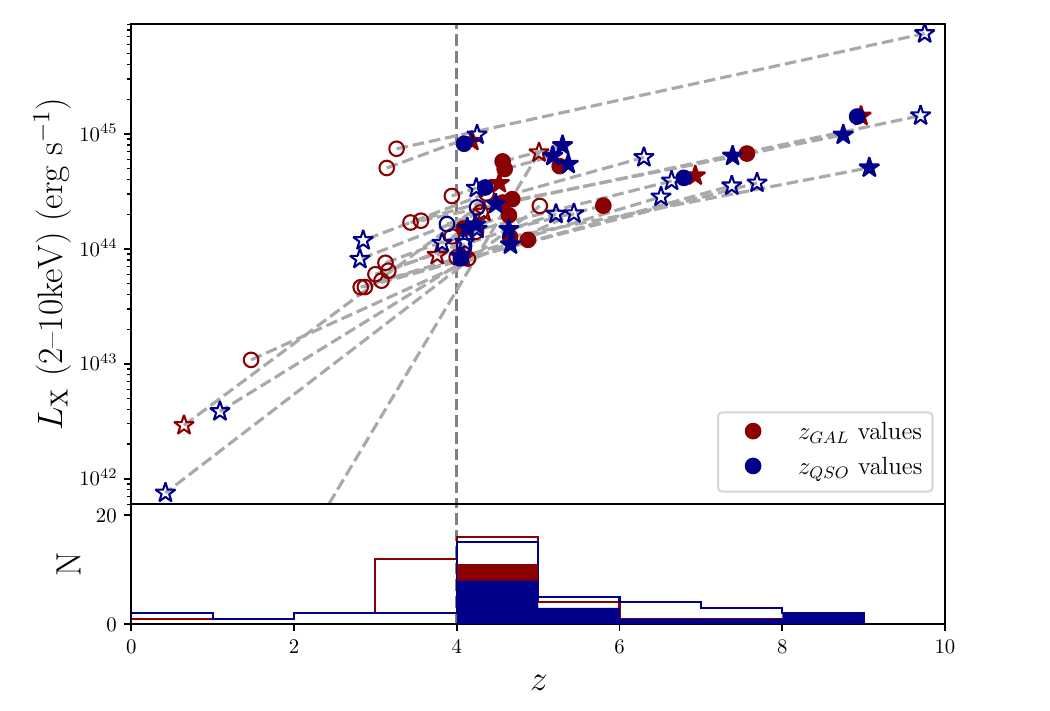}
    \caption[Luminosities of blind+extracted sample]{Rest-frame 2-10\,keV X-ray luminosity against redshift, obtained from the full band count rate, for all sources included in our high-redshift samples. Sources are shown at redshifts based on both the galaxy template ($z_{GAL}$) and AGN template ($z_{QSO}$) photometric redshift fits (red and blue respectively). Sources that would leave the high redshift sample given either their galaxy or AGN redshift fit are indicated by empty points. For each source the best-fitting photometric redshift is shown as a star. The corresponding number of sources with $z_{GAL}$ or $z_{QSO}$ redshift is shown by the histogram, for the most conservative sample (solid) and the most optimistic sample (empty).}
    \label{chpt3:fig:LXvsz}
\end{figure}

We determine $L_\mathrm{X}$ for the sources in our high-redshift samples using both 
$z_\mathrm{gal}$ and $z_\mathrm{QSO}$.
As can be seen in Figure \ref{chpt3:fig:LXvsz}, sources can move in and out of our redshift range depending on the choice of photometric redshift, as well as changing their estimated $L_\mathrm{X}$ (although the change in the number of sources at a given redshift has a more significant effect on our XLF measurements). In \S\ref{chpt3:xlf:redshift} we investigate the impact of this uncertainty on the measured XLF.

\subsection{XLF measurements from the blind source catalogue}\label{chpt3:xlf:blind}

For our initial measurements of the XLF, we first consider only the more secure (lower false probability threshold) blind X-ray detected sources in the primary (i.e. $z_\mathrm{best}>4$) sample, consisting of 21 sources.
Our sample size remains too small to perform a secure fit to a parametric model of the XLF. 
Thus, we divide these 21 $z\geq4$ sources into redshift bins of $z=4-5$, $5-7$ and $7-10$, corresponding to the highest redshift where 
parametric XLF models have been constrained in prior studies, the epoch of cosmic reionisation and extremely high redshifts, respectively. 
The sources within each of these redshift ranges are then split into $0.5\,\mathrm{dex}$ wide luminosity bins from $\log{L_\mathrm{X(2-10keV)}\mathrm{\,erg\,s^{-1}}}=43-46$, which allows the shape of the X-ray Luminosity Function to be seen across each redshift range from binned measurements.

We estimate the space density of AGN, in the form of the XLF ($\phi_\mathrm{obs}(L_{\mathrm{X},i},z_j)$), in each luminosity ($L_{\mathrm{X},i}$) and redshift ($z_j$) bin, using the $n_\mathrm{obs}/n_\mathrm{mdl}$ method of \citet{Miyaji2001}. This method accounts for changes in the X-ray sensitivity and the shape of the underlying XLF within our comparatively broad redshift and luminosity bins, providing a more robust estimate of the XLF compared to more simplistic methods (e.g. $1/V_\mathrm{max}$ methods) given the low source numbers within our sample. 

\begin{figure*}
    \centering
    \includegraphics[width=\linewidth]{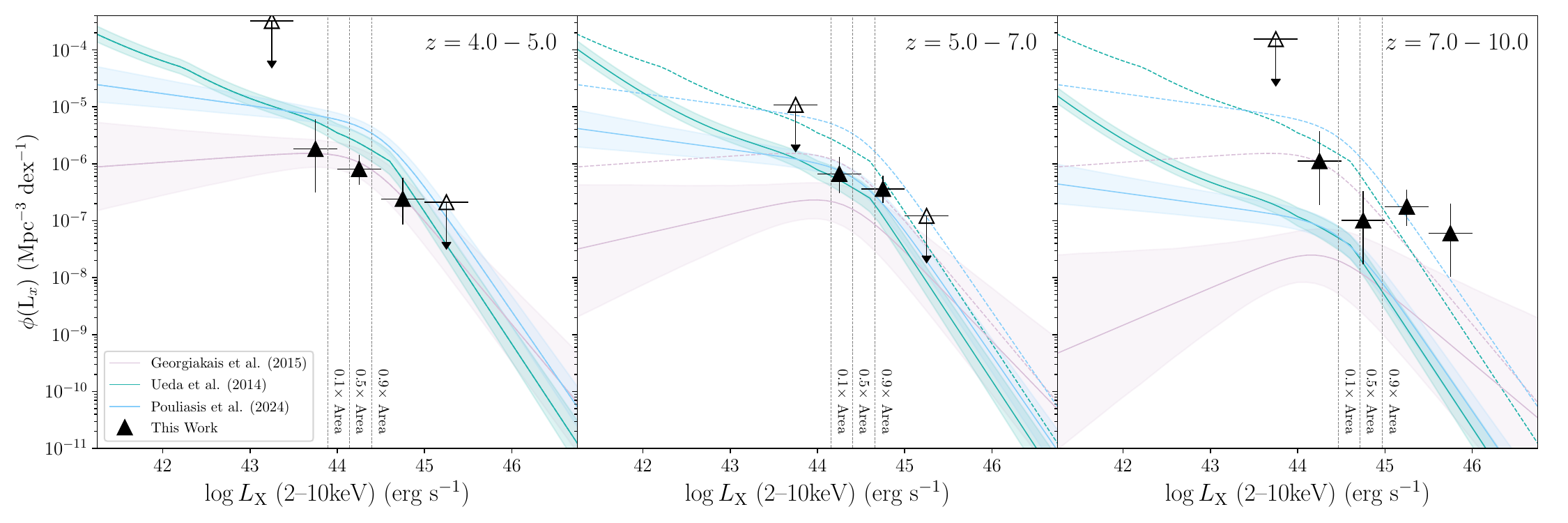}
    \caption[Blind sources full-band measurements of the XLF]{Binned measurements of the high redshift XLF of AGN (black triangles), based on the cross-matched blind source sample and the best-fitting photometric redshift estimates (i.e. the primary high-redshift sample).
        At $z=4-5$ (left hand panel) our measurements are consistent with the model XLF of \citet{Georgakakis2015} (pink), \citet{Ueda2014} (green) and \citet{Pouliasis2024} (blue), which were fit to data within the same redshift range. At $z=5-7$ our measured space densities are higher than predicted by the extrapolated model of \citet{Georgakakis2008} but consistent with \citet{Ueda2014} and \citet{Pouliasis2024}, while at $z=7-10$ (right hand panel) we measure significantly higher space densities than predicted by any of the model extrapolations. The luminosities corresponding to the {\it Chandra} sensitivity being achieved over 10\%, 50\% and 90\% of the total survey area are indicated by the vertical dashed lines. Upper limits are provided by the upper Poisson $1\sigma$ uncertainty on 0 sources, as given by \citet{Gehrels1986}.
    For comparison the XLF parametric models evaluated at $z=4.5$ (i.e. as shown in the $z=4-5$ to panel) are shown in the $z=5-7$ and $7-10$ panels by the dashed pink, green and blue lines. 
    }
    \label{chpt3:fig:blindXLF}
\end{figure*}

To obtain binned XLF measurements, the predicted value of the XLF, as given by a pre-existing parametric model evaluated at the centre of each bin, is scaled by the ratio of the number of observed sources in that bin to the number of sources predicted by the model:
\begin{equation}
    \phi_\mathrm{obs}(L_{\mathrm{X},i}, z_j)= \phi_\mathrm{mdl}(L_{\mathrm{X},i}, z_j)\frac{n_\mathrm{obs}}{n_\mathrm{mdl}}
    \label{chpt3:eq:NobsNmdl}
\end{equation}
where $n_\mathrm{obs}$ is the observed number of sources in each bin. The parameterised XLF model, $\phi_\mathrm{mdl}(L_{\mathrm{X}_i}, z_j)$, is taken to be the value of the Luminosity Dependent Density Evolution (LDDE) XLF model of \citet{Georgakakis2015} at the centre of the $z-L_{X}$ bin (i.e. at $L_{\mathrm{X},i}$ and $z_j$). 
Despite requiring a parametric model from previous XLF studies, 
the $n_\mathrm{obs}/n_\mathrm{mdl}$ method is only marginally sensitive to the specific choice of model; the larger uncertainties due to our low source numbers dominates here. Finally, the predicted number of sources in the bin, $n_\mathrm{model}$, is given by
\begin{equation}
    n_\mathrm{mdl}=\int_{z_1}^{z_2}\int_{\log{L_{X_1}}}^{\log{L_{X_2}}} \phi(L_X,z)
    A\big(c | L_X,z
    \big) \frac{\mathrm{d}V_\mathrm{co}}{\mathrm{d}z}\;\mathrm{d}\log{L_X}\;\mathrm{d}z ,
    \label{chpt3:eq:Nmdl}
\end{equation}
where $\phi(L_X,z)$ is the parametrised XLF model that varies across the $L_\mathrm{X}-z$ bin, $\frac{\mathrm{d}V_\mathrm{co}}{\mathrm{d}z}$ is the differential comoving volume at a given redshift, $z$, and $A(c | L_X ,z)$ is the sky area imaged by Chandra to a given count rate, $c$, corresponding to a given 2--10\,keV rest-frame luminosity, $L_X$, and redshift, as determined in \S\ref{chpt3:data:chandra:area} with a false probability threshold of $4\times10^{-6}$ (as used in the blind sample creation). 
For each redshift and luminosity interval, error bars are given by the $1\sigma$ equivalent uncertainties (i.e. 68.3\% confidence interval) on the number of sources based on Poisson uncertainties in the source numbers \citep{Gehrels1986} folded through the calculation of the observed XLF. Where there are no observed sources in a bin (i.e. $N_\mathrm{obs}=0$) we place an upper limit on the XLF based on the upper $1\sigma$-equivalent limit for a sample size of 0.
The resulting observed, binned XLF measurements are shown in Figure \ref{chpt3:fig:blindXLF}.

For comparison, we also plot the parametric models of \citet{Georgakakis2015} (LDDE model), \citet{Ueda2014} and \citet{Pouliasis2024} that are based on fits to data at redshifts of $z=3-5$, $z=0.002-5$ and $z=3-6$ respectively, extrapolated here when necessary to the central redshift of each bin; $z=4.5$, $z=6$ and $z=8.5$ respectively. The uncertainties in these models are obtained through Monte Carlo simulations using the quoted XLF model parameter values and their uncertainties.\footnote{We note that propagating the quoted uncertainties in each parameter when extrapolating these prior XLF measurements will not fully capture co-variances between parameters.}

At $z=4-5$ our observed XLF measurements are consistent with the parametric models, which have been fit to data including this redshift range. However, as the redshift increases, our measurements of the XLF can be seen to remain at high space densities, with only a slight decrease from the values at $z=4-5$ towards $z=7-10$, whilst the extrapolated models predict a clear drop in the space density. These results indicate a less steep evolution in the space density of AGN in the early Universe than has been predicted from parametric models, with more sources found at higher redshifts than previously expected.

Despite the obvious trends revealed by our measurements, there remain large uncertainties due to the small number of sources. To improve the constraints further, in the next sub-section we combine the full sample of blind detections and directly extracted sources to improve our constraints.

\subsection{XLF measurements from both blind and extracted X-ray samples}\label{chpt3:xlf:extracted}
In order to fully exploit the deep X-ray data available from the \emph{Chandra} COSMOS-Legacy survey, and improve the statistics of our measurements of the high-redshift AGN X-ray Luminosity Function, we extracted X-ray counts at the positions of all COSMOS2020 sources. A sample of X-ray AGN is then determined from the combination of the blind X-ray detected sources and the directly extracted X-ray measurements (see \S\ref{chpt3:data:extraction} for details of the sample creation).
By using the known positions of sources based on pre-existing, multi-wavelength imaging, we are able to increase our sample of high-redshift sources and include lower significance detections without introducing substantial contamination due to false-positive detections. 
Following the process detailed above (in section \ref{chpt3:xlf:blind}), we obtain the blind+extracted sample measurements of the XLF using this new X-ray AGN catalogue.

\begin{figure*}
    \centering
    \includegraphics[width=\linewidth]{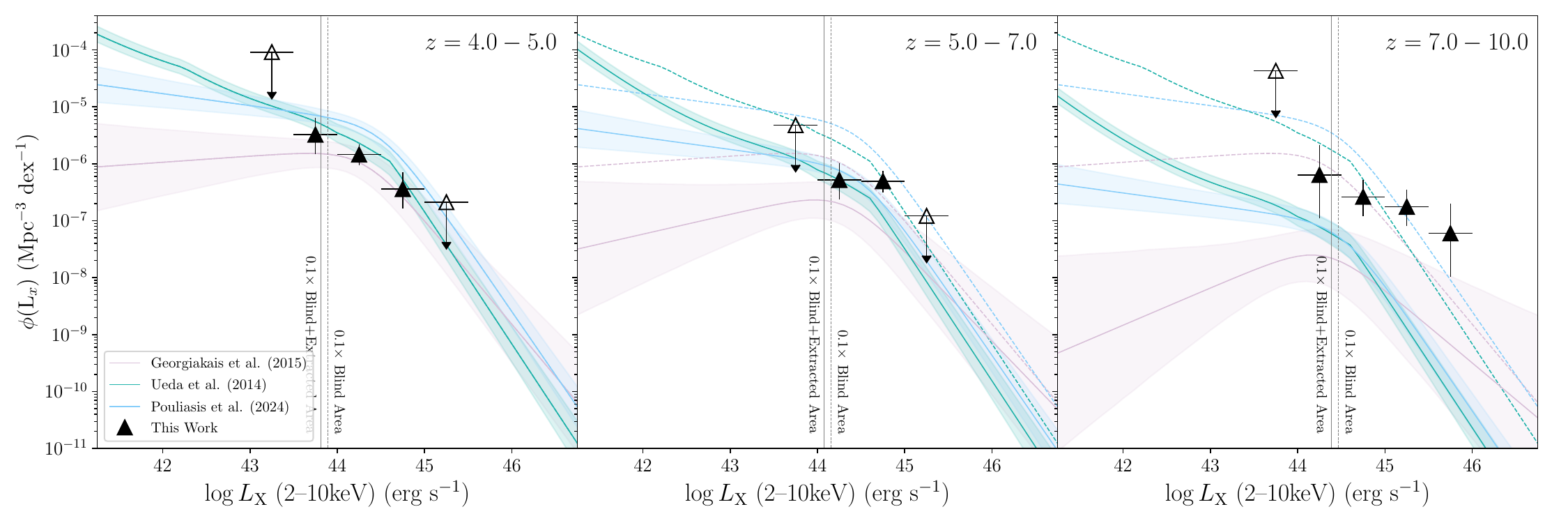}
    \caption[Blind+extracted full-band measurements of the XLF]{As for Figure \ref{chpt3:fig:blindXLF}, with sources from both the blind source detection and directly extracted counts. Consistent with models at $z=4-5$, our binned XLF measurements begin to hint at a higher than predicted space density at $z=5-7$ whilst clearly indicating the space density of AGN is higher than predicted by extrapolations of parametric models at $z=7-10$.}
    \label{chpt3:fig:extractedXLF}
\end{figure*}

The rest-frame 2-10\,keV luminosity ($L_\mathrm{X}$) of each source within this combined catalogue is based on the X-ray count rates from the blind source catalogue, where available; otherwise the directly extracted X-ray measurements are used. We measure the binned XLF based on this combined sample of $z\geq4$ X-ray detected sources as described above for the blind X-ray detected source sample only.
As a less stringent false probability threshold was used in the creation of the blind$+$extracted X-ray sample ($f_\mathrm{prob,thresh}=3.38\times10^{-5}$), the area curve for the {\it Chandra} imaging is recalculated (see \S\ref{chpt3:data:chandra:area} for details of the calculation). This new area curve is used to determine the predicted number of sources (Equation \ref{chpt3:eq:Nmdl}). The resulting measurements of the high-redshift XLF are shown in figure \ref{chpt3:fig:extractedXLF}.

With the less stringent false probability threshold possible for this combined sample, we have improved the sensitivity of the X-ray observations and increased the area of sky visible by Chandra at lower luminosities. This new sensitivity has reduced the upper limits on the space density at the faint-end of our observed XLF measurements, allowing for tighter constraints on the faint-end of the high-redshift XLF. 

Due to the larger sample size used here, we have increased the number of sources at each redshift interval, and thus reduced the uncertainty in the observed XLF measurements. As can be seen in figure \ref{chpt3:fig:extractedXLF}, the space density at $z=4-5$ increases slightly from that in figure \ref{chpt3:fig:blindXLF}, to be above that of \cite{Georgakakis2015}, whilst remaining consistent with all extrapolated models shown. 
These improved observed XLF measurements remain significantly higher than the model predictions at  $z=7-10$ and continue to indicate a higher space density at $z=5-7$ than previously predicted.
We also note the relatively flat slope of our $z=7-10$ measurements over the full $L_\mathrm{X}\approx 10^{44-46}$~erg~s$^{-1}$ range, with no evidence of a break at $L_\mathrm{X}\sim10^{44.5}$~erg~s$^{-1}$ seen in the model extrapolations.

\subsection{Impact on the XLF measurements from differing photometric redshift estimates}\label{chpt3:xlf:redshift}
Our adopted photometric redshifts were obtained by fitting different combinations of templates to the observed Spectral Energy Distributions (SEDs) of sources in the COSMOS2020 catalogue. The choice of templates and fitting process can impact the measured redshift, in particular the inclusion of an AGN component to the SED can drastically alter the resulting redshift measurement. The LePhare code was used to determine a redshift for each source in the COSMOS2020 catalogue both with and without an AGN component in the templates; these fits for all sources in our high-redshift samples are shown in Appendix \ref{apdx:SEDs}. As such, all sources within our blind+extracted X-ray sample have two possible photometric redshifts, with the choice of SED fit having the potential to dramatically alter the measured XLF. 

As can be seen in Figure \ref{chpt3:fig:LXvsz}, the photometric redshift of a source when found with AGN templates can alter whether that source falls within the high-redshift sample, compared to its photometric redshift as given by galaxy-only templates. 
The different fits can also move sources between our redshift bins and alter the estimated $L_\mathrm{X}$.
For the previous measurements of the XLF, in \S\ref{chpt3:xlf:blind} and \S\ref{chpt3:xlf:extracted}, we used the redshift of the best (lowest $\chi^2$) template fit, as described in \S\ref{chpt3:xlf:sample}. 
Here we investigate the impact of the different photometric redshifts in order to obtain the most conservative and most optimistic values for the XLF measurements based on our X-ray sample.

\begin{figure}
    \centering
    \includegraphics[width=\linewidth]{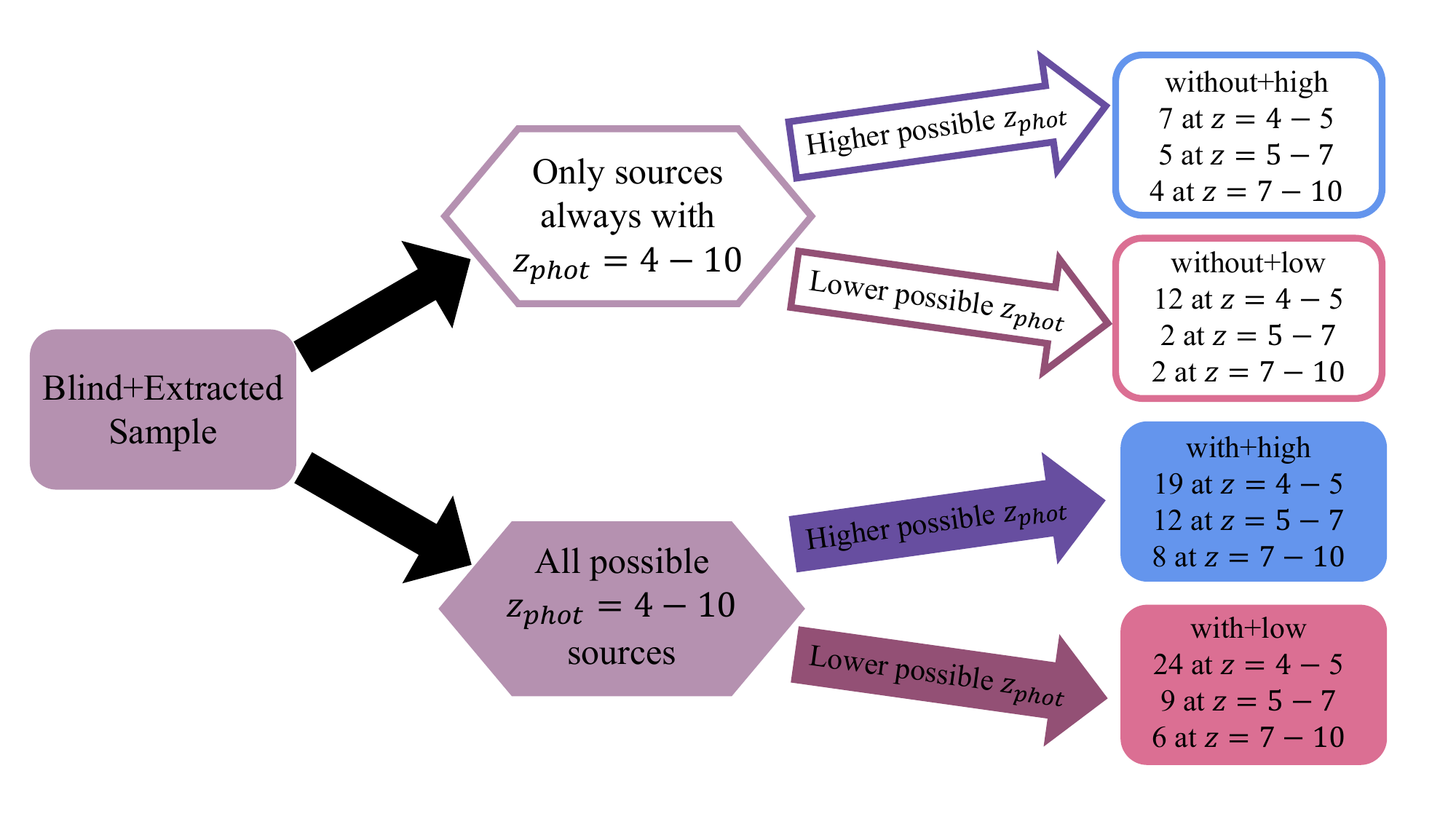}
    \caption[Redshift sample variant creation]{Flow chart illustrating the creation of the four photometric-redshift version dependant high-redshift samples, from our blind+extracted sample, with+low (solid pink), with+high (solid blue), without+low (empty pink) and without+high (empty blue). The number of sources in each redshift interval are shown for the four sample variants.}
    \label{chpt3:fig:redshiftsamples}
\end{figure}

Considering both forms of the photometric redshift, from the galaxy-only templates and the AGN templates, we generate four different high-redshift samples. The creation of these samples is summarised in Figure \ref{chpt3:fig:redshiftsamples}. Some of the sources within our X-ray sample only fall into the range of $z=4-10$ with either their galaxy or their AGN photometric redshift fits, but not the other. 
Thus, we consider here sources where \emph{either} the galaxy photometric redshift, $z_\mathrm{gal}$, \emph{or} the AGN photometric redshift, $z_\mathrm{QSO}$, is $\geq 4.0$, thus creating an inclusive sample of 39 possible $z\geq4.0$ sources  hereafter referred to as the ``with" sample.
In addition, we create a conservative high-redshift sample of 16 sources where we require that \emph{both} $z_\mathrm{gal}$ and $z_\mathrm{AGN}$ are $\geq 4.0$ (even if the redshift values themselves differ), and remove any sources that \emph{might} be at lower redshifts (referred to as the ``without" sample). 
For the sources where both $z_\mathrm{gal}$ and $z_\mathrm{AGN}$ places them within the $z=4-10$ redshift range, we also consider the cases where the redshift of each source is taken to be the smallest of the two redshift values and the case where it is taken to be the largest value, hereafter referred to as the ``low" and ``high" samples respectively (that can be defined from both the ``with'' and ``without'' samples). The resulting distribution of source numbers for the four samples (with+low, with+high, without+low, without+high) are shown in figure \ref{chpt3:fig:redshiftsamples}. 

\begin{figure*}
    \centering
    \includegraphics[width=\linewidth]{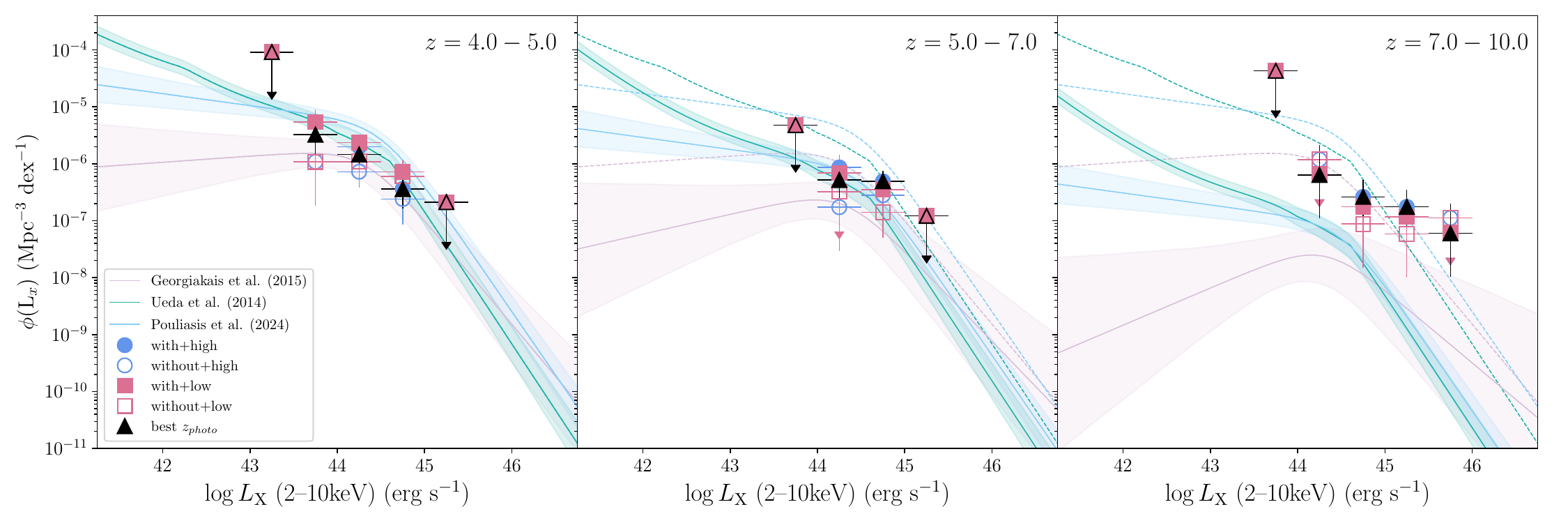}
    \caption[Measured XLF given by four versions of the blind and extracted Chandra-COSMOS2020 $z\geq4$ sample.]{Binned XLF measurements at $z\geq4$ determined from the blind and extracted sample using either the lowest (pink) or the highest (blue) possible photometric redshift, with (solid symbols) and without (empty symbols) sources for which only one photometric redshift measurement, given by the galaxy or AGN redshift fits, is $z\geq4$. For comparison, the measured XLF from the best redshift values (as found in \S\ref{chpt3:data:extraction}) is shown by the black points. Parametric models from \citet{Ueda2014}, \citet{Georgakakis2015} and \citet{Pouliasis2024} (green, pink and blue lines respectively) are shown extrapolated to the centre of each redshift range (solid) or evaluated at a fixed $z=4.5$ (dashed). Even for the lowest measured XLF, providing the most conservative result, we find a higher space density than predicted by the parametric models at $z=7-10$ and consistent with minimal evolution since $z\sim4$.}
    \label{chpt3:fig:zrangesxlf}
\end{figure*}

The observed XLF measurements are determined for all four high-redshift sample variants following the process detailed in \S\ref{chpt3:xlf:blind}. The without sample can be expected to produce the most conservative XLF measurements; the without+high sample producing the lowest XLF measurements in the $z=4-5$ range and the without+low giving the smallest measurements in the $z=7-10$ redshift range. Conversely, there will be a maximum XLF measurement in the $z=4-5$ and $z=7-10$ ranges when determined from the with+low and with+high samples respectively. The four forms of the XLF measurements are shown in figure \ref{chpt3:fig:zrangesxlf}.

At $z=4-5$ the different XLF measurements all remain consistent with the models, whilst the most optimistic values suggest the XLF tends to higher space densities at the faint end. Towards higher redshifts, however, even the most conservative measurement indicates a higher space density of AGN than predicted by the extrapolations of models. 

\section{Obscured AGN and obscuration-corrected measurements of the XLF}\label{chpt3:obscuration}
Obscuring material in both the host galaxy and torus of dusty material within the AGN will absorb lower energy, soft X-ray photons produced by the corona more readily than higher energy, hard photons, preventing them from escaping the system and thus reducing the observed soft-band counts from a source. As such, a more heavily obscured AGN will appear to produce relatively more counts in the hard-band than in the soft-band, as more low energy photons are absorbed. Thus, despite the same initial production of soft and hard photons, an obscured source will have a stronger signal at harder X-rays relative to its observed emissions in softer bands \citep[see][]{Hickox2018}.
In the early Universe, as fewer stars have formed, galaxies are expected to have a higher gas content \citep[see e.g.][for review]{Carilli2013} and thus early SMBHs are likely to be more heavily obscured due to galaxy-wide gas. This higher gas content may also increase the obscuration due to the torus around the SMBH as it is funnelled into the central regions of the galaxy, fuelling the growth of the SMBH that is expected to occur. Studies such as that of \citet{Vito2018} have found the obscured fraction of AGN within their samples to increase with redshift, consistent with these expectations.

A decrease in observed counts at Soft and Full band energies, due to obscuring material around the AGN, results in the intrinsic luminosity of sources being under estimated and can result in an AGN not being observed above the X-ray background. Thus, increased obscuration may result in AGN which have lower observed luminosities and thus reduce the measured space density of AGN. 
At the higher redshifts probed in this study, our observed X-ray bands probe relatively high rest-frame energies, mitigating the impact of obscuration, but sufficiently high levels of obscuration will still affect our observed count rates (and thus inferred luminosities), particularly for column densities in the Compton-thick regime.
In this section we investigate the effect of obscuration in the high-redshift blind+extracted sample on our measurements of the X-ray Luminosity Function (see \S\ref{chpt3:obscuration:hardxlf} and \S\ref{chpt3:obscuration:bayHR}). We will then perform a simple correction to the measurements, in order to account for any obscuration present in our sample (see \S\ref{chpt3:obscuration:correct}). 

\subsection{The Hard-Band XLF}\label{chpt3:obscuration:hardxlf}
Obscured sources are typically more easily detected at hard-band (2--7\,keV) energies than in the soft-band (0.5--2\,keV). However, the measurements of the XLF we performed above were done using the full-band (0.5--7\,keV) only in order to maximise our sample size and sensitivity. Given the high redshifts of our sample the observed hard-band imaging from Chandra will be probing very high energies, of $>10$~keV, and should thus be relatively immune to obscuration (for $\mathrm{N_H}<10^{24}$) compared to the full-band observations. Thus, a first step to assess whether obscuration in our high-redshift sources has a significant effect upon the observed XLF measurements is to calculate the observed XLF from source detections made in the hard-band, which provides a more reliable tracer of $\mathrm{L_X}$ at the expense of sensitivity and sample size.

Using the sample of blind detected and hard-band extracted sources, obtained in \S\ref{chpt3:data:extraction} with a false probability threshold of $1.74\times10^{-5}$ corresponding to a $5\%$ False Fraction in the $z\geq4$ sources, we calculate the observed XLF measurements given by the hard-band count rates. The measured hard-band count rates of our sources are converted directly to a rest-frame 2-10\,keV X-ray luminosity ($\mathrm{L_X}$, as in \S\ref{chpt3:xlf:sample}) and the measured XLF given by these hard-band observations is determined following the process detailed in \S\ref{chpt3:xlf:blind}. For these measurements, the sensitivity in the hard-band is calculated using the false probability threshold of $1.74\times10^{-5}$ (see \S\ref{chpt3:data:chandra:area}).

\begin{figure*}
    \centering
    \includegraphics[width=\linewidth]{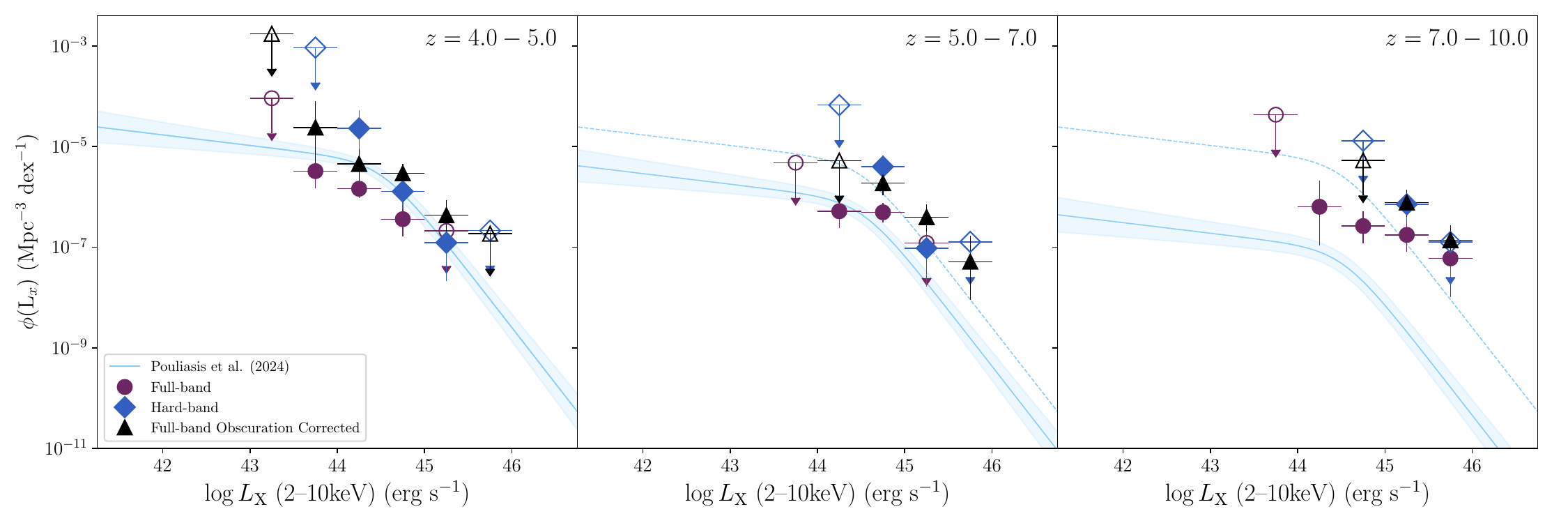}
    \caption[Full-band obscuration-corrected XLF]{ Full band obscuration corrected XLF measurements (black triangles; determined in \S\ref{chpt3:obscuration:hardxlf}) for the redshift bins of $z=4-5$, $5-7$ and $7-10$. The uncorrected full band (purple circles; from figure \ref{chpt3:fig:extractedXLF}), hard band (blue diamonds) and \citet{Pouliasis2024} model extrapolations are shown for comparison. Our corrected measurements can be seen to be more consistent with the hard band measurements than the uncorrected case, indicating an even higher space density of AGN at these high-redshifts than is predicted by the model extrapolations.}
    \label{chpt3:fig:correctedXLF}
\end{figure*}

The resulting measurements of the hard-band XLF are shown by the blue diamonds in figure \ref{chpt3:fig:correctedXLF}, with the original full-band measurements shown by the purple circles for comparison. The hard-band XLF measurements indicate a higher space density of AGN, in the early Universe, than is seen with the full-band measurements. This disparity between the full and hard band measurements can be seen to increase with redshift. This is indicative of an increasingly obscured population towards higher redshifts, as has been indicated by prior X-ray studies from \citet{Vito2018, Peca2002} and in recent discoveries of large obscured populations of AGN by JWST \citep{Yang2023, Maiolino2023, Greene2023} which have generally remained undetected at X-ray wavelengths. With our hard-band XLF measurements indicating significant obscuration within the high-redshift sample, our full-band measurements of the XLF can be expected to be significantly affected by this obscuration. However, obscuration may vary between sources rather than affecting the sample in a uniform manner. As such, in order to further analyse the obscuration within our sample and the impact on our XLF measurements, we must determine the hardness ratios of each source in the full-band sample. 

\subsection{Hardness Ratios}\label{chpt3:obscuration:bayHR}
A good indication of the amount of obscuration there is in an AGN, by material surrounding the black hole and accretion disk, is given by the ratio of the measured soft and hard counts of the source, known as the hardness ratio. Fewer soft-band photons will escape in a more obscured AGN and hence it will have a more positive hardness ratio than a less obscured AGN of the same intrinsic spectrum. 

For all sources within our high-redshift blind+extracted sample (summarised in section \ref{chpt3:xlf:sample}) we determine their Hardness Ratios, following the Bayesian Estimation method detailed in \citet{Park2006}. As detailed in \S\ref{chpt3:data:extraction}, the total hard and soft-band counts of each source are taken to be the 70\% PSF measurements obtained in the blind catalogue creation, where available, or extracted directly from the Chandra imaging at the position of the COSMOS2020 source. The posterior distribution of possible values of the Hardness Ratio, for each source in the sample, is then determined allowing for the Poisson nature of the hard and soft band counts. As such the Hardness Ratio of sources only detected in one energy band can still be determined and constraints can be placed on the level of obscuration of such a source. For this calculation we take the form of the Hardness Ratio ($HR$) to be,

\begin{equation}
    HR=\frac{H-S}{H+S},
\end{equation}

where $H$ and $S$ are the total number of source counts measured in the hard and soft-bands, respectively. The Hardness Ratio for each source within our high redshift sample is thus calculated from the measured total counts in the hard and soft bands, with the corresponding background and effective exposure (that re-weights all HRs to account for the telescope vignetting and combination of {\it Chandra} observations from different epochs), using a Gaussian quadrature algorithm \citep[as detailed in Appendix A.2 of][]{Park2006}. 

\begin{figure}
    \centering
    \includegraphics[width=\linewidth]{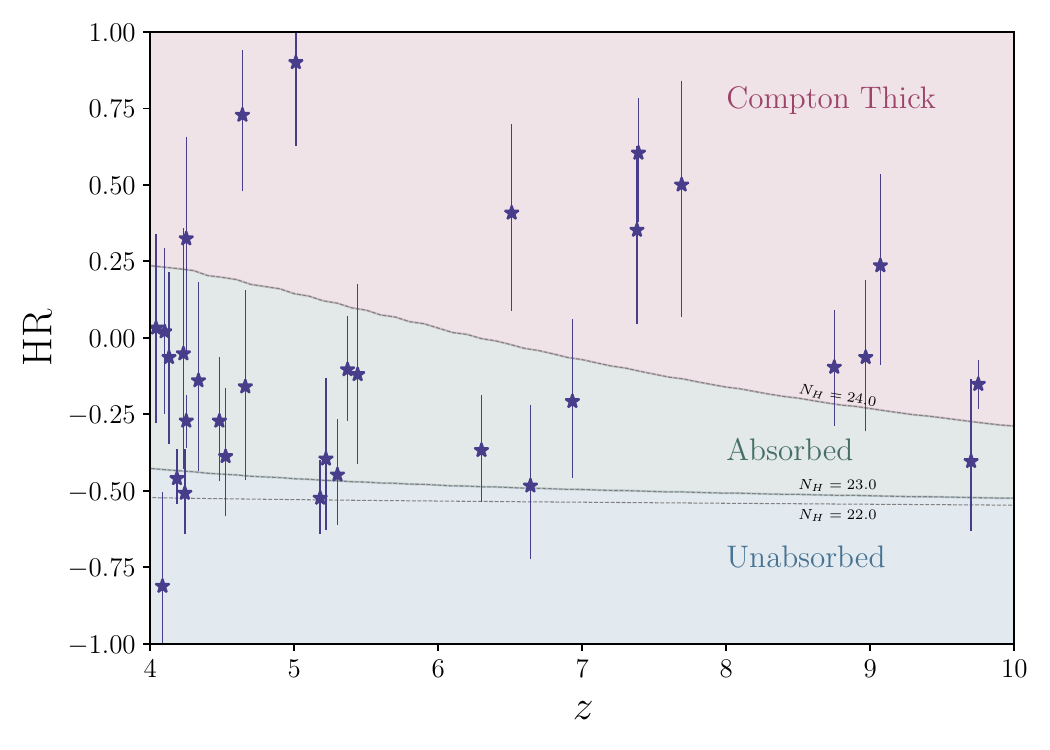}
    \caption[Redshift dependence of HR]{The Hardness Ratio (HR) with respect to the best-fitting redshift of all sources within our blind+extracted sample. The dotted lines indicated the expected value of HR based on the Borus spectral model \citep{Balokovic2018} with a column density of $\mathrm{N_H}=10^{22}\mathrm{cm^{-2}}$, $10^{23}\mathrm{cm^{-2}}$ and $10^{24}\mathrm{cm^{-2}}$, and are used to divide our sample into corresponding unabsorbed, absorbed and Compton Thick regions. It can be see that at $z>7$ the majority of sources lie within the Compton Thick region, however there remains a broad distribution of HR in the sample.}
    \label{chpt3:fig:BEHRz}
\end{figure}

The resulting hardness ratios for all $z\geq4$ sources within our blind+extracted sample are shown in figure \ref{chpt3:fig:BEHRz}. 
We find a weak trend in the overall hardness of the population, with the majority of $z>7$ sources lying at relatively high hardness ratios ($HR\gtrsim -0.25$).
As can be seen in figure \ref{chpt3:fig:BEHRz}, there is significant scatter across the redshift range of this study, indicating that a single obscuration correction applied to the XLF measurements would be inaccurate. However, it should be noted that the conversion between Hardness Ratio and the obscuration, $\mathrm{N_H}$, of a source is dependent upon the redshift. This is due to the shift in the X-ray spectrum towards lower energies with redshift, and as such two sources of the same $\mathrm{N_H}$ but different redshifts, will have different $HR$ values.

Previously our spectral models assumed negligible intrinsic obscuration, using the spectral model of \citet{Aird2015}. Here, in order to account for the effect of obscuration on our observed source counts we employ the Borus spectral model of \citet{Balokovic2018}, which includes an obscuring torus component. Using this Borus model (folded through the Chandra response at approximately the middle of the survey period as in section \ref{chpt3:xlf:sample} to convert count rates into intrinsic rest-frame 2--10~keV luminosities)\footnote{The degradation of Chandra's sensitivity across its lifetime has negligible impact upon our results \citep[see][]{Barlow-Hall2024}, and are incorporated into the exposure maps used to obtain source counts.} we determine the Hardness Ratio across the redshift range of $z=4-10$ corresponding to constant obscuration (i.e. constant values of $\mathrm{N_H}$). 
With increasing redshift, the observed hardness ratio for a source of given $\mathrm{N_H}$ decreases, with a more rapid drop at higher $\mathrm{N_H}$ values (see figure \ref{chpt3:fig:BEHRz}). For obscuration below $\mathrm{N_H}=10^{23}\mathrm{cm^{-2}}$ the Hardness Ratio remains roughly constant, with the track of $\mathrm{N_H}=10^{22}\mathrm{cm^{-2}}$ showing no redshift evolution. As such we take unabsorbed\footnote{This interval will potentially contain some sources with a low level of absorption, however for the purpose of this work they shall all be assumed to be unabsorbed, as in \citet{Vito2018}.} sources to be those of $\mathrm{N_H}<10^{23}\mathrm{cm^{-2}}$ (given by their measured HR), while we define ``absorbed'' sources as those with Hardness Ratios corresponding to $10^{23}\mathrm{cm^{-2}}\leq N_\mathrm{H}\leq10^{24}\mathrm{cm^{-2}}$ and Compton-thick sources as those with Hardness Ratios corresponding to $\mathrm{N_H}>10^{24}\mathrm{cm^{-2}}$, as indicated by the coloured regions in figure~\ref{chpt3:fig:BEHRz}.

\begin{figure*}
    \centering
    \includegraphics[width=\linewidth]{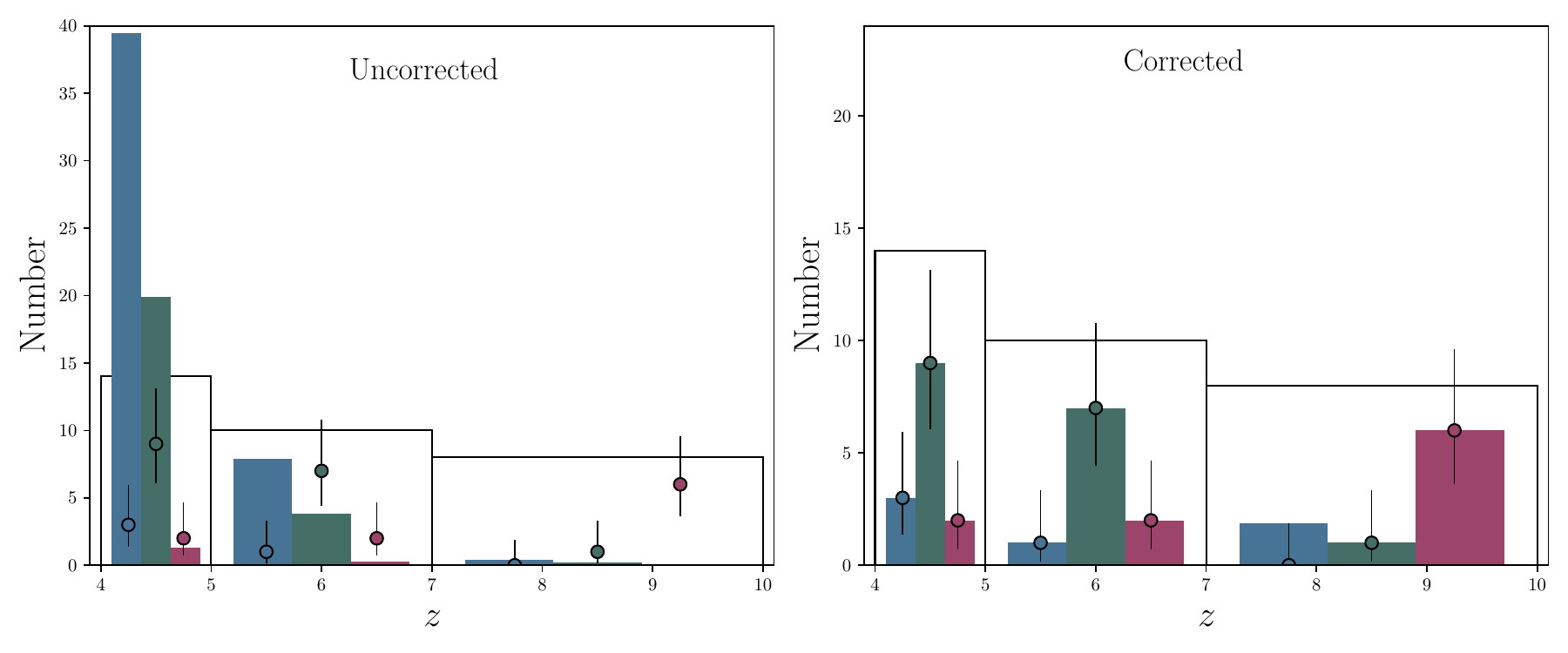}
    \caption[Obscuration uncorrected and corrected source numbers]{Histograms of model predicted source numbers (left) and corrected source numbers (right; see section \ref{chpt3:obscuration:correct}) for unabsorbed (blue), absorbed (green) and Compton-thick (red) samples, with the observed number of sources shown by the points. Open histograms show full sample of observed sources before $\mathrm{N_H}$ binning, within each redshift interval. It should be noted that within the highest redshift bin ($z=4-10$) the unabsorbed source number is corrected to the upper limit on 0 sources, as given by \citet{Gehrels1986}, and as such is above the observed number of sources.}
    \label{chpt3:fig:numbers}
\end{figure*}

With Hardness Ratios determined for all sources in the blind+extracted sample, we split the sample into unabsorbed, absorbed and Compton-thick based on their measured HR. As can be seen in figure \ref{chpt3:fig:BEHRz}, there are fewer unabsorbed sources (those that lie below the $\log{N_\mathrm{H}}=23$ line on the HR-z plot) with increasing redshift, which is consistent with a more obscured population of AGN at higher redshifts. The resulting numbers of sources in each redshift range ($z=4-5$, $z=5-7$ and $z=7-10$) and within each obscuration range (unabsorbed, absorbed and Compton-thick) are shown in Figure \ref{chpt3:fig:numbers}. The majority of our sources can be seen to fall within the absorbed and Compton-thick $\mathrm{N_H}$ ranges, with the majority of sources found to be absorbed \citep[see also][]{Vito2014}. 

It is important to note that the number of detected sources with different levels of obscuration will be strongly affected by selection biases. In particular, more obscured sources produce fewer observable counts and thus if they had the same \emph{intrinsic} space density as less obscured sources they would be under-represented in observed samples, although the extent of the effect will depend on the observed energy band. 
To explore this effect further, we determine the predicted number of sources in each redshift and obscuration bin, assuming all sources in a given obscuration bin have a space density given by the PDE XLF model from \citet{Pouliasis2024}. To convert an intrinsic rest-frame 2--10~keV luminosity to an observed full-band count rate, we adopt the Borus spectral model evaluated at  $\mathrm{N_H}=10^{22.5}\mathrm{cm^{-2}}$, $10^{23.5}\mathrm{cm^{-2}}$ and $10^{24.5}\mathrm{cm^{-2}}$, corresponding to the approximate centre of each obscuration interval ($\mathrm{N_H}<10^{23}$, $10^{23}\mathrm{cm^{-2}}\leq N_\mathrm{H}\leq10^{24}\mathrm{cm^{-2}}$ and $\mathrm{N_H}>10^{24}\mathrm{cm^{-2}}$ respectively). The corresponding count-rate is used to determine the area in equation \ref{chpt3:eq:Nmdl} and folded through the XLF model to produce an expected number of sources in our survey area.
We thus calculate the predicted number of sources within each redshift interval, summed across all possible luminosities\footnote{We take the minimum luminosity of the range of possible X-ray luminosities as that given by the count rate observable at 0.1\% of the total survey area.} for a given obscuration level. The resulting $\mathrm{N_H}$ dependent model predictions are shown by the coloured histograms in the left panel of figure \ref{chpt3:fig:numbers}.

The predicted number of sources overall decreases with redshift, as sources appear fainter at higher redshifts and thus are less likely to be detected, as well as due to the decrease in space density toward higher redshift in the \citet{Pouliasis2024} XLF model. 
The predicted numbers also drop rapidly with increasing $N_\mathrm{H}$ due to the suppression of the observed counts from such sources; only the brightest (and thus rarest) examples are expected to be detected.
However, in the \emph{observed} samples we find more absorbed ($N_\mathrm{H}\approx 10^{23-24}$~cm$^{-2}$, green data points in Figure~\ref{chpt3:fig:numbers}, left
) than unabsorbed ($N_\mathrm{H} \lesssim 10^{23}$~cm$^{-2}$, blue data points) at $z=4-5$ and $z=5-7$.
At $z=7-10$ we do not detect any unabsorbed sources, and the majority of the detected sources appear to be heavily obscured (Compton-thick), despite the strong selection biases \emph{against} finding such sources.

The hardness ratios measured for our sample, and the comparison of predicted to observed source numbers at different levels of obscuration, clearly indicates higher intrinsic space density of obscured sources at high redshifts and the need for corrections to the parametrisation of the XLF calculation to include both an $N_\mathrm{H}$ and redshift dependence.

\subsection{Obscuration Corrections to the XLF}\label{chpt3:obscuration:correct}

The study of the obscuration in the blind and extracted sample of high-redshift AGN, described above, indicates a significant redshift-dependent population of obscured sources, which is not accounted for in the model XLF. Thus, a correction to account for the disparity between the observed numbers of obscured sources and the predicted numbers (shown in the left hand plot of figure \ref{chpt3:fig:numbers}) needs to be applied to scale up the XLF and thus provide an accurate characterisation of the space density of high-$z$ AGN.

With the broad distribution in Hardness Ratio with redshift, a correction factor that depends on both redshift and $\mathrm{N_H}$ is required to suitably account for obscuration within our sample. Given the limited size and dynamic range of our samples, we shall assume that the same correction factor applies at all $L_\mathrm{X}$ (i.e. we assume no luminosity dependence in the obscured fraction and that the underlying shape of the XLF is the same across all levels of obscuration). We determine a scale factor, $\gamma(N_\mathrm{H,bin},z_\mathrm{bin})$, 
that must be applied to our fiducial XLF model, $\phi_\mathrm{fidcl}(L_\mathrm{X},z)$ (given by the \citealt{Pouliasis2024} PDE model) to recover a corrected model for the $N_\mathrm{H}$ and $z$ bin, 

\begin{equation}
    \phi_\mathrm{mdl,corr}(L_\mathrm{X},z | N_\mathrm{H})=\gamma(N_\mathrm{H,bin},z_\mathrm{bin})\phi_\mathrm{fidcl}(L_\mathrm{X},z).
\end{equation}

\begin{figure*}
    \centering
    \includegraphics[width=\linewidth]{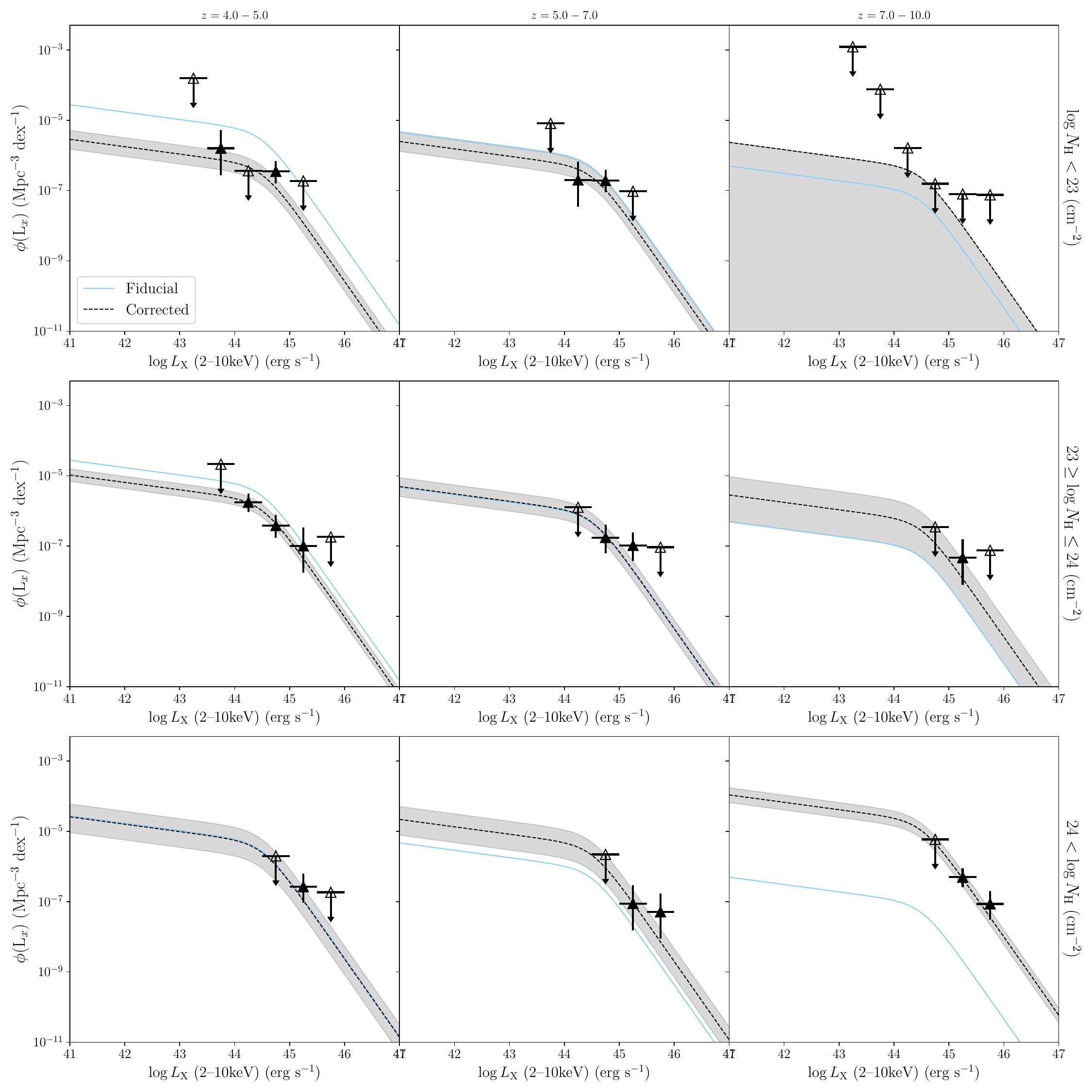}
    \caption[Uncorrected and corrected XLF for obscuration and redshift bins]{The \citet{Pouliasis2024} XLF model before (solid line) and after (dashed line) correction to our obscuration binned measurements, for $\mathrm{N_H}<10^{23}$ (top row), $10^{23}\mathrm{cm^{-2}}\leq N_\mathrm{H}\leq10^{24}\mathrm{cm^{-2}}$ (middle row) and $\mathrm{N_H}>10^{24}\mathrm{cm^{-2}}$ (bottom row). Our measured XLF are shown for comparison by the black triangles, with limits given by \citet{Gehrels1986} where no sources exist within our sample (open triangles). The Obscuration correction can be seen to reproduce the shape of the XLF well, as seen with our blind+extracted measurements. For the case of $\mathrm{N_H}<10^{23}$ at $z=7-10$ the blind+extracted sample contains no sources, thus the obscuration correction can only provide an upper limit on the number of sources, based on the survey area of the sample.}
    \label{chpt3:fig:individualCorrections}
\end{figure*}

This scale factor is defined such that we recover the observed number of sources in a given $N_\mathrm{H}$--$z$ bin after folding through the appropriate area curve, i.e. such that $n_\mathrm{obs}(N_\mathrm{H,bin},z_\mathrm{bin})=n_\mathrm{mdl}(N_\mathrm{H,bin},z_\mathrm{bin}|\phi_\mathrm{mdl,corr})$. As in equation \ref{chpt3:eq:Nmdl} for the calculation of the expected number of sources given an XLF model, we can write this as 

\begin{align}
    n_\mathrm{mdl}(N_\mathrm{H,bin},z_\mathrm{bin} | \phi_\mathrm{mdl,corr}) = \phantom{xxxxxxxxxxxxxxxxxxxxxxxxxxxxxxx}\nonumber\\
    \iint \Bigg[
     \gamma(N_\mathrm{H,bin},z_\mathrm{bin}) \phi_\mathrm{fidcl}(L_\mathrm{X},z) A\big(c | L_\mathrm{X}, z, N_\mathrm{H,bin}\big)\Bigg] \textstyle{\frac{dV_\mathrm{co}}{dz}} dz \;d\log L_\mathrm{X}\nonumber\\
 = \gamma(N_\mathrm{H,bin},z_\mathrm{bin}) \; n_\mathrm{mdl}(N_\mathrm{H,bin},z_\mathrm{bin}|\phi_\mathrm{fidcl}) 
    \label{chpt3:eq:Nmdl_NH}
\end{align}

with sky area $A\big(c| L_\mathrm{X}, z, N_\mathrm{H,bin}\big)\;$ at an observed count rate, $c$, that depends on the redshift ($z$), rest-frame 2--10\,keV luminosity ($L_\mathrm{X}$) and obscuring column density ($N_\mathrm{H,bin}$) and is calculated assuming the Borus spectral model \citep{Balokovic2018} and the \textit{Chandra} response. The correction factor can thus be written as, 

\begin{align}
   \gamma(N_\mathrm{H,bin},z_\mathrm{bin})&=
   \frac{n_\mathrm{mdl}(N_\mathrm{H,bin},z_\mathrm{bin}| \phi_\mathrm{mdl,corr})}{n_\mathrm{mdl}(N_\mathrm{H,bin},z_\mathrm{bin}|\phi_\mathrm{fidcl})} \nonumber\\
   &=
   \frac{n_\mathrm{obs}(N_\mathrm{H,bin},z_\mathrm{bin})}{n_\mathrm{mdl}(N_\mathrm{H,bin},z_\mathrm{bin}|\phi_\mathrm{fidcl})},
    \label{chpt3:eq:scale}
\end{align}

where $n_\mathrm{obs}(N_\mathrm{H,bin},z_\mathrm{bin})$ is the observed number of sources within a given obscuration range ($N_\mathrm{H,bin}$) for a given redshift bin, 
 and $n_\mathrm{mdl}(N_\mathrm{H,bin},z_\mathrm{bin}|\phi_\mathrm{fidcl})$ is the expected total number of sources (across all $L_\mathrm{X}$) in an $N_\mathrm{H}$--$z$ bin assuming our fiducial XLF model. We calculate the scale factor for each $z$ bin for the unabsorbed, absorbed and Compton-thick $N_\mathrm{H}$ bins, assuming the shape and redshift evolution of the XLF remains that of the fiducial model within the bin. The obscured column density is taken to be $N_\mathrm{H}=10^{22.5}$, $10^{23.5}$ and $10^{24.5}$~cm$^{-2}$, for unabsorbed, absorbed and Compton-thick sources, respectively, as a representative value for the conversion to count rates. Figure \ref{chpt3:fig:individualCorrections} shows the fiducial and corrected XLF models at $z=4-5$, $5-7$ and $7-10$, for unabsorbed, absorbed and Compton-thick sources. As previously in our measurements of the XLF, in the case of no observed sources, we take the Poisson upper limit from \citet{Gehrels1986} on 0 sources in place of the observed number.

Applying the resulting redshift and obscuration dependent correction factors, $\gamma(N_\mathrm{H,bin},z_\mathrm{bin})$, to the predicted numbers scales them to agree with the observed number of sources in each redshift and obscuration interval, as shown in the right hand plot of figure \ref{chpt3:fig:numbers}.

Our corrected XLF model matches the observed binned XLF well across all three redshift ranges and for the different levels of obscuration (see Figure~\ref{chpt3:fig:individualCorrections}). With increasing obscuration, an increasingly positive shift in the normalisation of the fiducial model is required to agree with the observations (in contrast, the fiducial model XLF over predicts the space density of unabsorbed sources and the corrected model is shifted down). It should be noted that at $z=7-10$ there are no unabsorbed sources within our sample, and as such the model can only be corrected to the upper limit on the binned XLF, as given by the Poisson uncertainty of \citet{Gehrels1986}, and the shape traced by these limits depends mainly on the sensitivity and area coverage of the Chandra data used here.

Having obtained the correction factors and confirmed they suitably adjust the model XLF for the obscuration observed, we can obtain the obscuration-corrected, binned measurements of the XLF across all $N_\mathrm{H}$ using the entire full-band blind$+$extracted sample.
Equation~\ref{chpt3:eq:NobsNmdl} is updated to
\begin{align}
    \phi_\mathrm{obs,corr}(L_{\mathrm{X},i}, z_j) = &\; \phi_\mathrm{mdl,corr}(L_{\mathrm{X},i}, z_j)\frac{n_\mathrm{obs}(L_{\mathrm{X},i}, z_j)}{n_\mathrm{mdl}(L_{\mathrm{X},i}, z_j)}\nonumber\\
        =& \; 
(\gamma_\mathrm{unabs}+\gamma_\mathrm{abs}+\gamma_\mathrm{ctk})\; \phi_\mathrm{fidcl}(L_{\mathrm{X},i}, z_j) \; \times \nonumber \\
 & \quad\quad \frac{n_\mathrm{obs}(L_{\mathrm{X},i}, z_j)}{n_\mathrm{mdl,unabs}+n_\mathrm{mdl,abs}+n_\mathrm{mdl,ctk}}
    \label{chpt3:eq:correctingXLF}
\end{align}

where $\gamma_\mathrm{unabs}$, $\gamma_\mathrm{abs}$ and $\gamma_\mathrm{ctk}$ are the obscuration correction factors for the $z$ bin, for the $N_\mathrm{H}$ bins of unabsorbed, absorbed and Compton-thick sources respectively (see table \ref{chpt3:tab:obscorrections} for values). The predicted number of unabsorbed ($n_\mathrm{mdl,unabs}$), absorbed ($n_\mathrm{mdl,abs}$) and Compton-thick ($n_\mathrm{mdl,ctk}$) sources in each luminosity--redshift bin are calculated using equation \ref{chpt3:eq:Nmdl_NH} but with the corresponding integration limits, and summed to recover the overall predicted number of sources in the $L_\mathrm{X}$--$z$ bin. The resulting observed XLF measurements are shown by the black triangles in figure \ref{chpt3:fig:correctedXLF}, with the full-band blind$+$extracted XLF before absorption correction (from \S\ref{chpt3:xlf:extracted}) and the hard-band blind XLF (from \S\ref{chpt3:obscuration:hardxlf}) plotted for comparison.

With the obscuration correction, our Full-band measurements more closely follow those of the Hard-band XLF measurements, showing a higher space density across all redshift ranges than the uncorrected Full-band measurements. At $z=4-5$ these corrected measurements remain broadly consistent with the prior model predictions, as in section \ref{chpt3:xlf:extracted}. However, even at $z=5-7$ our measurements suggest a higher space density than the extrapolated models, and the models drastically under-predict the sample observed at $z=7-10$.

\begin{table}
    \centering
    \begin{tabular}{c|ccc}
        z & $\gamma_{unabs}$ & $\gamma_{abs}$ & $\gamma_{ctk}$ \\ \hline
        4--5 & 0.10$^{+0.08}_{-0.05}$ & 0.38$^{+0.19}_{-0.13}$ & 0.9$^{+1.2}_{-0.6}$ \\
        5--7 & 0.5$^{+0.4}_{-0.3}$ & 1.1$^{+0.8}_{-0.5}$ & 4$^{+6}_{-3}$ \\
        7--10 & $<4.73$ & 5.7$^{13}_{-0.5}$ & 217$^{+130}_{-86}$ \\ \hline
    \end{tabular}
    \caption[Obscuration corrections to XLF model of \citet{Pouliasis2024}]{Obscuration corrections to the \citet{Pouliasis2024} PDE XLF model for unabsorbed ($\gamma_{unabs}$), absorbed ($\gamma_{abs}$) and Compton-thick ($\gamma_{ctk}$) sources determined here.}
    \label{chpt3:tab:obscorrections}
\end{table}

Across the full redshift range of our sample ($z=4-10$) and within each redshift bin ($z=4-5$, $5-7$ and $7-10$) we use our obscuration correction factors ($\gamma_\mathrm{unabs}$, $\gamma_\mathrm{abs}$ and $\gamma_\mathrm{ctk}$) to determine the intrinsic fraction of obscured (i.e. $\mathrm{N_H}>10^{23}$) AGN, $f_\mathrm{obsc}$,  

\begin{equation}
    f_\mathrm{obsc}=\frac{\gamma_
    \mathrm{abs}+\gamma_\mathrm{ctk}}{\gamma_\mathrm{unabs}+\gamma_
    \mathrm{abs}+\gamma_\mathrm{ctk}},
\end{equation}

and the intrinsic fraction of Compton-thick AGN, $f_\mathrm{ctk}$,

\begin{equation}
    f_\mathrm{ctk}=\frac{\gamma_\mathrm{ctk}}{\gamma_\mathrm{unabs}+\gamma_
    \mathrm{abs}+\gamma_\mathrm{ctk}}.
\end{equation}

Uncertainties are determined through random posterior draws of possible ``true'' values of the fraction. From this the fraction of obscured and Compton-thick AGN across the redshift range of $z=4-10$ is $0.982^{+0.007}_{-0.008}$ and $0.91^{+0.03}_{-0.03}$ respectively. 

\begin{figure}
    \centering
    \includegraphics[width=\linewidth]{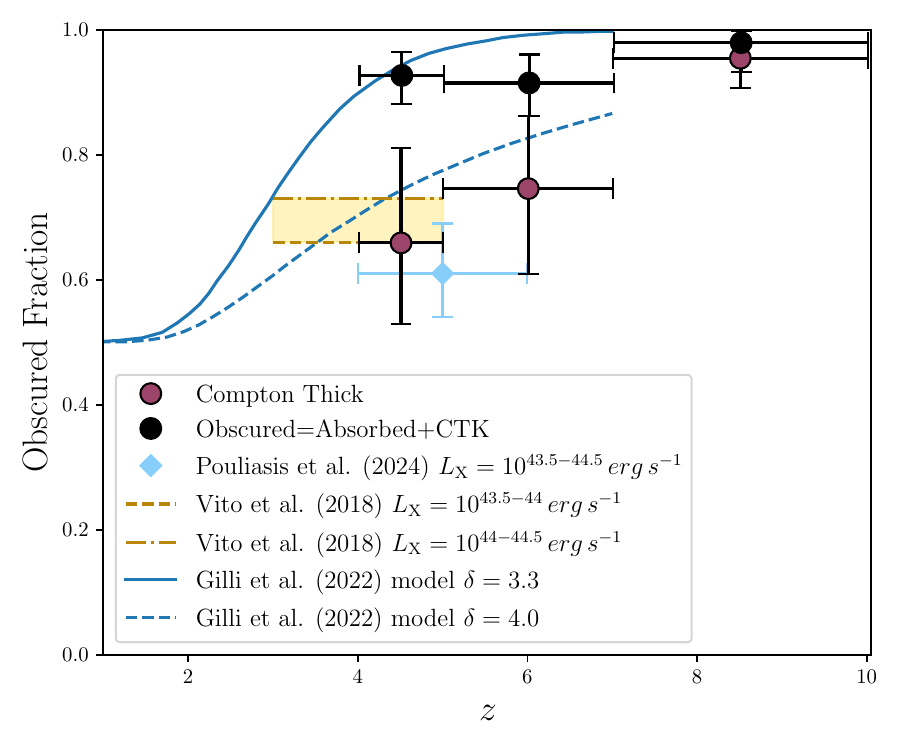}
    \caption{Fraction of the blind$+$extracted sample that are Compton Thick (pink points) and the fraction that are obscured (black points), across the full luminosity range of the sample. Previous observational measurements of the obscured fraction from \citet{Vito2018} for $L_\mathrm{X}=10^{43.5-44}$~erg~s$^{-1}$ (dashed) and $10^{44-44.5}$~erg~s$^{-1}$ (dot-dashed) are shown by the gold region and the measurement from \citet{Pouliasis2024}) is shown by the red diamond. Our Compton Thick fraction is consistent with these observational measurements at $z=4-5$, but shows a clear increase with redshift. The high obscured fraction measured from our blind$+$extracted sample is indicative of the large fracton of obscured sources in the early Universe which has been predicted by the models of \citet{Gilli2022} based on the expected column densities of gas within the host galaxies of high-$z$ AGN.}
    \label{fig:obsFrac}
\end{figure}

Across the three redshift bins ($z=4-5$, $5-7$ and $7-10$) used in this work, we find the fraction of obscured ($\mathrm{N_H}>10^{23}$) sources increases with redshift with the fraction of Compton Thick sources becoming increasingly dominant towards earlier times; increasing from $0.66_{-0.13}^{+0.15}$ at $z=4-5$ to $0.80^{+0.13}_{-0.04}$ at $z=5-7$ and $0.97^{+0.02}_{-0.01}$ at $z=7-10$. This can be seen in figure \ref{fig:obsFrac}. Whilst the measured fraction of Compton Thick sources is roughly consistent with previous results \citep[][]{Pouliasis2024, Vito2018}, this fraction rapidly increases with redshift as found by \citet{Gilli2022} who suggest such levels of obscuration may be due to the high gas content throughout the host galaxies of early AGN.

\section{Discussion}\label{chpt3:discussion}
\subsection{Comparisons to previous measurements of the XLF at high redshift}\label{chpt3:discussion:comparexlf}
The AGN X-ray Luminosity Function (XLF) has previously only been measured to $z\sim6$ \citep[e.g.][]{Georgakakis2015, Pouliasis2024, Vito2018}, to which parametric models were then fitted, due to the limitations on high-redshift X-ray surveys (i.e. the small area observable to the required depth with current telescopes, observed decline in space density of AGN with increasing redshifts, and the difficulty of identifying the counterparts to X-ray AGN at high redshifts) and the XLF beyond $z=6$ has remained poorly constrained. With new deep near-IR imaging from UltraVISTA, incorporated within the COSMOS2020 multi-wavelength catalogue, it is possible to measure redshifts for sources at $z>6$ that we have matched with X-ray sources in the \textit{Chandra}-COSMOS Legacy imaging and used to identify additional lower significance X-ray sources. We have thus obtained measurements of the XLF at moderate luminosities out to $z=10$ that demonstrate substantially higher space densities of early X-ray AGN than predicted by the extrapolations of prior parametric models. 

The work performed in this paper relies on photometric redshift measurements. As investigated in section \ref{chpt3:xlf:redshift}, the choice of templates used in the photometric redshift fitting adds an uncertainty to the XLF measurements. As such we determined the uncertainty range in our XLF measurements due to the uncertainties intrinsic to photometric redshifts, shown in figure \ref{chpt3:fig:zrangesxlf}.
We note that even in our most pessimistic scenario based on the differing photometric redshift solutions, we still find a significantly higher space density of X-ray AGN at $z=7-10$ than expected based on prior model extrapolations. 
We also find evidence that a large fraction of the early X-ray AGN population is obscured; applying obscuration corrections to the full-band XLF measurements (see \S\ref{chpt3:obscuration}) further increases our measurements of the space density of high-$z$ X-ray AGN. We now combine our obscuration corrected XLF and photometric redshift uncertainties, in order to determine the most optimistic and most conservative measurements of the XLF from our sample, which we then compare with constraints from prior studies.

\begin{figure*}
    \centering
    \includegraphics[width=\linewidth]{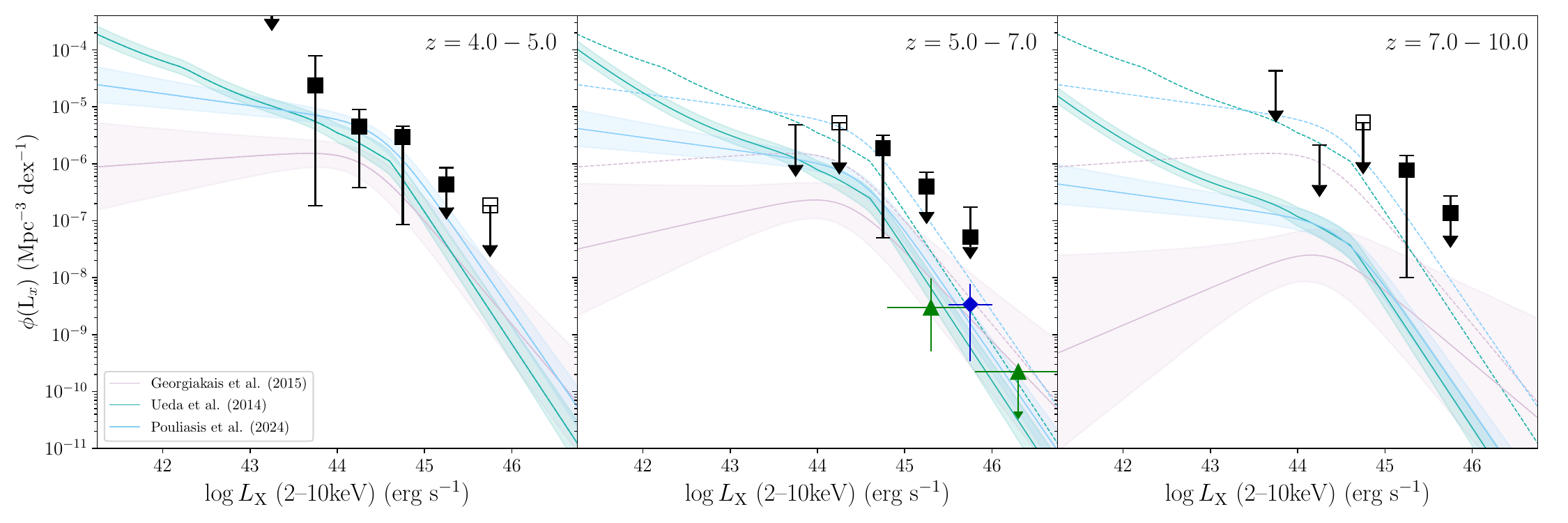}
    \caption[XLF comparisons]{The bounds of the XLF measured in this paper, with lower bounds given by the lower bounds of the photometry dependent XLF determined in \S\ref{chpt3:xlf:redshift} and the upper bounds given by the upper uncertainty on the obscuration corrected XLF measurements (see section \ref{chpt3:obscuration:correct}). Extrapolated models of XLF fit to redshift $z\sim3-6$ data are plotted at the central redshift of each bin (i.e. $z=4.5$, $z=6$ and $z=8.5$) for comparison, with the dotted lines indicating the space density given by the models at $z=4-5$. The bright-end XLF measurements of \citet{Barlow-Hall2023} and \citet{Wolf2021} are shown by the green triangles and the blue diamond respectively.}
    \label{chpt3:fig:xlfrange}
\end{figure*}

For our most conservative estimate, we take the lower bound of the uncertainty range shown in figure \ref{chpt3:fig:zrangesxlf} when considering the impact of photometric redshift uncertainties (N.B. these measurements do \emph{not} include any correction for obscuration). We then take the upper uncertainty on the obscuration-corrected measurements as our most optimistic, upper limit. The resulting range in our observed XLF measurements, at the redshifts of $z=4-5$, $5-7$ and $7-10$, are shown in figure \ref{chpt3:fig:xlfrange}.

We note that applying the obscuration corrections (as in section \ref{chpt3:obscuration:correct}) whilst also considering the range of possible redshifts for each source (as in section \ref{chpt3:xlf:redshift}) would provide a more comprehensive uncertainty range of our XLF measurements. However, as there is significant obscuration across the full redshift range of the sample, this would increase the disparity between the model predictions and the most optimistic space densities at $z=7-10$, whilst the most conservative measurements of the XLF would remain the value of the lower uncertainties found in section \ref{chpt3:xlf:redshift} for the uncorrected XLF. As such, with the obscuration corrections only applied to the best redshift measurements, the results shown in figure \ref{chpt3:fig:xlfrange} may still be relatively conservative, rather than showing strict upper limits on the high-redshift AGN space density. 
We also recall that a number of X-ray sources in the blind source catalogue either lack reliable counterparts or have reliable counterparts but lack reliable photometric redshift estimates. Some of these X-ray sources may fall within our redshift range and would increase our upper limit further still. 

In Figure~\ref{chpt3:fig:xlfrange} we compare these bounds on our XLF measurements with both parametric model extrapolations.
Within the redshift range of $z=4-5$, at which previous XLF models have been fitted to survey data, we find our measurements of the XLF to be consistent with the parametric models (as is to be expected) and thus suitably representative of the AGN population at the end of cosmic reionisation (see left hand panel of figure \ref{chpt3:fig:xlfrange}). However, the upper limits on our measurements, which account for the obscuration within the sample, suggest that the space density of fainter AGN could be substantially higher than predicted based on prior model extrapolations. Our higher space density estimates are due to the potentially substantial fraction of the population at these redshifts that are obscured or Compton-thick ($\gtrsim90$\%) that may not be fully accounted for in the prior studies or their parametrisations.

At higher redshifts (centre panel of figure \ref{chpt3:fig:xlfrange}), our measurements remain broadly consistent with the parametric model extrapolations at $z=5-7$. Whilst the most conservative measurements are clearly consistent with the model predictions, our most optimistic values indicate that the true space density of AGN may be substantially higher. 
We also compare here with direct constraints on the bright end of the XLF by \citet{Barlow-Hall2023} using \textit{Swift} 
 (ExSeSS) and by \citet{Wolf2021} with eROSITA (eFEDs, determined for a narrower redshift interval of $z=5.7-6.4$). 
 We find that these while these previous estimates are consistent with our limits, 
our best estimate (solid black square) and upper bound indicate that the space density of $L_\mathrm{X}\sim10^{45-46}$~erg~s$^{-1}$ AGN at $z\sim5-7$ could be up to a factor $\sim100$ higher than 
that measured by \citet{Barlow-Hall2023} and \citet{Wolf2021}.
It should be noted that the measurement of \citet{Barlow-Hall2023} will be incomplete, in particular for obscured AGN, and should thus be considered a lower limit, which could explain the substantial discrepancy with our new Chandra/COSMOS measurements.

Despite the agreement between our blind+extracted measurements and the extrapolations of parametric models out to $z=7$, the extrapolated models can be seen to under-predict the observed XLF at $z=7-10$ (see right hand plot of \ref{chpt3:fig:xlfrange}). Although the most conservative measurements of the observed XLF at this very high redshift is within $1\sigma$ of the \citet{Georgakakis2015} LDDE model predictions, the AGN space-density in the very early universe is likely much higher than previously expected. Although previous studies \citep[e.g][]{Brusa2009,Vito2018} suggest the need for a high-redshift decline in the AGN space density, the results we have presented here indicate that there is no such decline in the AGN space density at $z\gtrsim7$.

The space density of AGN we have determined here indicates a large number of actively accreting supermassive black holes within the very early Universe, which \emph{are} detectable
even with current X-ray telescopes, even if the high levels of obscuration (becoming predominantly Compton-thick at the highest redshifts that we probe) mean that their observable X-ray flux is strongly suppressed relative to their intrinsic luminosities.
Our findings suggest that significant black hole growth may be occurring at early times, which we investigate further in \S\ref{chpt3:discussion:earlygrowth}. This early AGN population within the first galaxies would likely also have a significant impact upon the evolution of galaxies in the early Universe. In addition, as contamination of galaxy observations by AGN will significantly impact any measurements of galaxy properties, we stress the importance of accounting for the impact of AGN light on any measurements of galaxy properties at $z\gtrsim4$ \citep[as is accounted for in e.g.][]{Graaff2024, Trinca2024, Bowler2020}.

\subsection{Comparison with the bolometric QLF and new AGN populations identified with JWST}\label{chpt3:discussion:qlf}
New, sensitive IR imaging enabled by JWST (along with follow-up spectroscopy of faint sources) has enabled the identification of surprising large 
populations of moderate- to low-luminosity AGN at redshifts above $z\sim4$ \citep[e.g.][]{Greene2023, Matthee2024, Kokorev2024, Larson2023}, with some studies claiming to have found early AGN out to $z\sim10$ \citep{Kovacs2024}. These early JWST results indicate a relatively high AGN occupation fraction within the first galaxies, with the majority of JWST studies pointing towards unexpectedly high AGN space densities within the early Universe \citep[see e.g.][]{Greene2023, Maiolino2023}.

\begin{figure*}
    \centering
    \includegraphics[width=\linewidth]{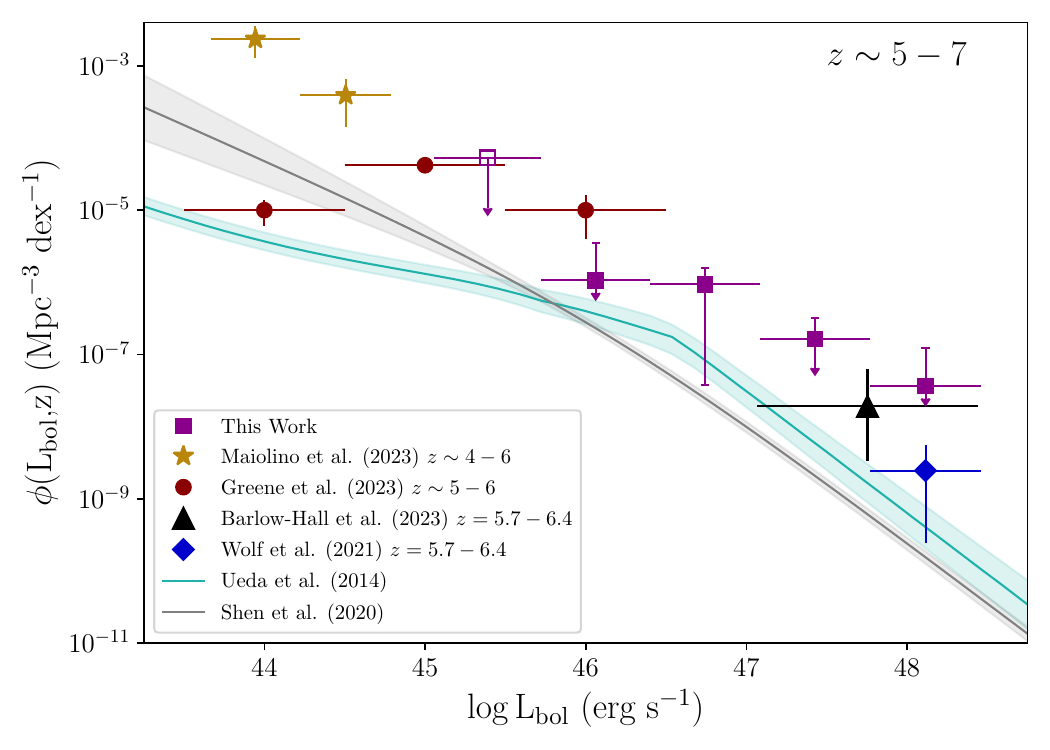}
    \caption[QLF comparisons to JWST results]{
    Our estimates of the bolometric QLF (purple squares) based on our final XLF measurements and converted to bolometric estimates based on the corrections from \citet{Shen2020}.
        The solid squares indicate our best estimates, while the uncertainties show the range 
    from the maximum obscuration-corrected measurement to the minimum unobscured measurements (as shown in figure \ref{chpt3:fig:xlfrange}, with arrows indicating conservative upper limits given the range of possible photometric redshifts). Bright-end $z=5.7-6.4$ X-ray measurements are shown from the results of \citet{Barlow-Hall2023} (black triangle) and \citet{Wolf2021} (blue diamond). The JWST measurements at $z=5-6$ from \citet{Greene2023} are shown by the red circles, and the $z=4-6$ UVLF measurements of \citet{Maiolino2023}, similarly converted to bolometric luminosities using the conversions of \citet{Shen2020}, are shown by the gold stars. For comparison, the bolometric QLF model of \citet{Shen2020} and bolometric converted XLF model of \citet{Ueda2014} are plotted at $z=5.7-6.4$ (grey and turquoise respectively). Our constraints on the QLF can be seen to bridge the gap between the latest JWST observations and the bright-end X-ray measurements, 
    as well as being consistent with substantially higher space densities of low-to-moderate luminosity AGN in the early Universe than predicated by extrapolations of previous models.
            }
    \label{chpt3:fig:jwst_qlf}
\end{figure*}

In order to compare our X-ray measurements of the $z\gtrsim4$ AGN space density to the JWST findings, we convert our measurements of the XLF into into estimates of the bolometric Quasar Luminosity Function (QLF), adopting the luminosity-dependent bolometric corrections of \citet{Shen2020}. Applying the bolometric correction to the full-band, obscuration-corrected XLF measurements obtained above (see section \ref{chpt3:obscuration:correct}), we compare our new QLF estimates at $z=5-7$ to measurements based on JWST data by \citet{Maiolino2023} (from UV luminosity function measurements at $z\sim4-6$) and \citet{Greene2023} (see figure \ref{chpt3:fig:jwst_qlf}). 
\citet{Greene2023} performed their measurements of the QLF on the space density of Little Red Dots, based on photometric selection of Red Compact sources and spectroscopic broad lines, and thus may not capture all low-luminosity AGN. In contrast, \citet{Maiolino2023} use a less stringent selection of AGN, based simply on the identification of a broad H$\alpha$ or H$\beta$ line, and thus measure a higher space density than \citet{Greene2023} at low-luminosities. In addition to the JWST measurements, we also show the previous constraints on the bright-end QLF based on X-ray measurements from \citet{Barlow-Hall2023} and from a bolometric conversion of the \citet{Wolf2021} measurements of the XLF \citep[converted using the][bolometric correction]{Shen2020}.

As can be seen in figure \ref{chpt3:fig:jwst_qlf}, our obscuration-corrected measurements of the AGN space density follow a consistent trend between the early JWST results and X-ray observations at the bright-end of the luminosity function. As noted by \citet{Scholtz2023}, the majority of AGN detected by JWST are found to have no detected X-ray emission even when image stacking is applied \citep[see][]{Maiolino2024, Yue2024}. This lack of X-ray detection points towards a potentially very high obscuration fraction \citep[of the order of $100\times$ the fraction of unobscured sources, as in][and references there in]{Maiolino2024}, as is consistent with the obscured population inferred from our blind$+$extracted sample here($\sim220\times$, given by the obscuration correction to the model XLF). Despite this, other studies with JWST have successfully identified high-$z$ AGN with X-ray detections \citep[][]{Kocevski2024, Bogdan2023, Napolitano2025}. \citet{Pouliasis2024} argued that X-ray selection is likely to miss all but a small fraction of the heavily obscured AGN population; precisely those sources being found by JWST. With our blind$+$extracted sample pushing the limit of current X-ray imaging, and probing higher luminosity AGN than the JWST imaging of studies such as \citet{Maiolino2023, Greene2023}, it is likely that the AGN found by JWST are simply lower (intrinsic) luminosity equivalents to the population that we have identified here, which also exhibit high (often Compton-thick) levels of obscuration that suppresses their X-ray emission and makes them difficult to detect even in the very deepest X-ray imaging (or stacks). In order to investigate this further, spectra and JWST imaging of our high redshift sample are required to determine their physical nature and compare their properties with the population of compact AGN \citep[known as Little Red Dots, e.g.][]{Kocevski2024} found by JWST.

The moderate luminosity measurements of the AGN space density we have obtained in this work, clearly point towards a reconciliation of the JWST AGN population and those identified through X-ray surveys. With improved redshift measurements from the mid-IR imaging of JWST, even fainter X-ray sources could be identified within the Chandra Deep Fields (following the method detailed in section \ref{chpt3:data:extraction}) and thus place further constraints on the faint-end of the XLF, where the majority of the AGN population resides. In order to truly reconcile these measurements, the impact of Compton-thick levels of obscuration needs to be carefully taken into account. Only with even deeper X-ray observations than currently feasible with Chandra do we stand a chance of identifying JWST AGN at X-ray wavelengths, assuming they have some intrinsic X-ray emission due to the AGN, requiring the next generation of X-ray observatories such as AXIS \citep{Reynolds2023, Cappelluti2024}.

\subsection{The Implications on Early Black Hole Growth}\label{chpt3:discussion:earlygrowth}
Our measurements indicate a substantially higher space density of moderate-luminosity AGN at $z\gtrsim6$ than expected given earlier X-ray studies and in line with recent JWST findings.
Previous measurements of X-ray selected AGN out to redshifts of $z\sim6$ found a decline in the space density of AGN with increasing redshift from which studies such as \citet{Vito2014, Aird2015} determined the Black Hole Accretion Density (BHAD) to drop more rapidly with redshift than the Star Formation Rate Density (SFRD) of galaxies \citep[see also e.g.][]{Pouliasis2024, Zhang2023, Habouzit2021}. This trend suggests the build-up of stellar mass within galaxies may have occurred before their central SMBH underwent significant accretion-driven mass build-up. 
Using our XLF measurements we determine the corresponding BHAD in each of our redshift bins ($z=4-5$, $5-7$ and $7-10$) following the process detailed in\citet{Yang2021} and \citet{Pouliasis2024}, whereby
\begin{align}
    \mathrm{BHAD} &= \frac{1-\epsilon}{\epsilon c^2} \int L_\mathrm{bol} \times \phi_\mathrm{bol} (L_\mathrm{bol},z) d \log L_\mathrm{bol}\nonumber\\
    &= \frac{1-\epsilon}{\epsilon c^2} \int L_\mathrm{bol} \times \phi_\mathrm{corr} (L_X,z) 
    \frac{d \log L_X}{d\log L_\mathrm{bol}}    
   d\log L_\mathrm{bol}
\end{align}
where we take the radiative efficiency, $\epsilon$, to be 0.1 \citep[as in][]{Pouliasis2024}, $\phi_\mathrm{corr}(L_\mathrm{X}, z)$ is our obscuration-corrected XLF determined in \S\ref{chpt3:obscuration}, and the $\frac{d \log L_X}{d\log L_\mathrm{bol}}$ accounts for the change of variables between X-ray and bolometric luminosities.  
As our model assumes a declining XLF within each redshift bin, we calculate the BHAD across each redshift bin, for the bolometric luminosity range of $43<\log L_\mathrm{bol}<49$ 
then average over the redshift bin to obtain the BHAD at the given redshifts. 

\begin{figure}
    \centering
    \includegraphics[width=\linewidth]{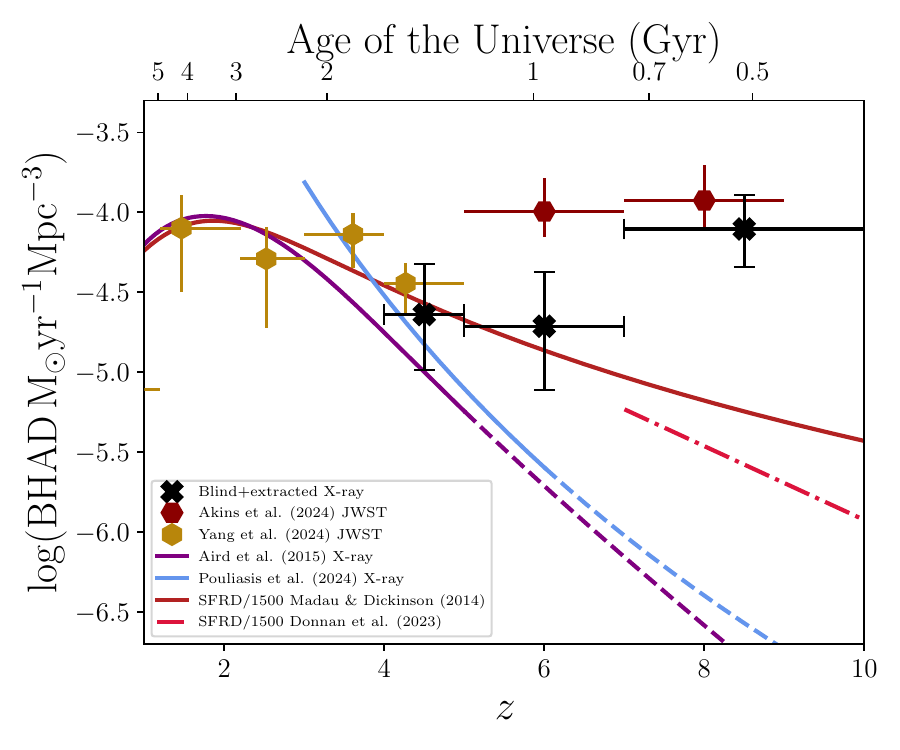}
    \caption{Evolution of the BHAD with redshift, as given by our measurements of the $z=4-10$ XLF (black crosses). Previous X-ray based measurements by \citet{Aird2015} and \citet{Pouliasis2024} are shown by the purple and blue lines respectively and can be seen to follow a rapid decline with increasing redshift. For comparison we show the SFRD of \citet[solid red]{Madau2014}, and a more recent estimate of the high redshift SFRD as measured by \citet[dot-dashed crimson]{Donnan2023}, both rescaled by a factor of $1/1500$ to compare with the BHAD. Our measurements of the BHAD can be seen to increase toward higher redshift and are consistent with recent JWST measurements based on mid-IR selected AGN \citep[][shown by the gold hexagons]{Yang2023} and Little Red Dots \citep[][shown by the red hexagons]{Akins2024}. These measurements suggest early SMBHs underwent significant mass growth prior to the build-up of stellar mass within their host galaxies.}
    \label{fig:bhad}
\end{figure}

Our resulting BHAD measurements (see figure \ref{fig:bhad}) show an increase with redshift from the end of cosmic reionisation ($z\sim5$) out to $z\gtrsim10$. Our measurements are consistent with the BHAD that is estimated based on AGN populations identified with JWST \citep[][]{Yang2023} and in particular the high BHAD inferred from the large populations of Little Red Dots identified with JWST out to $z\sim10$ \citep[red hexagons in figure~\ref{fig:bhad}]{Akins2024}. Unlike the X-ray measurements of \citet{Vito2014, Aird2015}, the BHAD found here indicates that significant Black Hole mass assembly may have occurred \emph{prior} to stellar mass build up. As such, BHs may build up a significant fraction of their mass before the stellar mass of their host galaxy is assembled, as has been postulated in the work of \citet{Guetzoyan2024, Terrazas2024}. Mass measurements of early SMBHs also find over-massive SMBHs \citep[e.g.][]{Maiolino2023, Kocevski2024}, i.e. they are more massive than their host galaxies, which requires this BH mass build-up to precede that of the host galaxy's stellar mass.

The high hardness ratios that we measure for blind$+$extracted sample (see \S\ref{chpt3:obscuration:correct}, above) indicates the majority of the moderate-luminosity AGN population at these redshifts is heavily obscured and may be almost entirely Compton-thick at $z\gtrsim7$ (as indicated by the obscured and Compton-thick fractions in Figure \ref{fig:obsFrac}). The trend we observed here is consistent with that predicted by \citet{Gilli2022} through extensive modelling of the the redshift evolution in the obscuration of AGN due to the column density of gas in the Interstellar Medium (ISM) of galaxies (as well as the obscuring torus associated with the AGN itself). \citet{Gilli2022} found the ISM density increases steeply with redshift as $(1+z)^{3.3-4}$from a Milky Way-like ISM at $z=0$ to $>100\times$ higher column density at $z=3$, and up to Compton-thick ISM densities at $z>6$. The high fraction of Compton-thick sources found for our sample, $0.982^{+0.007}_{-0.008}$ for $z=4-10$, thus appears to be consistent with a high abundance of gas within the early Universe, in agreement with the gas mass fractions expected within high redshift galaxies \citep[see review of][]{Carilli2013}. Thus, our results are indicative of highly obscured AGN due to the dense ISM of host galaxies, as has been proposed as the mechanism responsible for the lack of X-ray emissions in Little Red Dots. 
The high space density and obscuration fraction we have measured also indicates a substantial build-up of SMBHs within heavily obscured environments in the early Universe.
Such highly abundant gas that may fuel both AGN and star formation within the first galaxies, with early AGN activity potentially impacting the evolution of their host galaxies. Whilst assessing the mechanisms that may drive this early SMBH growth and any co-evolution with their host galaxies is beyond the scope of this work, our observation findings indicate the exciting prospects for the future of our understanding of galaxy formation within the early Universe.

\section{Conclusions}\label{chpt3:conclusion}
Searches for high-redshift moderate-luminosity X-ray AGN, corresponding to the majority of the BH mass growth, have been limited by the lack of sufficiently deep near-IR data to reliably measure photometric redshifts as well as the sensitivity of current X-ray surveys. As such, the X-ray Luminosity Function (XLF) has been poorly constrained at high redshifts, with only a few studies fitting parametric models based on samples spanning to $z\sim6$ \citep[e.g.][]{Georgakakis2015, Pouliasis2024}. In this work we have presented new observational measurements of the XLF at $z=4-10$ by combining the deep NIR imaging from the UltraVISTA survey \citep[incorporated within the COSMOS2020 photometric catalogue of][]{Weaver2022} with the Chandra legacy imaging from \citet{Civano2016}. 

In order to measure the high-redshift XLF we have obtained a sample of $z\geq4$ X-ray detected sources within the COSMOS field, with photometric redshifts from 
the COSMOS2020 catalogue. We first performed a cross-match of blind X-ray detections in the Chandra COSMOS Legacy data with the COSMOS2020 catalogue, producing a sample of 21 (likely) high-redshift X-ray sources. We then pushed the limits of the Chandra imaging by extracting X-ray counts at the positions of high-redshift sources within the COSMOS2020 catalogue, allowing us to increase our sample size with an additional 11 (likely) high-redshift X-ray sources that were not identified by the initial X-ray soure detection procedure and pushing our sample to somewhat fainter flux limits.  

We use our resulting samples of high-redshift AGN to measure the XLF at $z=4-10$ following the $n_\mathrm{obs}/n_\mathrm{mdl}$ method of \citet{Miyaji2001}. The XLF was measured for the blind sample only and then the blind+extracted sample using the best fit photometric redshifts of each source and subsequently testing the upper and lower bounds on the XLF measurements given the possible $z_\mathrm{photo}$ of our sources. We then investigated the effect of obscuration within the AGN sample on the measured XLF and applied an obscuration correction to our XLF measurements.

These new, improved XLF measurements at moderate luminosities place important constraints on the space density of AGN where previously only extrapolated predictions from XLF models at lower redshifts were available. Our conclusions are as follows:
\begin{enumerate}
        
    \item We have measured the XLF at redshifts of $z=4-5$, $z=5-7$ and $z=7-10$, 
            spanning a luminosity range of $L_\mathrm{X}\approx 10^{43-46}$~erg~s$^{-1}$, and placing upper limits on bins where no sources are found.
        These XLF measurements, using either the blind-only sample or the blind+extracted sample, are consistent with the XLF models at $z=4-5$ (where the models were fit to pre-existing samples). However, our measurements indicate a higher space density of X-ray AGN toward higher redshifts than predicted by the model extrapolations. At $z=7-10$ our measurements clearly show a space density closer to that of the models at $z=4-5$, suggesting that any decline in the space density of AGN toward high redshift decline is far weaker than lower redshift measurements had previously indicated.

    \item We also considered the impact of differing photometric redshift estimates for sources in our sample, 
        and calculated the range of XLF measurements when considering the the maximum and minimum number of X-ray sources in our sample at each redshift and luminosity interval. The resulting measurements provide a more conservative uncertainty on our measured XLF, and remain consistent with the model predictions at $z=4-5$ and $z=5-7$. However, the upper bound of these limits indicate a potentially higher space density of AGN than previously expected at $z=5-7$ and $z=7-10$. Even our lower bounds hint at a higher space density of AGN than expected from the model extrapolations at $z=7-10$, the highest redshifts investigated here. 

    \item Performing measurements of the XLF using only sources detected at hard X-ray energies (2--7~keV, corresponding to rest-frame energies $\gtrsim$10~keV for the redshifts considered here), we found that obscuration likely has a significant impact on measurements of the XLF. The hard-band measurements indicate a much higher space density of AGN than our initial full-band measurements. The hardness ratios of our X-ray sample, determined through a Bayesian estimation method in order to account for the low photon counts, also indicate a significant fraction of our sample are likely to be heavily obscured or Compton-thick AGN.

    \item Dividing the X-ray sample into unabsorbed, absorbed and Compton-thick AGN, based on their observed hardness ratios, we have estimated $N_{H}$- and redshift-dependent corrections to the XLF measurements. We find that the obscuration-corrected XLF, based on our full-band sample, indicates a substantially higher space density than the uncorrected measurements and thus further increases the discrepancy with the model extrapolations. 
    
    \item From the obscuration corrections we calculate the intrinsic fraction of obscured ($N_\mathrm{H}=10^{23-24}\mathrm{cm^2}$) and Compton-thick ($N_\mathrm{H}>10^{24}\mathrm{cm^2}$) AGN.     We find the fraction of heavily obscured (absorbed and Compton-thick) sources increasingly dominates over the fraction of unobscured sources ($N_\mathrm{H}\leq10^{23}\mathrm{cm^2}$), with increasing redshift. The obscured fraction of AGN across the full redshift range of $z=4-10$ is $0.982^{+0.007}_{-0.008}$. Notably the Compton-thick fraction of AGN dominates the population, rising from $0.66_{-0.13}^{+0.15}$ at $z=4-5$ to $0.80^{+0.13}_{-0.04}$ and $0.97^{+0.02}_{-0.01}$ at $z=5-7$ and $z=7-10$ respectively.

    \item Considering the full range of uncertainty in the XLF we have measured,     we find that the space density of AGN in the early Universe \emph{could} be substantially higher than previously expected. This larger population of early AGN could be expected to have a significant effect on the growth of the first galaxies.

    \item Our obscuration-corrected XLF measurements indicate a rising space density towards lower luminosities, which appears to be consistent with the high space densities of lower luminosity AGN being discovered within deep JWST fields. Thus, our measurements point towards a reconciliation between X-ray measurements and the latest JWST findings at $z\geq6$, and suggests that the apparent X-ray weakness of many of the high-redshift AGN found by JWST may be due to high levels of obscuration that suppress the X-ray emission.

    \item Given our obscuration-corrected XLF measurements we calculate the corresponding Black Hole Accretion rate Density (BHAD) at $z=4-5$, $z=5-7$ and $z=7-10$. These measurements show an mild increase BHAD toward higher redshift, at $z>4$, and are consistent with recent estimates of high levels of black hole growth in the early Universe from JWST studies. Thus, substantial Black Hole growth  may have occurred before the build-up of the bulk of stellar mass in such galaxies.
\end{enumerate}

Whilst the results obtained here point towards a reconciliation between the new JWST AGN observations and bright-end X-ray measurements, further studies using larger area X-ray surveys and redshift measurements possible with both the mid-IR imaging of JWST and the larger sky areas covered by \textit{Euclid} offer the potential to constrain the space density of AGN at fainter luminosities and to improve the constraints found here. Furthermore, our measurements of the XLF presents exciting possibilities for the population studies that will be possible with future X-rays telescopes, such as ESA's \textit{NewAthena} X-ray Observatory mission \citet{Cruise2024} that will enable wide-area, deep X-ray surveys that will be highly effective at unveiling the prevalent population of high-redshift, heavily obscured, moderate-luminosity AGN out to $z\gtrsim6$ suggested by our measurements. 

\section*{Acknowledgements}
The authors thank A. Georgakais and D. Alexander for their helpful comments.
CLBH acknowledges support from an STFC PhD studentship.
CLBH and JA acknowledge support from a UKRI Future Leaders Fellowship (grant codes MR/T020989/1 and MR/Y019539/1). 
For the purpose of open access, the author has applied a Creative Commons Attribution (CC BY) licence to any Author Accepted Manuscript version arising from this submission.
We have benefited from the publicly available programming language {\sc Python}, including {\sc NumPy} \& {\sc SciPy} \citep[][]{VanDerWalt2011, Virtanen2020}, {\sc Matplotlib} \citep[][]{Hunter2007}, {\sc Astropy} \citep[][]{Robitaille2013} and the {\sc Topcat} analysis program \citep{Taylor2013}.

\input{ms.bbl}

\appendix

\section{COSMOS2020 LePhare Redshift Fits of the $\mathbf{z\geq4}$ Blind+Extracted Sources}\label{apdx:SEDs}
In this appendix I provide plots of the redshift photometric fits of my sample of 39 blind+extracted $z\geq4$ sources as performed by \citet{Weaver2022} using LePhare with both galaxy only (red) and AGN (blue) Spectral Energy Density (SED) models (see figure \ref{apdx:SEDs,fig:seds}). The galaxy templates used by \citet{Weaver2022} are those of \citet{Polletta2007} \citet{Ilbert2009}, \citet{Bruzual2003} and \citet{Onodera2012}, and the AGN templates consist of a galaxy SED with a QSO component of varying dominance over the galaxy SED \citep[see][for a detailed explanation of the creation of the AGN templates]{Salvato2009}. For each source the probability distribution of the redshift with the galaxy template fits is shown in the inset plot with the redshift of the AGN fit shown by the blue line for comparison and the $z<4$ region shaded. UltraVISTA Ks-band imaging for all of these sources is also shown, with the location of the source as given by the FARMER catalogue indicated by the orange cross, for which all sources can be seen to have a Ks-band detection as is required of my sample.

The shape of the SEDs of all my blind+extracted $z\geq4$ X-ray sources are suitably reproduced by the LePhare fits performed by \citet{Weaver2022}. For those sources which are within close proximity to a brighter object within the Ks-band image shown, the measured photometry at the blue end of the SEDs may be contaminated in certain filters resulting in the scatter of photometry in those sources. It can be seen that, by eye, there are cases where the galaxy only model provides a better fit to the measured photometry, the majority of which have the galaxy redshift ($z_{GAL}$) around $z\sim4$. For a number of my sources the measured photometry rises at IR wavelengths, as does the AGN sed model, above that predicted by the galaxy only models. This trend is indicative of high-redshift AGN appearing as red sources with rising power-law slope, as is being seen by JWST \citep[e.g. see][]{Matthee2024, Maiolino2024, Yue2024, Brooks2024}.

\begin{figure*}
    \centering
    \includegraphics[width=\linewidth]{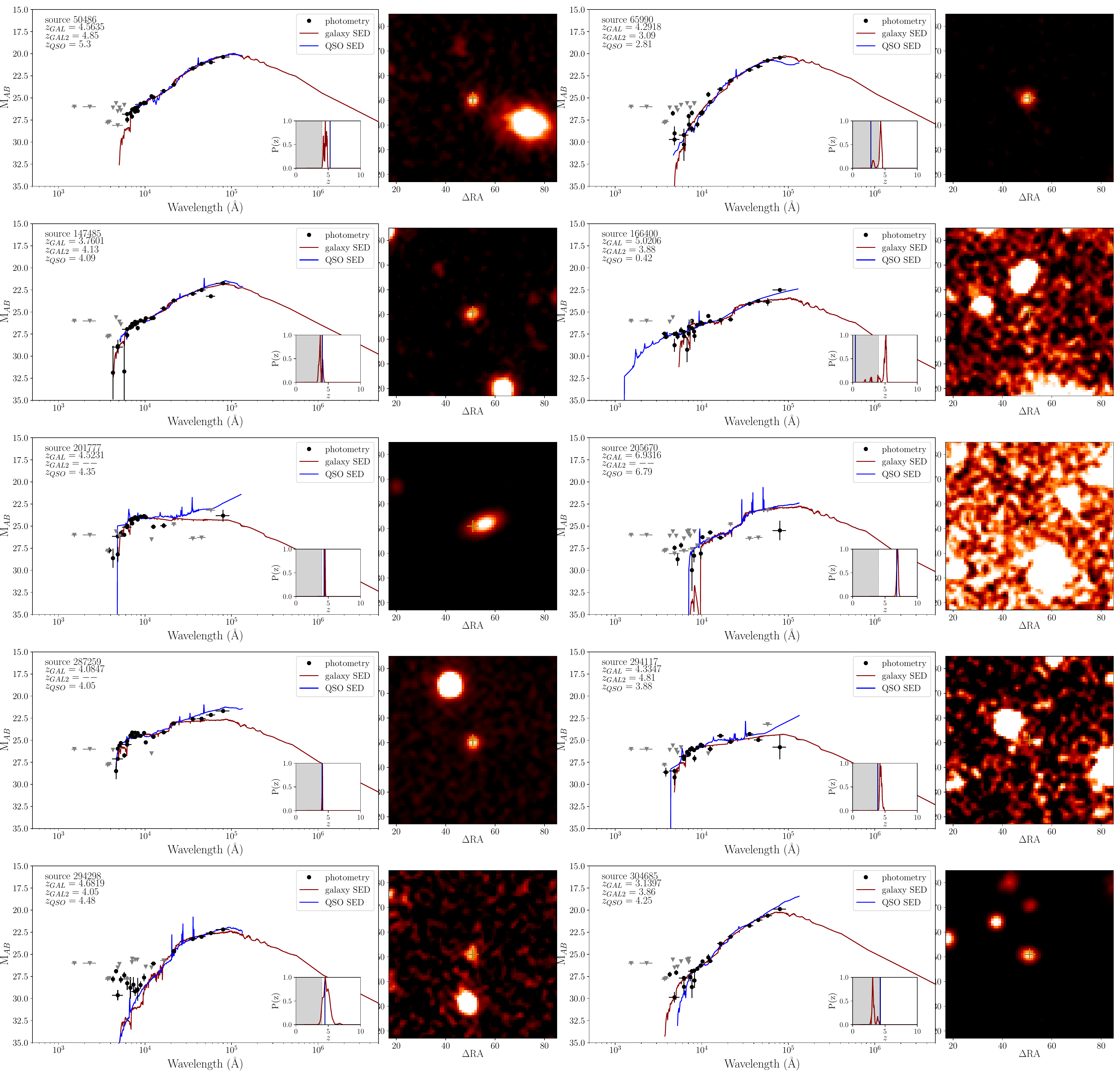}
    \caption[Galaxy and AGN photometric fits of blind+extracted $z>4$ sources]{The LePhare photometric redshifts fits of all 39 potentially $z\geq4$ sources in my blind+extracted sample, as given by galaxy templates (red) and AGN templates (blue). The photometric measurements of each source are shown by the black points, with the measurement in filters for which there are none detections plotted as an upper limit given by the filter depth (grey triangles). The spectral shape measured by the photometry of each source is reproduced by the fitted models. Some sources showing a slight rise in brightness at IR wavelengths above that predicted by the galaxy models, indicative of high redshift AGN being red sources with rising power-law slope \citep[as has been found with JWST;][]{Matthee2024, Maiolino2024} The probability distribution of possible redshifts based on the galaxy template fits is shown in the inset plot, with the best AGN redshift fit indicated by the blue line, and the region outside the redshift range of my blind+extracted sample shaded. The Ks-band UltraVISTA imaging of each source is shown, with the COSMOS2020 position indicated by the gold cross and a detection clearly seen in all sources which have been fit by both a galaxy and AGN template.}
    \label{apdx:SEDs,fig:seds}
\end{figure*}

\begin{figure*}
    \centering
    \includegraphics[width=\linewidth]{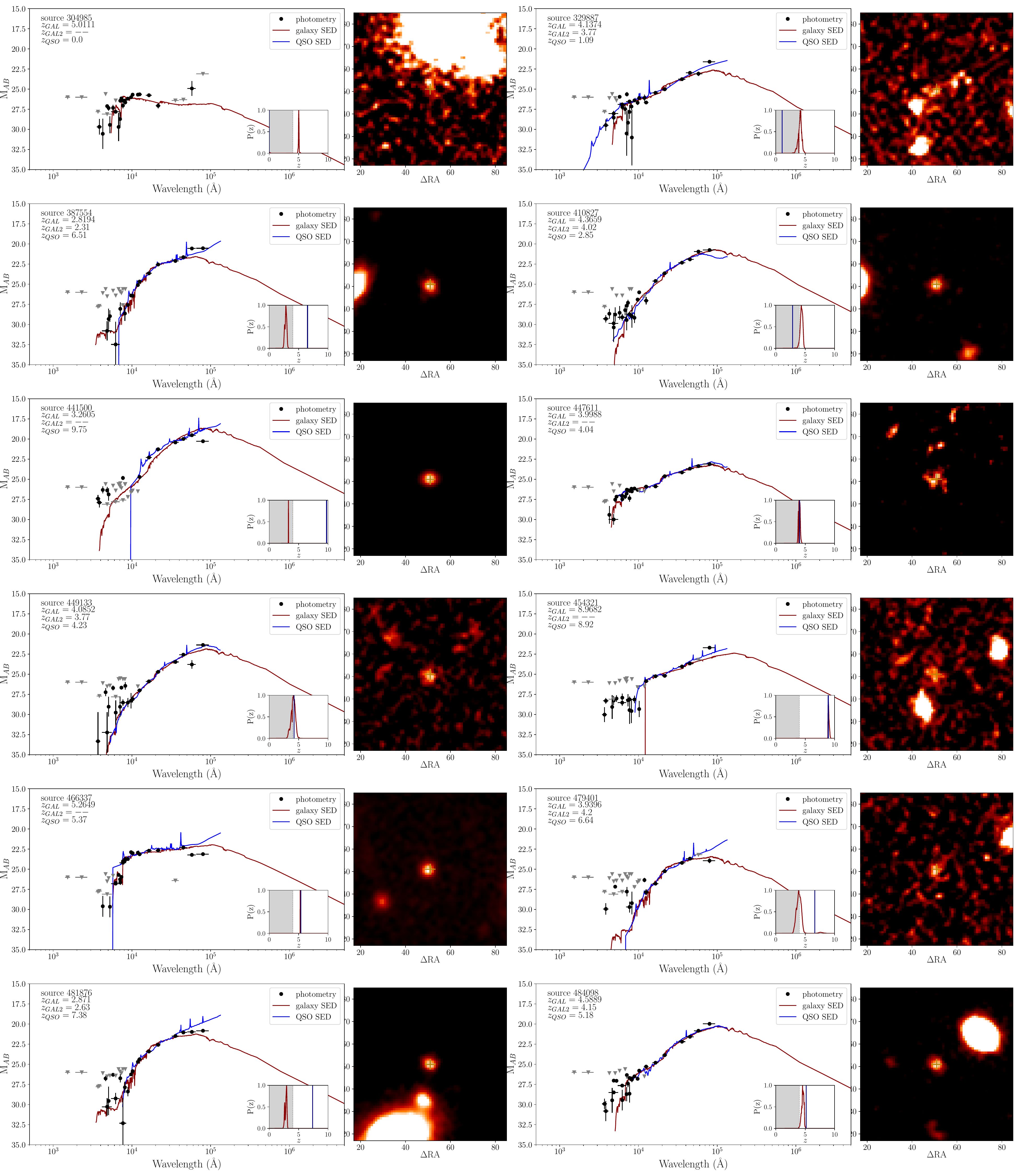}
    \contcaption{}
\end{figure*}

\begin{figure*}
    \centering
    \includegraphics[width=\linewidth]{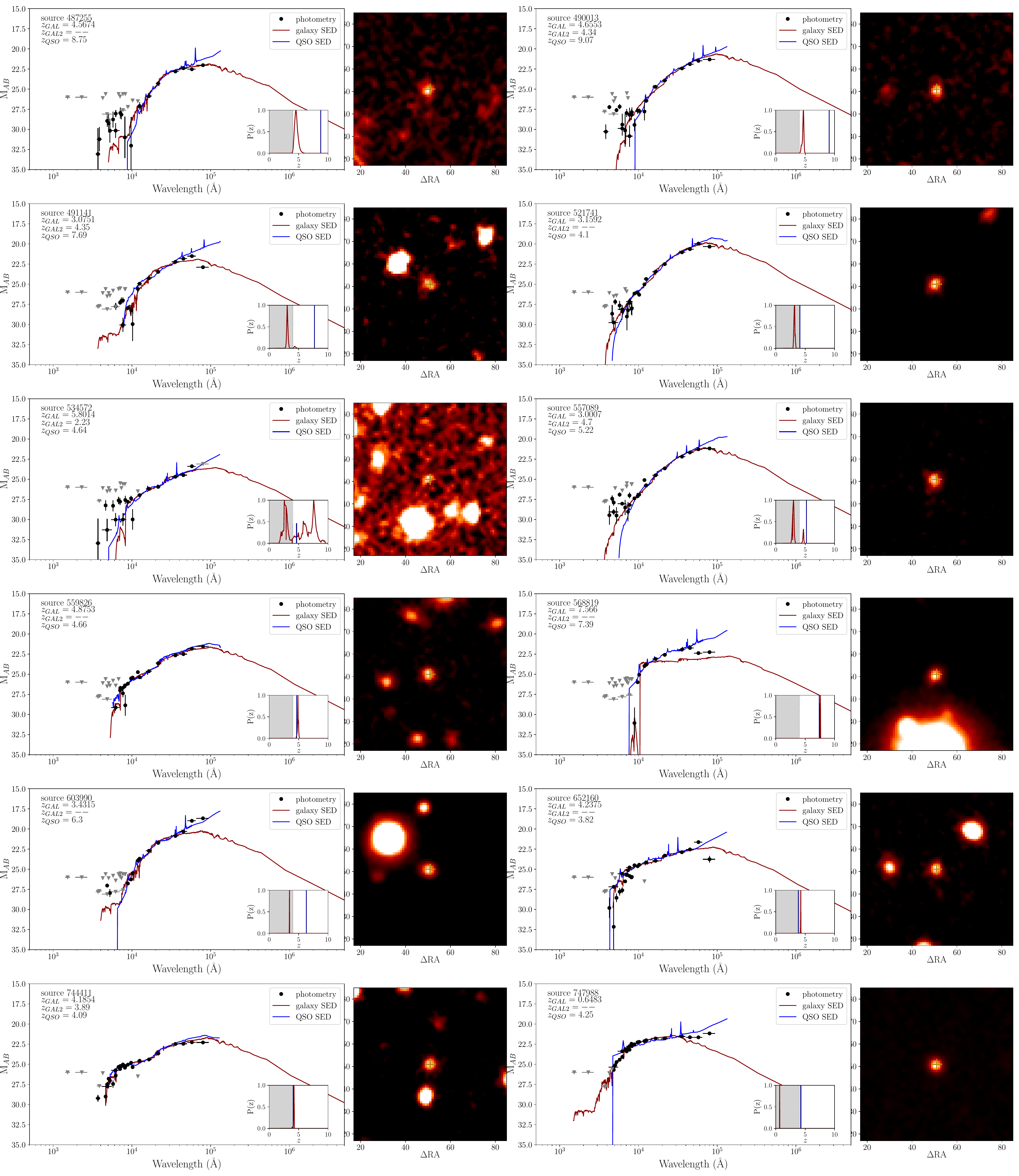}
    \contcaption{}
\end{figure*}

\begin{figure*}
    \centering
    \includegraphics[width=\linewidth]{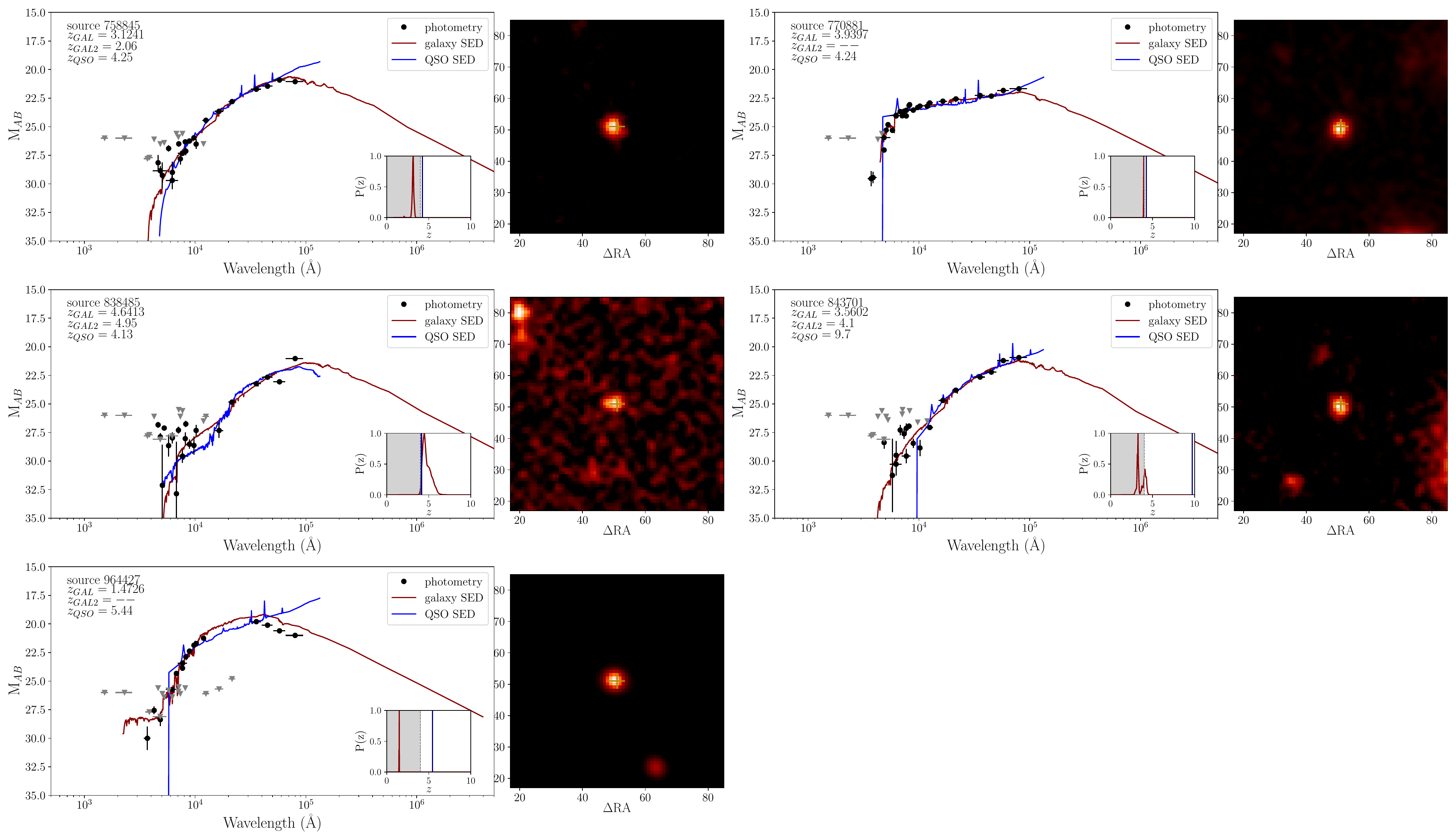}
    \contcaption{}
\end{figure*}

\bsp	\label{lastpage}

\begin{thebibliography}{}
\makeatletter
\relax
\def\mn@urlcharsother{\let\do\@makeother \do\$\do\&\do\#\do\^\do\_\do\%\do\~}
\def\mn@doi{\begingroup\mn@urlcharsother \@ifnextchar [ {\mn@doi@}
  {\mn@doi@[]}}
\def\mn@doi@[#1]#2{\def\@tempa{#1}\ifx\@tempa\@empty \href
  {http://dx.doi.org/#2} {doi:#2}\else \href {http://dx.doi.org/#2} {#1}\fi
  \endgroup}
\def\mn@eprint#1#2{\mn@eprint@#1:#2::\@nil}
\def\mn@eprint@arXiv#1{\href {http://arxiv.org/abs/#1} {{\tt arXiv:#1}}}
\def\mn@eprint@dblp#1{\href {http://dblp.uni-trier.de/rec/bibtex/#1.xml}
  {dblp:#1}}
\def\mn@eprint@#1:#2:#3:#4\@nil{\def\@tempa {#1}\def\@tempb {#2}\def\@tempc
  {#3}\ifx \@tempc \@empty \let \@tempc \@tempb \let \@tempb \@tempa \fi \ifx
  \@tempb \@empty \def\@tempb {arXiv}\fi \@ifundefined
  {mn@eprint@\@tempb}{\@tempb:\@tempc}{\expandafter \expandafter \csname
  mn@eprint@\@tempb\endcsname \expandafter{\@tempc}}}

\bibitem[\protect\citeauthoryear{{Aihara} et~al.,}{{Aihara}
  et~al.}{2019}]{Aihara2019}
{Aihara} H.,  et~al., 2019, \mn@doi [\pasj] {10.1093/pasj/psz103}, \href
  {https://ui.adsabs.harvard.edu/abs/2019PASJ...71..114A} {71, 114}

\bibitem[\protect\citeauthoryear{{Aird} et~al.,}{{Aird}
  et~al.}{2010}]{Aird2010}
{Aird} J.,  et~al., 2010, \mn@doi [\mnras] {10.1111/j.1365-2966.2009.15829.x},
  \href {https://ui.adsabs.harvard.edu/abs/2010MNRAS.401.2531A} {401, 2531}

\bibitem[\protect\citeauthoryear{{Aird}, {Coil}, {Georgakakis}, {Nandra},
  {Barro}  \& {P{\'e}rez-Gonz{\'a}lez}}{{Aird} et~al.}{2015}]{Aird2015}
{Aird} J.,  {Coil} A.~L.,  {Georgakakis} A.,  {Nandra} K.,  {Barro} G.,
  {P{\'e}rez-Gonz{\'a}lez} P.~G.,  2015, \mn@doi [\mnras]
  {10.1093/mnras/stv1062}, \href
  {https://ui.adsabs.harvard.edu/abs/2015MNRAS.451.1892A} {451, 1892}

\bibitem[\protect\citeauthoryear{{Aird}, {Coil}  \& {Georgakakis}}{{Aird}
  et~al.}{2017}]{Aird2017}
{Aird} J.,  {Coil} A.~L.,   {Georgakakis} A.,  2017, \mn@doi [\mnras]
  {10.1093/mnras/stw2932}, \href
  {https://ui.adsabs.harvard.edu/abs/2017MNRAS.465.3390A} {465, 3390}

\bibitem[\protect\citeauthoryear{{Aird}, {Coil}  \& {Georgakakis}}{{Aird}
  et~al.}{2018}]{Aird2018}
{Aird} J.,  {Coil} A.~L.,   {Georgakakis} A.,  2018, \mn@doi [\mnras]
  {10.1093/mnras/stx2700}, \href
  {https://ui.adsabs.harvard.edu/abs/2018MNRAS.474.1225A} {474, 1225}

\bibitem[\protect\citeauthoryear{{Aird}, {Coil}  \& {Georgakakis}}{{Aird}
  et~al.}{2019}]{Aird2019}
{Aird} J.,  {Coil} A.~L.,   {Georgakakis} A.,  2019, \mn@doi [\mnras]
  {10.1093/mnras/stz125}, \href
  {https://ui.adsabs.harvard.edu/abs/2019MNRAS.484.4360A} {484, 4360}

\bibitem[\protect\citeauthoryear{{Akins} et~al.,}{{Akins}
  et~al.}{2025}]{Akins2024}
{Akins} H.~B.,  et~al., 2025, \mn@doi [\apjl] {10.3847/2041-8213/adab76}, \href
  {https://ui.adsabs.harvard.edu/abs/2025ApJ...980L..29A} {980, L29}

\bibitem[\protect\citeauthoryear{{Alexander} \& {Hickox}}{{Alexander} \&
  {Hickox}}{2012}]{Alexander2012}
{Alexander} D.~M.,  {Hickox} R.~C.,  2012, \mn@doi [\nar]
  {10.1016/j.newar.2011.11.003}, \href
  {https://ui.adsabs.harvard.edu/abs/2012NewAR..56...93A} {56, 93}

\bibitem[\protect\citeauthoryear{{Ananna} et~al.,}{{Ananna}
  et~al.}{2019}]{Ananna2019}
{Ananna} T.~T.,  et~al., 2019, \mn@doi [\apj] {10.3847/1538-4357/aafb77}, \href
  {https://ui.adsabs.harvard.edu/abs/2019ApJ...871..240A} {871, 240}

\bibitem[\protect\citeauthoryear{{Arevalo Gonzalez} et~al.,}{{Arevalo Gonzalez}
  et~al.}{2025}]{Gonzalez2025}
{Arevalo Gonzalez} F.,  et~al., 2025, \mn@doi [arXiv e-prints]
  {10.48550/arXiv.2501.09585}, \href
  {https://ui.adsabs.harvard.edu/abs/2025arXiv250109585A} {p. arXiv:2501.09585}

\bibitem[\protect\citeauthoryear{{Arnouts} et~al.,}{{Arnouts}
  et~al.}{2002}]{Arnouts2002}
{Arnouts} S.,  et~al., 2002, \mn@doi [\mnras]
  {10.1046/j.1365-8711.2002.04988.x}, \href
  {https://ui.adsabs.harvard.edu/abs/2002MNRAS.329..355A} {329, 355}

\bibitem[\protect\citeauthoryear{{Ashby} et~al.,}{{Ashby}
  et~al.}{2018}]{Ashby2018}
{Ashby} M.~L.~N.,  et~al., 2018, \mn@doi [\apjs] {10.3847/1538-4365/aad4fb},
  \href {https://ui.adsabs.harvard.edu/abs/2018ApJS..237...39A} {237, 39}

\bibitem[\protect\citeauthoryear{{Assef} et~al.,}{{Assef}
  et~al.}{2011}]{Assef2011}
{Assef} R.~J.,  et~al., 2011, \mn@doi [\apj] {10.1088/0004-637X/728/1/56},
  \href {https://ui.adsabs.harvard.edu/abs/2011ApJ...728...56A} {728, 56}

\bibitem[\protect\citeauthoryear{{Astropy Collaboration} et~al.,}{{Astropy
  Collaboration} et~al.}{2013}]{Robitaille2013}
{Astropy Collaboration} et~al., 2013, \mn@doi [\aap]
  {10.1051/0004-6361/201322068}, \href
  {https://ui.adsabs.harvard.edu/abs/2013A&A...558A..33A} {558, A33}

\bibitem[\protect\citeauthoryear{{Balokovi{\'c}} et~al.,}{{Balokovi{\'c}}
  et~al.}{2018}]{Balokovic2018}
{Balokovi{\'c}} M.,  et~al., 2018, \mn@doi [\apj] {10.3847/1538-4357/aaa7eb},
  \href {https://ui.adsabs.harvard.edu/abs/2018ApJ...854...42B} {854, 42}

\bibitem[\protect\citeauthoryear{{Barlow-Hall}}{{Barlow-Hall}}{2024}]{Barlow-Hall2024}
{Barlow-Hall} C.,  2024, Phd thesis, The University of Edinburgh

\bibitem[\protect\citeauthoryear{{Barlow-Hall}, {Delaney}, {Aird}, {Evans},
  {Osborne}  \& {Watson}}{{Barlow-Hall} et~al.}{2023}]{Barlow-Hall2023}
{Barlow-Hall} C.~L.,  {Delaney} J.,  {Aird} J.,  {Evans} P.~A.,  {Osborne}
  J.~P.,   {Watson} M.~G.,  2023, \mn@doi [\mnras] {10.1093/mnras/stad100},
  \href {https://ui.adsabs.harvard.edu/abs/2023MNRAS.519.6055B} {519, 6055}

\bibitem[\protect\citeauthoryear{{Bertin} \& {Arnouts}}{{Bertin} \&
  {Arnouts}}{1996}]{Bertin1996}
{Bertin} E.,  {Arnouts} S.,  1996, \mn@doi [\aaps] {10.1051/aas:1996164}, \href
  {https://ui.adsabs.harvard.edu/abs/1996A&AS..117..393B} {117, 393}

\bibitem[\protect\citeauthoryear{{Bogd{\'a}n} et~al.,}{{Bogd{\'a}n}
  et~al.}{2024}]{Bogdan2023}
{Bogd{\'a}n} {\'A}.,  et~al., 2024, \mn@doi [Nature Astronomy]
  {10.1038/s41550-023-02111-9}, \href
  {https://ui.adsabs.harvard.edu/abs/2024NatAs...8..126B} {8, 126}

\bibitem[\protect\citeauthoryear{{Bowler}, {Jarvis}, {Dunlop}, {McLure},
  {McLeod}, {Adams}, {Milvang-Jensen}  \& {McCracken}}{{Bowler}
  et~al.}{2020}]{Bowler2020}
{Bowler} R.~A.~A.,  {Jarvis} M.~J.,  {Dunlop} J.~S.,  {McLure} R.~J.,  {McLeod}
  D.~J.,  {Adams} N.~J.,  {Milvang-Jensen} B.,   {McCracken} H.~J.,  2020,
  \mn@doi [\mnras] {10.1093/mnras/staa313}, \href
  {https://ui.adsabs.harvard.edu/abs/2020MNRAS.493.2059B} {493, 2059}

\bibitem[\protect\citeauthoryear{{Boyle} \& {Terlevich}}{{Boyle} \&
  {Terlevich}}{1998}]{Boyle1998}
{Boyle} B.~J.,  {Terlevich} R.~J.,  1998, \mn@doi [\mnras]
  {10.1046/j.1365-8711.1998.01264.x}, \href
  {https://ui.adsabs.harvard.edu/abs/1998MNRAS.293L..49B} {293, L49}

\bibitem[\protect\citeauthoryear{{Boyle}, {Shanks}, {Croom}, {Smith}, {Miller},
  {Loaring}  \& {Heymans}}{{Boyle} et~al.}{2000}]{Boyle2000}
{Boyle} B.~J.,  {Shanks} T.,  {Croom} S.~M.,  {Smith} R.~J.,  {Miller} L.,
  {Loaring} N.,   {Heymans} C.,  2000, \mn@doi [\mnras]
  {10.1046/j.1365-8711.2000.03730.x}, \href
  {https://ui.adsabs.harvard.edu/abs/2000MNRAS.317.1014B} {317, 1014}

\bibitem[\protect\citeauthoryear{{Brammer}, {van Dokkum}  \& {Coppi}}{{Brammer}
  et~al.}{2008}]{Brammer2008}
{Brammer} G.~B.,  {van Dokkum} P.~G.,   {Coppi} P.,  2008, \mn@doi [\apj]
  {10.1086/591786}, \href
  {https://ui.adsabs.harvard.edu/abs/2008ApJ...686.1503B} {686, 1503}

\bibitem[\protect\citeauthoryear{{Brandt} \& {Alexander}}{{Brandt} \&
  {Alexander}}{2015}]{Brandt2015}
{Brandt} W.~N.,  {Alexander} D.~M.,  2015, \mn@doi [\aapr]
  {10.1007/s00159-014-0081-z}, \href
  {https://ui.adsabs.harvard.edu/abs/2015A&ARv..23....1B} {23, 1}

\bibitem[\protect\citeauthoryear{{Brandt} \& {Hasinger}}{{Brandt} \&
  {Hasinger}}{2005}]{Brandt2005}
{Brandt} W.~N.,  {Hasinger} G.,  2005, \mn@doi [\araa]
  {10.1146/annurev.astro.43.051804.102213}, \href
  {https://ui.adsabs.harvard.edu/abs/2005ARA&A..43..827B} {43, 827}

\bibitem[\protect\citeauthoryear{{Brightman} \& {Nandra}}{{Brightman} \&
  {Nandra}}{2011}]{Brightman2011}
{Brightman} M.,  {Nandra} K.,  2011, \mn@doi [\mnras]
  {10.1111/j.1365-2966.2011.18207.x}, \href
  {https://ui.adsabs.harvard.edu/abs/2011MNRAS.413.1206B} {413, 1206}

\bibitem[\protect\citeauthoryear{{Brinch} et~al.,}{{Brinch}
  et~al.}{2023}]{Brinch2023}
{Brinch} M.,  et~al., 2023, \mn@doi [\apj] {10.3847/1538-4357/ac9d96}, \href
  {https://ui.adsabs.harvard.edu/abs/2023ApJ...943..153B} {943, 153}

\bibitem[\protect\citeauthoryear{{Brooks} et~al.,}{{Brooks}
  et~al.}{2024}]{Brooks2024}
{Brooks} M.,  et~al., 2024, \mn@doi [arXiv e-prints]
  {10.48550/arXiv.2410.07340}, \href
  {https://ui.adsabs.harvard.edu/abs/2024arXiv241007340B} {p. arXiv:2410.07340}

\bibitem[\protect\citeauthoryear{{Brusa} et~al.,}{{Brusa}
  et~al.}{2009}]{Brusa2009}
{Brusa} M.,  et~al., 2009, \mn@doi [\apj] {10.1088/0004-637X/693/1/8}, \href
  {https://ui.adsabs.harvard.edu/abs/2009ApJ...693....8B} {693, 8}

\bibitem[\protect\citeauthoryear{{Bruzual} \& {Charlot}}{{Bruzual} \&
  {Charlot}}{2003}]{Bruzual2003}
{Bruzual} G.,  {Charlot} S.,  2003, \mn@doi [\mnras]
  {10.1046/j.1365-8711.2003.06897.x}, \href
  {https://ui.adsabs.harvard.edu/abs/2003MNRAS.344.1000B} {344, 1000}

\bibitem[\protect\citeauthoryear{{Buchner} et~al.,}{{Buchner}
  et~al.}{2015}]{Buchner2015}
{Buchner} J.,  et~al., 2015, \mn@doi [\apj] {10.1088/0004-637X/802/2/89}, \href
  {https://ui.adsabs.harvard.edu/abs/2015ApJ...802...89B} {802, 89}

\bibitem[\protect\citeauthoryear{{Buchner}, {Salvato}, {Budava\'ri}  \&
  {Fotopoulou}}{{Buchner} et~al.}{2021}]{Buchner2021}
{Buchner} J.,  {Salvato} M.,  {Budava\'ri} T.,   {Fotopoulou} S.,  2021, {nway:
  Bayesian cross-matching of astronomical catalogs}, Astrophysics Source Code
  Library, record ascl:2102.014

\bibitem[\protect\citeauthoryear{{Cameron}, {Morris}, {Virani}, {Wolk},
  {Blackwell}, {Minow}  \& {O'Dell}}{{Cameron} et~al.}{2002}]{Cameron2002}
{Cameron} R.~A.,  {Morris} D.~C.,  {Virani} S.~N.,  {Wolk} S.~J.,  {Blackwell}
  W.~C.,  {Minow} J.~I.,   {O'Dell} S.~L.,  2002, in {Bohlender} D.~A.,
  {Durand} D.,   {Handley} T.~H.,  eds,  Astronomical Society of the Pacific
  Conference Series Vol. 281, Astronomical Data Analysis Software and Systems
  XI. p.~132

\bibitem[\protect\citeauthoryear{{Capak} et~al.,}{{Capak}
  et~al.}{2007}]{Capak2007}
{Capak} P.,  et~al., 2007, \mn@doi [\apjs] {10.1086/519081}, \href
  {https://ui.adsabs.harvard.edu/abs/2007ApJS..172...99C} {172, 99}

\bibitem[\protect\citeauthoryear{{Cappelluti} et~al.,}{{Cappelluti}
  et~al.}{2024}]{Cappelluti2024}
{Cappelluti} N.,  et~al., 2024, \mn@doi [Universe] {10.3390/universe10070276},
  \href {https://ui.adsabs.harvard.edu/abs/2024Univ...10..276C} {10, 276}

\bibitem[\protect\citeauthoryear{{Carilli} \& {Walter}}{{Carilli} \&
  {Walter}}{2013}]{Carilli2013}
{Carilli} C.~L.,  {Walter} F.,  2013, \mn@doi [\araa]
  {10.1146/annurev-astro-082812-140953}, \href
  {https://ui.adsabs.harvard.edu/abs/2013ARA&A..51..105C} {51, 105}

\bibitem[\protect\citeauthoryear{{Civano} et~al.,}{{Civano}
  et~al.}{2016}]{Civano2016}
{Civano} F.,  et~al., 2016, \mn@doi [\apj] {10.3847/0004-637X/819/1/62}, \href
  {https://ui.adsabs.harvard.edu/abs/2016ApJ...819...62C} {819, 62}

\bibitem[\protect\citeauthoryear{{Coupon}, {Czakon}, {Bosch}, {Komiyama},
  {Medezinski}, {Miyazaki}  \& {Oguri}}{{Coupon} et~al.}{2018}]{Coupon2018}
{Coupon} J.,  {Czakon} N.,  {Bosch} J.,  {Komiyama} Y.,  {Medezinski} E.,
  {Miyazaki} S.,   {Oguri} M.,  2018, \mn@doi [\pasj] {10.1093/pasj/psx047},
  \href {https://ui.adsabs.harvard.edu/abs/2018PASJ...70S...7C} {70, S7}

\bibitem[\protect\citeauthoryear{{Cruise} et~al.,}{{Cruise}
  et~al.}{2025}]{Cruise2024}
{Cruise} M.,  et~al., 2025, \mn@doi [Nature Astronomy]
  {10.1038/s41550-024-02416-3}, \href
  {https://ui.adsabs.harvard.edu/abs/2025NatAs...9...36C} {9, 36}

\bibitem[\protect\citeauthoryear{{Delaney}, {Aird}, {Evans}, {Barlow-Hall},
  {Osborne}  \& {Watson}}{{Delaney} et~al.}{2023}]{Delaney2023}
{Delaney} J.~N.,  {Aird} J.,  {Evans} P.~A.,  {Barlow-Hall} C.,  {Osborne}
  J.~P.,   {Watson} M.~G.,  2023, \mn@doi [\mnras] {10.1093/mnras/stac3703},
  \href {https://ui.adsabs.harvard.edu/abs/2023MNRAS.521.1620D} {521, 1620}

\bibitem[\protect\citeauthoryear{{Donnan} et~al.,}{{Donnan}
  et~al.}{2023a}]{Donnan2023}
{Donnan} C.~T.,  et~al., 2023a, \mn@doi [\mnras] {10.1093/mnras/stac3472},
  \href {https://ui.adsabs.harvard.edu/abs/2023MNRAS.518.6011D} {518, 6011}

\bibitem[\protect\citeauthoryear{{Donnan}, {McLeod}, {McLure}, {Dunlop},
  {Carnall}, {Cullen}  \& {Magee}}{{Donnan} et~al.}{2023b}]{Donnan2023a}
{Donnan} C.~T.,  {McLeod} D.~J.,  {McLure} R.~J.,  {Dunlop} J.~S.,  {Carnall}
  A.~C.,  {Cullen} F.,   {Magee} D.,  2023b, \mn@doi [\mnras]
  {10.1093/mnras/stad471}, \href
  {https://ui.adsabs.harvard.edu/abs/2023MNRAS.520.4554D} {520, 4554}

\bibitem[\protect\citeauthoryear{{Donnan} et~al.,}{{Donnan}
  et~al.}{2024}]{Donnan2024}
{Donnan} C.~T.,  et~al., 2024, \mn@doi [\mnras] {10.1093/mnras/stae2037}, \href
  {https://ui.adsabs.harvard.edu/abs/2024MNRAS.533.3222D} {533, 3222}

\bibitem[\protect\citeauthoryear{{Elvis} et~al.,}{{Elvis}
  et~al.}{2009}]{Elvis2009}
{Elvis} M.,  et~al., 2009, \mn@doi [\apjs] {10.1088/0067-0049/184/1/158}, \href
  {https://ui.adsabs.harvard.edu/abs/2009ApJS..184..158E} {184, 158}

\bibitem[\protect\citeauthoryear{{Fan} et~al.,}{{Fan} et~al.}{2001}]{Fan2001}
{Fan} X.,  et~al., 2001, \mn@doi [\aj] {10.1086/324111}, \href
  {https://ui.adsabs.harvard.edu/abs/2001AJ....122.2833F} {122, 2833}

\bibitem[\protect\citeauthoryear{{Finkelstein} et~al.,}{{Finkelstein}
  et~al.}{2022}]{Finkelstein2022}
{Finkelstein} S.~L.,  et~al., 2022, \mn@doi [\apjl] {10.3847/2041-8213/ac966e},
  \href {https://ui.adsabs.harvard.edu/abs/2022ApJ...940L..55F} {940, L55}

\bibitem[\protect\citeauthoryear{{Fruscione}, {Burke}  \&
  {Siemiginowska}}{{Fruscione} et~al.}{2009}]{Fruscione2009}
{Fruscione} A.,  {Burke} D.,   {Siemiginowska} A.,  2009, Chandra News, \href
  {https://ui.adsabs.harvard.edu/abs/2009ChNew..16...26F} {16, 26}

\bibitem[\protect\citeauthoryear{{Gaia Collaboration} et~al.,}{{Gaia
  Collaboration} et~al.}{2016}]{GaiaCollaboration2016}
{Gaia Collaboration} et~al., 2016, \mn@doi [\aap]
  {10.1051/0004-6361/201629512}, \href
  {https://ui.adsabs.harvard.edu/abs/2016A&A...595A...2G} {595, A2}

\bibitem[\protect\citeauthoryear{{Gehrels}}{{Gehrels}}{1986}]{Gehrels1986}
{Gehrels} N.,  1986, \mn@doi [\apj] {10.1086/164079}, \href
  {https://ui.adsabs.harvard.edu/abs/1986ApJ...303..336G} {303, 336}

\bibitem[\protect\citeauthoryear{{Georgakakis}, {Nandra}, {Laird}, {Aird}  \&
  {Trichas}}{{Georgakakis} et~al.}{2008}]{Georgakakis2008}
{Georgakakis} A.,  {Nandra} K.,  {Laird} E.~S.,  {Aird} J.,   {Trichas} M.,
  2008, \mn@doi [\mnras] {10.1111/j.1365-2966.2008.13423.x}, \href
  {https://ui.adsabs.harvard.edu/abs/2008MNRAS.388.1205G} {388, 1205}

\bibitem[\protect\citeauthoryear{{Georgakakis} et~al.,}{{Georgakakis}
  et~al.}{2015}]{Georgakakis2015}
{Georgakakis} A.,  et~al., 2015, \mn@doi [\mnras] {10.1093/mnras/stv1703},
  \href {https://ui.adsabs.harvard.edu/abs/2015MNRAS.453.1946G} {453, 1946}

\bibitem[\protect\citeauthoryear{{Georgakakis}, {Aird}, {Schulze}, {Dwelly},
  {Salvato}, {Nandra}, {Merloni}  \& {Schneider}}{{Georgakakis}
  et~al.}{2017}]{Georgakakis2017}
{Georgakakis} A.,  {Aird} J.,  {Schulze} A.,  {Dwelly} T.,  {Salvato} M.,
  {Nandra} K.,  {Merloni} A.,   {Schneider} D.~P.,  2017, \mn@doi [\mnras]
  {10.1093/mnras/stx1602}, \href
  {https://ui.adsabs.harvard.edu/abs/2017MNRAS.471.1976G} {471, 1976}

\bibitem[\protect\citeauthoryear{{Gilli}, {Comastri}  \& {Hasinger}}{{Gilli}
  et~al.}{2007}]{Gilli2007}
{Gilli} R.,  {Comastri} A.,   {Hasinger} G.,  2007, \mn@doi [\aap]
  {10.1051/0004-6361:20066334}, \href
  {https://ui.adsabs.harvard.edu/abs/2007A&A...463...79G} {463, 79}

\bibitem[\protect\citeauthoryear{{Gilli} et~al.,}{{Gilli}
  et~al.}{2009}]{Gilli2009}
{Gilli} R.,  et~al., 2009, \mn@doi [\aap] {10.1051/0004-6361:200810821}, \href
  {https://ui.adsabs.harvard.edu/abs/2009A&A...494...33G} {494, 33}

\bibitem[\protect\citeauthoryear{{Gilli} et~al.,}{{Gilli}
  et~al.}{2022}]{Gilli2022}
{Gilli} R.,  et~al., 2022, \mn@doi [\aap] {10.1051/0004-6361/202243708}, \href
  {https://ui.adsabs.harvard.edu/abs/2022A&A...666A..17G} {666, A17}

\bibitem[\protect\citeauthoryear{{Grant}, {Bautz}, {Plucinsky}  \&
  {Ford}}{{Grant} et~al.}{2024}]{Grant2024}
{Grant} C.~E.,  {Bautz} M.~W.,  {Plucinsky} P.~P.,   {Ford} P.~G.,  2024, in
  {den Herder} J.-W.~A.,  {Nikzad} S.,   {Nakazawa} K.,  eds,  Society of
  Photo-Optical Instrumentation Engineers (SPIE) Conference Series Vol. 13093,
  Space Telescopes and Instrumentation 2024: Ultraviolet to Gamma Ray. p.
  130931E (\mn@eprint {arXiv} {2406.18395}), \mn@doi{10.1117/12.3018498}

\bibitem[\protect\citeauthoryear{{Greene} et~al.,}{{Greene}
  et~al.}{2024}]{Greene2023}
{Greene} J.~E.,  et~al., 2024, \mn@doi [\apj] {10.3847/1538-4357/ad1e5f}, \href
  {https://ui.adsabs.harvard.edu/abs/2024ApJ...964...39G} {964, 39}

\bibitem[\protect\citeauthoryear{{Guetzoyan}, {Aird}, {Georgakakis}, {Coil},
  {Barlow-Hall}, {Hickox}, {Rankine}  \& {Terrazas}}{{Guetzoyan}
  et~al.}{2025}]{Guetzoyan2024}
{Guetzoyan} P.,  {Aird} J.,  {Georgakakis} A.,  {Coil} A.~L.,  {Barlow-Hall}
  C.,  {Hickox} R.~C.,  {Rankine} A.~L.,   {Terrazas} B.~A.,  2025, \mn@doi
  [\mnras] {10.1093/mnras/stae2564}, \href
  {https://ui.adsabs.harvard.edu/abs/2025MNRAS.536...79G} {536, 79}

\bibitem[\protect\citeauthoryear{{HI4PI Collaboration} et~al.,}{{HI4PI
  Collaboration} et~al.}{2016}]{HI4PICollaboration2016}
{HI4PI Collaboration} et~al., 2016, \mn@doi [\aap]
  {10.1051/0004-6361/201629178}, \href
  {https://ui.adsabs.harvard.edu/abs/2016A&A...594A.116H} {594, A116}

\bibitem[\protect\citeauthoryear{{Habouzit} et~al.,}{{Habouzit}
  et~al.}{2021}]{Habouzit2021}
{Habouzit} M.,  et~al., 2021, \mn@doi [\mnras] {10.1093/mnras/stab496}, \href
  {https://ui.adsabs.harvard.edu/abs/2021MNRAS.503.1940H} {503, 1940}

\bibitem[\protect\citeauthoryear{{Hainline} et~al.,}{{Hainline}
  et~al.}{2025}]{Hainline2024}
{Hainline} K.~N.,  et~al., 2025, \mn@doi [\apj] {10.3847/1538-4357/ad9920},
  \href {https://ui.adsabs.harvard.edu/abs/2025ApJ...979..138H} {979, 138}

\bibitem[\protect\citeauthoryear{{Heckman} \& {Best}}{{Heckman} \&
  {Best}}{2014}]{Heckman2014}
{Heckman} T.~M.,  {Best} P.~N.,  2014, \mn@doi [\araa]
  {10.1146/annurev-astro-081913-035722}, \href
  {https://ui.adsabs.harvard.edu/abs/2014ARA&A..52..589H} {52, 589}

\bibitem[\protect\citeauthoryear{{Hickox} \& {Alexander}}{{Hickox} \&
  {Alexander}}{2018}]{Hickox2018}
{Hickox} R.~C.,  {Alexander} D.~M.,  2018, \mn@doi [\araa]
  {10.1146/annurev-astro-081817-051803}, \href
  {https://ui.adsabs.harvard.edu/abs/2018ARA&A..56..625H} {56, 625}

\bibitem[\protect\citeauthoryear{{Hunter}}{{Hunter}}{2007}]{Hunter2007}
{Hunter} J.~D.,  2007, \mn@doi [Computing in Science and Engineering]
  {10.1109/MCSE.2007.55}, \href
  {https://ui.adsabs.harvard.edu/abs/2007CSE.....9...90H} {9, 90}

\bibitem[\protect\citeauthoryear{{Ikeda}, {Awaki}  \& {Terashima}}{{Ikeda}
  et~al.}{2009}]{Ikeda2009}
{Ikeda} S.,  {Awaki} H.,   {Terashima} Y.,  2009, \mn@doi [\apj]
  {10.1088/0004-637X/692/1/608}, \href
  {https://ui.adsabs.harvard.edu/abs/2009ApJ...692..608I} {692, 608}

\bibitem[\protect\citeauthoryear{{Ilbert} et~al.,}{{Ilbert}
  et~al.}{2006}]{Ilbert2006}
{Ilbert} O.,  et~al., 2006, \mn@doi [\aap] {10.1051/0004-6361:20065138}, \href
  {https://ui.adsabs.harvard.edu/abs/2006A&A...457..841I} {457, 841}

\bibitem[\protect\citeauthoryear{{Ilbert} et~al.,}{{Ilbert}
  et~al.}{2009}]{Ilbert2009}
{Ilbert} O.,  et~al., 2009, \mn@doi [\apj] {10.1088/0004-637X/690/2/1236},
  \href {https://ui.adsabs.harvard.edu/abs/2009ApJ...690.1236I} {690, 1236}

\bibitem[\protect\citeauthoryear{{Ilbert} et~al.,}{{Ilbert}
  et~al.}{2013}]{Ilbert2013}
{Ilbert} O.,  et~al., 2013, \mn@doi [\aap] {10.1051/0004-6361/201321100}, \href
  {https://ui.adsabs.harvard.edu/abs/2013A&A...556A..55I} {556, A55}

\bibitem[\protect\citeauthoryear{{Inayoshi}, {Visbal}  \& {Haiman}}{{Inayoshi}
  et~al.}{2020}]{Inayoshi2020}
{Inayoshi} K.,  {Visbal} E.,   {Haiman} Z.,  2020, \mn@doi [\araa]
  {10.1146/annurev-astro-120419-014455}, \href
  {https://ui.adsabs.harvard.edu/abs/2020ARA&A..58...27I} {58, 27}

\bibitem[\protect\citeauthoryear{{Inayoshi}, {Kimura}  \& {Noda}}{{Inayoshi}
  et~al.}{2024}]{Inayoshi2024}
{Inayoshi} K.,  {Kimura} S.~S.,   {Noda} H.,  2024, \mn@doi [arXiv e-prints]
  {10.48550/arXiv.2412.03653}, \href
  {https://ui.adsabs.harvard.edu/abs/2024arXiv241203653I} {p. arXiv:2412.03653}

\bibitem[\protect\citeauthoryear{{Kalfountzou}, {Civano}, {Elvis}, {Trichas}
  \& {Green}}{{Kalfountzou} et~al.}{2014}]{Kalfountzou2014}
{Kalfountzou} E.,  {Civano} F.,  {Elvis} M.,  {Trichas} M.,   {Green} P.,
  2014, \mn@doi [\mnras] {10.1093/mnras/stu1745}, \href
  {https://ui.adsabs.harvard.edu/abs/2014MNRAS.445.1430K} {445, 1430}

\bibitem[\protect\citeauthoryear{{Kocevski} et~al.,}{{Kocevski}
  et~al.}{2018}]{Kocevski2018}
{Kocevski} D.~D.,  et~al., 2018, \mn@doi [\apjs] {10.3847/1538-4365/aab9b4},
  \href {https://ui.adsabs.harvard.edu/abs/2018ApJS..236...48K} {236, 48}

\bibitem[\protect\citeauthoryear{{Kocevski} et~al.,}{{Kocevski}
  et~al.}{2024}]{Kocevski2024}
{Kocevski} D.~D.,  et~al., 2024, \mn@doi [arXiv e-prints]
  {10.48550/arXiv.2404.03576}, \href
  {https://ui.adsabs.harvard.edu/abs/2024arXiv240403576K} {p. arXiv:2404.03576}

\bibitem[\protect\citeauthoryear{{Kokorev} et~al.,}{{Kokorev}
  et~al.}{2024}]{Kokorev2024}
{Kokorev} V.,  et~al., 2024, \mn@doi [\apj] {10.3847/1538-4357/ad4265}, \href
  {https://ui.adsabs.harvard.edu/abs/2024ApJ...968...38K} {968, 38}

\bibitem[\protect\citeauthoryear{{Kormendy} \& {Ho}}{{Kormendy} \&
  {Ho}}{2013}]{Kormendy2013}
{Kormendy} J.,  {Ho} L.~C.,  2013, \mn@doi [\araa]
  {10.1146/annurev-astro-082708-101811}, \href
  {https://ui.adsabs.harvard.edu/abs/2013ARA&A..51..511K} {51, 511}

\bibitem[\protect\citeauthoryear{{Kormendy} \& {Richstone}}{{Kormendy} \&
  {Richstone}}{1995}]{Kormendy1995}
{Kormendy} J.,  {Richstone} D.,  1995, \mn@doi [\araa]
  {10.1146/annurev.aa.33.090195.003053}, \href
  {https://ui.adsabs.harvard.edu/abs/1995ARA&A..33..581K} {33, 581}

\bibitem[\protect\citeauthoryear{{Kov{\'a}cs} et~al.,}{{Kov{\'a}cs}
  et~al.}{2024}]{Kovacs2024}
{Kov{\'a}cs} O.~E.,  et~al., 2024, \mn@doi [\apjl] {10.3847/2041-8213/ad391f},
  \href {https://ui.adsabs.harvard.edu/abs/2024ApJ...965L..21K} {965, L21}

\bibitem[\protect\citeauthoryear{{Lacy}, {Ridgway}, {Sajina}, {Petric},
  {Gates}, {Urrutia}  \& {Storrie-Lombardi}}{{Lacy} et~al.}{2015}]{Lacy2015}
{Lacy} M.,  {Ridgway} S.~E.,  {Sajina} A.,  {Petric} A.~O.,  {Gates} E.~L.,
  {Urrutia} T.,   {Storrie-Lombardi} L.~J.,  2015, \mn@doi [\apj]
  {10.1088/0004-637X/802/2/102}, \href
  {https://ui.adsabs.harvard.edu/abs/2015ApJ...802..102L} {802, 102}

\bibitem[\protect\citeauthoryear{{Laigle} et~al.,}{{Laigle}
  et~al.}{2016}]{Laigle2016}
{Laigle} C.,  et~al., 2016, \mn@doi [\apjs] {10.3847/0067-0049/224/2/24}, \href
  {https://ui.adsabs.harvard.edu/abs/2016ApJS..224...24L} {224, 24}

\bibitem[\protect\citeauthoryear{{Laird} et~al.,}{{Laird}
  et~al.}{2009}]{Laird2009}
{Laird} E.~S.,  et~al., 2009, \mn@doi [\apjs] {10.1088/0067-0049/180/1/102},
  \href {https://ui.adsabs.harvard.edu/abs/2009ApJS..180..102L} {180, 102}

\bibitem[\protect\citeauthoryear{{Laloux} et~al.,}{{Laloux}
  et~al.}{2023}]{Laloux2023}
{Laloux} B.,  et~al., 2023, \mn@doi [\mnras] {10.1093/mnras/stac3255}, \href
  {https://ui.adsabs.harvard.edu/abs/2023MNRAS.518.2546L} {518, 2546}

\bibitem[\protect\citeauthoryear{{Laloux} et~al.,}{{Laloux}
  et~al.}{2024}]{Laloux2024}
{Laloux} B.,  et~al., 2024, \mn@doi [\mnras] {10.1093/mnras/stae1649}, \href
  {https://ui.adsabs.harvard.edu/abs/2024MNRAS.532.3459L} {532, 3459}

\bibitem[\protect\citeauthoryear{{Lang}, {Hogg}  \& {Mykytyn}}{{Lang}
  et~al.}{2016}]{Lang2016}
{Lang} D.,  {Hogg} D.~W.,   {Mykytyn} D.,  2016, {The Tractor: Probabilistic
  astronomical source detection and measurement}, Astrophysics Source Code
  Library, record ascl:1604.008

\bibitem[\protect\citeauthoryear{{Larson} et~al.,}{{Larson}
  et~al.}{2023}]{Larson2023}
{Larson} R.~L.,  et~al., 2023, \mn@doi [\apjl] {10.3847/2041-8213/ace619},
  \href {https://ui.adsabs.harvard.edu/abs/2023ApJ...953L..29L} {953, L29}

\bibitem[\protect\citeauthoryear{{Le F{\`e}vre} et~al.,}{{Le F{\`e}vre}
  et~al.}{2020}]{LeFevre2020}
{Le F{\`e}vre} O.,  et~al., 2020, \mn@doi [\aap] {10.1051/0004-6361/201936965},
  \href {https://ui.adsabs.harvard.edu/abs/2020A&A...643A...1L} {643, A1}

\bibitem[\protect\citeauthoryear{{Luo} et~al.,}{{Luo} et~al.}{2017}]{Luo2017}
{Luo} B.,  et~al., 2017, \mn@doi [\apjs] {10.3847/1538-4365/228/1/2}, \href
  {https://ui.adsabs.harvard.edu/abs/2017ApJS..228....2L} {228, 2}

\bibitem[\protect\citeauthoryear{{Lusso}, {Valiante}  \& {Vito}}{{Lusso}
  et~al.}{2023}]{Lusso2023}
{Lusso} E.,  {Valiante} R.,   {Vito} F.,  2023, in , Handbook of X-ray and
  Gamma-ray Astrophysics.
p.~122, \mn@doi{10.1007/978-981-16-4544-0_116-1}

\bibitem[\protect\citeauthoryear{{Madau} \& {Dickinson}}{{Madau} \&
  {Dickinson}}{2014}]{Madau2014}
{Madau} P.,  {Dickinson} M.,  2014, \mn@doi [\araa]
  {10.1146/annurev-astro-081811-125615}, \href
  {https://ui.adsabs.harvard.edu/abs/2014ARA&A..52..415M} {52, 415}

\bibitem[\protect\citeauthoryear{{Madau} \& {Haardt}}{{Madau} \&
  {Haardt}}{2024}]{Madau2024}
{Madau} P.,  {Haardt} F.,  2024, \mn@doi [\apjl] {10.3847/2041-8213/ad90e1},
  \href {https://ui.adsabs.harvard.edu/abs/2024ApJ...976L..24M} {976, L24}

\bibitem[\protect\citeauthoryear{{Magdziarz} \& {Zdziarski}}{{Magdziarz} \&
  {Zdziarski}}{1995}]{Magdziarz1995}
{Magdziarz} P.,  {Zdziarski} A.~A.,  1995, \mn@doi [\mnras]
  {10.1093/mnras/273.3.837}, \href
  {https://ui.adsabs.harvard.edu/abs/1995MNRAS.273..837M} {273, 837}

\bibitem[\protect\citeauthoryear{{Maiolino} et~al.,}{{Maiolino}
  et~al.}{2024}]{Maiolino2023}
{Maiolino} R.,  et~al., 2024, \mn@doi [\aap] {10.1051/0004-6361/202347640},
  \href {https://ui.adsabs.harvard.edu/abs/2024A&A...691A.145M} {691, A145}

\bibitem[\protect\citeauthoryear{Maiolino et~al.,}{Maiolino
  et~al.}{2025a}]{Maiolino2025}
Maiolino R.,  et~al., 2025a, \mn@doi [Monthly Notices of the Royal Astronomical
  Society] {10.1093/mnras/staf359}, 538, 1921

\bibitem[\protect\citeauthoryear{{Maiolino} et~al.,}{{Maiolino}
  et~al.}{2025b}]{Maiolino2024}
{Maiolino} R.,  et~al., 2025b, \mn@doi [\mnras] {10.1093/mnras/staf359}, \href
  {https://ui.adsabs.harvard.edu/abs/2025MNRAS.538.1921M} {538, 1921}

\bibitem[\protect\citeauthoryear{{Marchesi} et~al.,}{{Marchesi}
  et~al.}{2016}]{Marchesi2016}
{Marchesi} S.,  et~al., 2016, \mn@doi [\apj] {10.3847/0004-637X/827/2/150},
  \href {https://ui.adsabs.harvard.edu/abs/2016ApJ...827..150M} {827, 150}

\bibitem[\protect\citeauthoryear{{Matthee} et~al.,}{{Matthee}
  et~al.}{2024}]{Matthee2024}
{Matthee} J.,  et~al., 2024, \mn@doi [\apj] {10.3847/1538-4357/ad2345}, \href
  {https://ui.adsabs.harvard.edu/abs/2024ApJ...963..129M} {963, 129}

\bibitem[\protect\citeauthoryear{{McCracken} et~al.,}{{McCracken}
  et~al.}{2012}]{McCracken2012}
{McCracken} H.~J.,  et~al., 2012, \mn@doi [\aap] {10.1051/0004-6361/201219507},
  \href {https://ui.adsabs.harvard.edu/abs/2012A&A...544A.156M} {544, A156}

\bibitem[\protect\citeauthoryear{{McGreer} et~al.,}{{McGreer}
  et~al.}{2013}]{McGreer2013}
{McGreer} I.~D.,  et~al., 2013, \mn@doi [\apj] {10.1088/0004-637X/768/2/105},
  \href {https://ui.adsabs.harvard.edu/abs/2013ApJ...768..105M} {768, 105}

\bibitem[\protect\citeauthoryear{{Miyaji}, {Hasinger}  \& {Schmidt}}{{Miyaji}
  et~al.}{2001}]{Miyaji2001}
{Miyaji} T.,  {Hasinger} G.,   {Schmidt} M.,  2001, \mn@doi [\aap]
  {10.1051/0004-6361:20010102}, \href
  {https://ui.adsabs.harvard.edu/abs/2001A&A...369...49M} {369, 49}

\bibitem[\protect\citeauthoryear{{Moneti} et~al.,}{{Moneti}
  et~al.}{2023}]{Moneti2023}
{Moneti} A.,  et~al., 2023, {VizieR Online Data Catalog: The fourth UltraVISTA
  data release (DR4) (Moneti+, 2019)}, VizieR On-line Data Catalog: II/373.
  Originally published in: 2012A\&A...544A.156M

\bibitem[\protect\citeauthoryear{{Muzzin} et~al.,}{{Muzzin}
  et~al.}{2013}]{Muzzin2013}
{Muzzin} A.,  et~al., 2013, \mn@doi [\apjs] {10.1088/0067-0049/206/1/8}, \href
  {https://ui.adsabs.harvard.edu/abs/2013ApJS..206....8M} {206, 8}

\bibitem[\protect\citeauthoryear{{Nandra} et~al.,}{{Nandra}
  et~al.}{2015}]{Nandra2015}
{Nandra} K.,  et~al., 2015, \mn@doi [\apjs] {10.1088/0067-0049/220/1/10}, \href
  {https://ui.adsabs.harvard.edu/abs/2015ApJS..220...10N} {220, 10}

\bibitem[\protect\citeauthoryear{{Napolitano} et~al.,}{{Napolitano}
  et~al.}{2025}]{Napolitano2025}
{Napolitano} L.,  et~al., 2025, \mn@doi [\aap] {10.1051/0004-6361/202452090},
  \href {https://ui.adsabs.harvard.edu/abs/2025A&A...693A..50N} {693, A50}

\bibitem[\protect\citeauthoryear{O'Dell et~al.,}{O'Dell
  et~al.}{2007}]{ODell2007a}
O'Dell S.~L.,  et~al., 2007, in {{UV}}, {{X-Ray}}, and {{Gamma-Ray Space
  Instrumentation}} for {{Astronomy XV}}. SPIE, pp 19--30,
  \mn@doi{10.1117/12.734594}, \url
  {https://www.spiedigitallibrary.org/conference-proceedings-of-spie/6686/668603/Managing-radiation-degradation-of-CCDs-on-the-Chandra-X-ray/10.1117/12.734594.full}

\bibitem[\protect\citeauthoryear{{Onodera} et~al.,}{{Onodera}
  et~al.}{2012}]{Onodera2012}
{Onodera} M.,  et~al., 2012, \mn@doi [\apj] {10.1088/0004-637X/755/1/26}, \href
  {https://ui.adsabs.harvard.edu/abs/2012ApJ...755...26O} {755, 26}

\bibitem[\protect\citeauthoryear{{Orofino}, {Ferrara}  \&
  {Gallerani}}{{Orofino} et~al.}{2018}]{Orofino2018}
{Orofino} M.~C.,  {Ferrara} A.,   {Gallerani} S.,  2018, \mn@doi [\mnras]
  {10.1093/mnras/sty1482}, \href
  {https://ui.adsabs.harvard.edu/abs/2018MNRAS.480..681O} {480, 681}

\bibitem[\protect\citeauthoryear{{Padovani} et~al.,}{{Padovani}
  et~al.}{2017}]{Padovani2017}
{Padovani} P.,  et~al., 2017, \mn@doi [\aapr] {10.1007/s00159-017-0102-9},
  \href {https://ui.adsabs.harvard.edu/abs/2017A&ARv..25....2P} {25, 2}

\bibitem[\protect\citeauthoryear{{Page} et~al.,}{{Page}
  et~al.}{1996}]{Page1996}
{Page} M.~J.,  et~al., 1996, \mn@doi [\mnras] {10.1093/mnras/281.2.579}, \href
  {https://ui.adsabs.harvard.edu/abs/1996MNRAS.281..579P} {281, 579}

\bibitem[\protect\citeauthoryear{{Park}, {Kashyap}, {Siemiginowska}, {van Dyk},
  {Zezas}, {Heinke}  \& {Wargelin}}{{Park} et~al.}{2006}]{Park2006}
{Park} T.,  {Kashyap} V.~L.,  {Siemiginowska} A.,  {van Dyk} D.~A.,  {Zezas}
  A.,  {Heinke} C.,   {Wargelin} B.~J.,  2006, \mn@doi [\apj] {10.1086/507406},
  \href {https://ui.adsabs.harvard.edu/abs/2006ApJ...652..610P} {652, 610}

\bibitem[\protect\citeauthoryear{{Peca} et~al.,}{{Peca}
  et~al.}{2023}]{Peca2002}
{Peca} A.,  et~al., 2023, \mn@doi [\apj] {10.3847/1538-4357/acac28}, \href
  {https://ui.adsabs.harvard.edu/abs/2023ApJ...943..162P} {943, 162}

\bibitem[\protect\citeauthoryear{{Pickles}}{{Pickles}}{1998}]{Pickles1998}
{Pickles} A.~J.,  1998, \mn@doi [\pasp] {10.1086/316197}, \href
  {https://ui.adsabs.harvard.edu/abs/1998PASP..110..863P} {110, 863}

\bibitem[\protect\citeauthoryear{{Polletta} et~al.,}{{Polletta}
  et~al.}{2007}]{Polletta2007}
{Polletta} M.,  et~al., 2007, \mn@doi [\apj] {10.1086/518113}, \href
  {https://ui.adsabs.harvard.edu/abs/2007ApJ...663...81P} {663, 81}

\bibitem[\protect\citeauthoryear{{Pouliasis} et~al.,}{{Pouliasis}
  et~al.}{2024}]{Pouliasis2024}
{Pouliasis} E.,  et~al., 2024, \mn@doi [\aap] {10.1051/0004-6361/202348479},
  \href {https://ui.adsabs.harvard.edu/abs/2024A&A...685A..97P} {685, A97}

\bibitem[\protect\citeauthoryear{{Rees}}{{Rees}}{1984}]{Rees1984}
{Rees} M.~J.,  1984, \mn@doi [\araa] {10.1146/annurev.aa.22.090184.002351},
  \href {https://ui.adsabs.harvard.edu/abs/1984ARA&A..22..471R} {22, 471}

\bibitem[\protect\citeauthoryear{{Reines} \& {Comastri}}{{Reines} \&
  {Comastri}}{2016}]{Reines2016}
{Reines} A.~E.,  {Comastri} A.,  2016, \mn@doi [\pasa] {10.1017/pasa.2016.46},
  \href {https://ui.adsabs.harvard.edu/abs/2016PASA...33...54R} {33, e054}

\bibitem[\protect\citeauthoryear{{Ren} \& {Trenti}}{{Ren} \&
  {Trenti}}{2021}]{Ren2021}
{Ren} K.,  {Trenti} M.,  2021, \mn@doi [\apj] {10.3847/1538-4357/ac2e02}, \href
  {https://ui.adsabs.harvard.edu/abs/2021ApJ...923..110R} {923, 110}

\bibitem[\protect\citeauthoryear{{Reynolds} et~al.,}{{Reynolds}
  et~al.}{2023}]{Reynolds2023}
{Reynolds} C.~S.,  et~al., 2023, in {Siegmund} O.~H.,  {Hoadley} K.,  eds,
  Society of Photo-Optical Instrumentation Engineers (SPIE) Conference Series
  Vol. 12678, UV, X-Ray, and Gamma-Ray Space Instrumentation for Astronomy
  XXIII. p. 126781E (\mn@eprint {arXiv} {2311.00780}),
  \mn@doi{10.1117/12.2677468}

\bibitem[\protect\citeauthoryear{{Ross} et~al.,}{{Ross}
  et~al.}{2013}]{Ross2013}
{Ross} N.~P.,  et~al., 2013, \mn@doi [\apj] {10.1088/0004-637X/773/1/14}, \href
  {https://ui.adsabs.harvard.edu/abs/2013ApJ...773...14R} {773, 14}

\bibitem[\protect\citeauthoryear{{Salvato} et~al.,}{{Salvato}
  et~al.}{2009}]{Salvato2009}
{Salvato} M.,  et~al., 2009, \mn@doi [\apj] {10.1088/0004-637X/690/2/1250},
  \href {https://ui.adsabs.harvard.edu/abs/2009ApJ...690.1250S} {690, 1250}

\bibitem[\protect\citeauthoryear{{Salvato} et~al.,}{{Salvato}
  et~al.}{2011}]{Salvato2011}
{Salvato} M.,  et~al., 2011, \mn@doi [\apj] {10.1088/0004-637X/742/2/61}, \href
  {https://ui.adsabs.harvard.edu/abs/2011ApJ...742...61S} {742, 61}

\bibitem[\protect\citeauthoryear{{Salvato} et~al.,}{{Salvato}
  et~al.}{2018}]{Salvato2018}
{Salvato} M.,  et~al., 2018, \mn@doi [\mnras] {10.1093/mnras/stx2651}, \href
  {https://ui.adsabs.harvard.edu/abs/2018MNRAS.473.4937S} {473, 4937}

\bibitem[\protect\citeauthoryear{{Sanders} et~al.,}{{Sanders}
  et~al.}{2007}]{Sanders2007}
{Sanders} D.~B.,  et~al., 2007, \mn@doi [\apjs] {10.1086/517885}, \href
  {https://ui.adsabs.harvard.edu/abs/2007ApJS..172...86S} {172, 86}

\bibitem[\protect\citeauthoryear{{Sawicki} et~al.,}{{Sawicki}
  et~al.}{2019}]{Sawicki2019}
{Sawicki} M.,  et~al., 2019, \mn@doi [\mnras] {10.1093/mnras/stz2522}, \href
  {https://ui.adsabs.harvard.edu/abs/2019MNRAS.489.5202S} {489, 5202}

\bibitem[\protect\citeauthoryear{{Schlafly} \& {Finkbeiner}}{{Schlafly} \&
  {Finkbeiner}}{2011}]{Schlafly2011}
{Schlafly} E.~F.,  {Finkbeiner} D.~P.,  2011, \mn@doi [\apj]
  {10.1088/0004-637X/737/2/103}, \href
  {https://ui.adsabs.harvard.edu/abs/2011ApJ...737..103S} {737, 103}

\bibitem[\protect\citeauthoryear{{Schmidt}}{{Schmidt}}{1968}]{Schmidt1968}
{Schmidt} M.,  1968, \mn@doi [\apj] {10.1086/149446}, \href
  {https://ui.adsabs.harvard.edu/abs/1968ApJ...151..393S} {151, 393}

\bibitem[\protect\citeauthoryear{{Schmidt}, {Schneider}  \& {Gunn}}{{Schmidt}
  et~al.}{1995}]{Schmidt1995}
{Schmidt} M.,  {Schneider} D.~P.,   {Gunn} J.~E.,  1995, \mn@doi [\aj]
  {10.1086/117497}, \href
  {https://ui.adsabs.harvard.edu/abs/1995AJ....110...68S} {110, 68}

\bibitem[\protect\citeauthoryear{{Scholtz} et~al.,}{{Scholtz}
  et~al.}{2025}]{Scholtz2023}
{Scholtz} J.,  et~al., 2025, \mn@doi [\aap] {10.1051/0004-6361/202348804},
  \href {https://ui.adsabs.harvard.edu/abs/2025A&A...697A.175S} {697, A175}

\bibitem[\protect\citeauthoryear{{Scoville} et~al.,}{{Scoville}
  et~al.}{2007}]{Scoville2007}
{Scoville} N.,  et~al., 2007, \mn@doi [\apjs] {10.1086/516585}, \href
  {https://ui.adsabs.harvard.edu/abs/2007ApJS..172....1S} {172, 1}

\bibitem[\protect\citeauthoryear{{Sharma} \& {Sharma}}{{Sharma} \&
  {Sharma}}{2024}]{Sharma2024}
{Sharma} R.,  {Sharma} M.,  2024, \mn@doi [\mnras] {10.1093/mnras/stae1007},
  \href {https://ui.adsabs.harvard.edu/abs/2024MNRAS.531.3287S} {531, 3287}

\bibitem[\protect\citeauthoryear{{Shen}, {Hopkins}, {Faucher-Gigu{\`e}re},
  {Alexander}, {Richards}, {Ross}  \& {Hickox}}{{Shen} et~al.}{2020}]{Shen2020}
{Shen} X.,  {Hopkins} P.~F.,  {Faucher-Gigu{\`e}re} C.-A.,  {Alexander} D.~M.,
  {Richards} G.~T.,  {Ross} N.~P.,   {Hickox} R.~C.,  2020, \mn@doi [\mnras]
  {10.1093/mnras/staa1381}, \href
  {https://ui.adsabs.harvard.edu/abs/2020MNRAS.495.3252S} {495, 3252}

\bibitem[\protect\citeauthoryear{{Shuntov} et~al.,}{{Shuntov}
  et~al.}{2022}]{Shuntov2022}
{Shuntov} M.,  et~al., 2022, \mn@doi [\aap] {10.1051/0004-6361/202243136},
  \href {https://ui.adsabs.harvard.edu/abs/2022A&A...664A..61S} {664, A61}

\bibitem[\protect\citeauthoryear{{Sillassen} et~al.,}{{Sillassen}
  et~al.}{2022}]{Sillassen2022}
{Sillassen} N.~B.,  et~al., 2022, \mn@doi [\aap] {10.1051/0004-6361/202244661},
  \href {https://ui.adsabs.harvard.edu/abs/2022A&A...665L...7S} {665, L7}

\bibitem[\protect\citeauthoryear{{Somerville} \& {Dav{\'e}}}{{Somerville} \&
  {Dav{\'e}}}{2015}]{Somerville2015}
{Somerville} R.~S.,  {Dav{\'e}} R.,  2015, \mn@doi [\araa]
  {10.1146/annurev-astro-082812-140951}, \href
  {https://ui.adsabs.harvard.edu/abs/2015ARA&A..53...51S} {53, 51}

\bibitem[\protect\citeauthoryear{{Szalay}, {Connolly}  \& {Szokoly}}{{Szalay}
  et~al.}{1999}]{Szalay1999}
{Szalay} A.~S.,  {Connolly} A.~J.,   {Szokoly} G.~P.,  1999, \mn@doi [\aj]
  {10.1086/300689}, \href
  {https://ui.adsabs.harvard.edu/abs/1999AJ....117...68S} {117, 68}

\bibitem[\protect\citeauthoryear{{Taniguchi} et~al.,}{{Taniguchi}
  et~al.}{2015}]{Taniguchi2015}
{Taniguchi} Y.,  et~al., 2015, \mn@doi [\pasj] {10.1093/pasj/psv106}, \href
  {https://ui.adsabs.harvard.edu/abs/2015PASJ...67..104T} {67, 104}

\bibitem[\protect\citeauthoryear{Taylor}{Taylor}{2013}]{Taylor2013}
Taylor M.,  2013, Starlink User Note, 253

\bibitem[\protect\citeauthoryear{{Terrazas}, {Aird}  \& {Coil}}{{Terrazas}
  et~al.}{2024}]{Terrazas2024}
{Terrazas} B.~A.,  {Aird} J.,   {Coil} A.~L.,  2024, \mn@doi [arXiv e-prints]
  {10.48550/arXiv.2411.08838}, \href
  {https://ui.adsabs.harvard.edu/abs/2024arXiv241108838T} {p. arXiv:2411.08838}

\bibitem[\protect\citeauthoryear{{Trakhtenbrot} et~al.,}{{Trakhtenbrot}
  et~al.}{2016}]{Trakhtenbrot2016}
{Trakhtenbrot} B.,  et~al., 2016, \mn@doi [\apj] {10.3847/0004-637X/825/1/4},
  \href {https://ui.adsabs.harvard.edu/abs/2016ApJ...825....4T} {825, 4}

\bibitem[\protect\citeauthoryear{{Trinca}, {Schneider}, {Valiante}, {Graziani},
  {Ferrotti}, {Omukai}  \& {Chon}}{{Trinca} et~al.}{2024}]{Trinca2024}
{Trinca} A.,  {Schneider} R.,  {Valiante} R.,  {Graziani} L.,  {Ferrotti} A.,
  {Omukai} K.,   {Chon} S.,  2024, \mn@doi [\mnras] {10.1093/mnras/stae651},
  \href {https://ui.adsabs.harvard.edu/abs/2024MNRAS.529.3563T} {529, 3563}

\bibitem[\protect\citeauthoryear{{Ueda}, {Akiyama}, {Hasinger}, {Miyaji}  \&
  {Watson}}{{Ueda} et~al.}{2014}]{Ueda2014}
{Ueda} Y.,  {Akiyama} M.,  {Hasinger} G.,  {Miyaji} T.,   {Watson} M.~G.,
  2014, \mn@doi [\apj] {10.1088/0004-637X/786/2/104}, \href
  {https://ui.adsabs.harvard.edu/abs/2014ApJ...786..104U} {786, 104}

\bibitem[\protect\citeauthoryear{{Virtanen} et~al.,}{{Virtanen}
  et~al.}{2020}]{Virtanen2020}
{Virtanen} P.,  et~al., 2020, \mn@doi [Nature Methods]
  {10.1038/s41592-019-0686-2}, \href
  {https://ui.adsabs.harvard.edu/abs/2020NatMe..17..261V} {17, 261}

\bibitem[\protect\citeauthoryear{{Vito}, {Gilli}, {Vignali}, {Comastri},
  {Brusa}, {Cappelluti}  \& {Iwasawa}}{{Vito} et~al.}{2014}]{Vito2014}
{Vito} F.,  {Gilli} R.,  {Vignali} C.,  {Comastri} A.,  {Brusa} M.,
  {Cappelluti} N.,   {Iwasawa} K.,  2014, \mn@doi [\mnras]
  {10.1093/mnras/stu2004}, \href
  {https://ui.adsabs.harvard.edu/abs/2014MNRAS.445.3557V} {445, 3557}

\bibitem[\protect\citeauthoryear{{Vito} et~al.,}{{Vito}
  et~al.}{2018}]{Vito2018}
{Vito} F.,  et~al., 2018, \mn@doi [\mnras] {10.1093/mnras/stx2486}, \href
  {https://ui.adsabs.harvard.edu/abs/2018MNRAS.473.2378V} {473, 2378}

\bibitem[\protect\citeauthoryear{{Weaver} et~al.,}{{Weaver}
  et~al.}{2022}]{Weaver2022}
{Weaver} J.~R.,  et~al., 2022, \mn@doi [\apjs] {10.3847/1538-4365/ac3078},
  \href {https://ui.adsabs.harvard.edu/abs/2022ApJS..258...11W} {258, 11}

\bibitem[\protect\citeauthoryear{{Weaver}, {Zalesky}, {Allen}  \&
  {Taamoli}}{{Weaver} et~al.}{2023}]{Weaver2023}
{Weaver} J.,  {Zalesky} L.,  {Allen} N.,   {Taamoli} S.,  2023, {The Farmer:
  Photometry routines for deep multi-wavelength galaxy surveys}, Astrophysics
  Source Code Library, record ascl:2312.016

\bibitem[\protect\citeauthoryear{{Weisskopf}, {Brinkman}, {Canizares},
  {Garmire}, {Murray}  \& {Van Speybroeck}}{{Weisskopf}
  et~al.}{2002}]{Weisskopf2002}
{Weisskopf} M.~C.,  {Brinkman} B.,  {Canizares} C.,  {Garmire} G.,  {Murray}
  S.,   {Van Speybroeck} L.~P.,  2002, \mn@doi [\pasp] {10.1086/338108}, \href
  {https://ui.adsabs.harvard.edu/abs/2002PASP..114....1W} {114, 1}

\bibitem[\protect\citeauthoryear{{Wise}}{{Wise}}{1997}]{Wise1997}
{Wise} M.,  1997, Chandra News, \href
  {https://ui.adsabs.harvard.edu/abs/1997ChNew...5...22W} {5, 22}

\bibitem[\protect\citeauthoryear{{Wolf} et~al.,}{{Wolf}
  et~al.}{2021}]{Wolf2021}
{Wolf} J.,  et~al., 2021, \mn@doi [\aap] {10.1051/0004-6361/202039724}, \href
  {https://ui.adsabs.harvard.edu/abs/2021A&A...647A...5W} {647, A5}

\bibitem[\protect\citeauthoryear{{Yang} et~al.,}{{Yang}
  et~al.}{2021}]{Yang2021}
{Yang} J.,  et~al., 2021, \mn@doi [\apj] {10.3847/1538-4357/ac2b32}, \href
  {https://ui.adsabs.harvard.edu/abs/2021ApJ...923..262Y} {923, 262}

\bibitem[\protect\citeauthoryear{{Yang} et~al.,}{{Yang}
  et~al.}{2023}]{Yang2023}
{Yang} G.,  et~al., 2023, \mn@doi [\apjl] {10.3847/2041-8213/acd639}, \href
  {https://ui.adsabs.harvard.edu/abs/2023ApJ...950L...5Y} {950, L5}

\bibitem[\protect\citeauthoryear{{Yue} et~al.,}{{Yue} et~al.}{2024}]{Yue2024}
{Yue} M.,  et~al., 2024, \mn@doi [\apj] {10.3847/1538-4357/ad3914}, \href
  {https://ui.adsabs.harvard.edu/abs/2024ApJ...966..176Y} {966, 176}

\bibitem[\protect\citeauthoryear{{Zhang}, {Behroozi}, {Volonteri}, {Silk},
  {Fan}, {Hopkins}, {Yang}  \& {Aird}}{{Zhang} et~al.}{2023}]{Zhang2023}
{Zhang} H.,  {Behroozi} P.,  {Volonteri} M.,  {Silk} J.,  {Fan} X.,  {Hopkins}
  P.~F.,  {Yang} J.,   {Aird} J.,  2023, \mn@doi [\mnras]
  {10.1093/mnras/stac2633}, \href
  {https://ui.adsabs.harvard.edu/abs/2023MNRAS.518.2123Z} {518, 2123}

\bibitem[\protect\citeauthoryear{{Zhu}, {Li}, {Li}, {Maji}, {Yajima},
  {Schneider}  \& {Hernquist}}{{Zhu} et~al.}{2022}]{Zhu2020}
{Zhu} Q.,  {Li} Y.,  {Li} Y.,  {Maji} M.,  {Yajima} H.,  {Schneider} R.,
  {Hernquist} L.,  2022, \mn@doi [\mnras] {10.1093/mnras/stac1556}, \href
  {https://ui.adsabs.harvard.edu/abs/2022MNRAS.514.5583Z} {514, 5583}

\bibitem[\protect\citeauthoryear{{Zubovas} \& {King}}{{Zubovas} \&
  {King}}{2021}]{Zubovas2021}
{Zubovas} K.,  {King} A.,  2021, \mn@doi [\mnras] {10.1093/mnras/stab004},
  \href {https://ui.adsabs.harvard.edu/abs/2021MNRAS.501.4289Z} {501, 4289}

\bibitem[\protect\citeauthoryear{{de Graaff} et~al.,}{{de Graaff}
  et~al.}{2025}]{Graaff2024}
{de Graaff} A.,  et~al., 2025, \mn@doi [\aap] {10.1051/0004-6361/202452186},
  \href {https://ui.adsabs.harvard.edu/abs/2025A&A...697A.189D} {697, A189}

\bibitem[\protect\citeauthoryear{{van der Walt}, {Colbert}  \&
  {Varoquaux}}{{van der Walt} et~al.}{2011}]{VanDerWalt2011}
{van der Walt} S.,  {Colbert} S.~C.,   {Varoquaux} G.,  2011, \mn@doi
  [Computing in Science and Engineering] {10.1109/MCSE.2011.37}, \href
  {https://ui.adsabs.harvard.edu/abs/2011CSE....13b..22V} {13, 22}

\makeatother
\end{thebibliography}
\end{document}